\documentclass[%
 aip,
 amsmath,amssymb,
 reprint,%
]{revtex4-1}

\usepackage{graphicx}
\usepackage{dcolumn}
\usepackage{bm}

\usepackage[utf8]{inputenc}
\usepackage[T1]{fontenc}
\usepackage{mathptmx}

\newcommand{\Zn}{\text{Z}}

\newcommand{\Hn}{\text{H}}
\newcommand{\gn}{\text{g}}
\newcommand{\mx}{\text{max}}

\newcommand{\vv}{\text{v}}

\newcommand{\ef}{\text{eff}}

\newcommand{\Qn}{\text{Q}}

\usepackage{url}
\usepackage[version=3]{mhchem}
\usepackage{color}
\usepackage{amsmath}
\usepackage{lipsum}
\usepackage{amssymb}
\usepackage[version=3]{mhchem}
\usepackage{mathrsfs}      
\usepackage{booktabs}   
\usepackage{subfigure}
\usepackage{float}
\usepackage{appendix}
\usepackage{comment}

\usepackage{epsfig}
\usepackage{dcolumn}%
\usepackage{relsize}

\begin{document}

\title{Consistent Kinetic-Continuum Recombination Model for High Temperature Reacting Flows} 

\author{Narendra Singh}
 \email{singh455@umn.edu.}
\author{Thomas Schwartzentruber}%
\affiliation{ 
Department of Aerospace Engineering and Mechanics, 
University of Minnesota, Minneapolis, MN 55455}%


\date{\today}

\begin{abstract}
A recombination reaction model for high-temperature chemical kinetics is derived from \textit{ab initio} simulations data. 
A kinetic recombination rate model is derived using a recently developed \textit{ab initio} state-specific dissociation model \cite{singh2019consistentI} and the principle of microscopic reversibility. When atoms recombine, the kinetic rate model shows that product molecules have high favoring for high vibrational energy states.  A continuum recombination rate model is then derived analytically from the kinetic recombination rate model. Similarly, the expression for the average vibrational energy of recombining molecules is also derived analytically. Finally, a simple model for non-Boltzmann vibrational energy distribution functions is derived. The distribution model includes both depletion of energy states due to dissociation and re-population of states due to recombination where a Boltzmann distribution is recovered in chemical equilibrium.
Isothermal relaxation simulations using the continuum dissociation and recombination model are performed and the results are compared with the state-of-the-art model.


 

\end{abstract}

\maketitle

\section{Introduction}
In hypersonic flows, recombination of the shock-heated dissociated gas alters the species concentrations which affect surface chemical reactions and heating rates. The chemical energy release to the bulk fluid in a recombination reaction depends on the specific internal energy state of the formed molecule. The state-specificity of recombination rates also affects the rovibrational population of the gas. For instance, recombination reactions re-populate the rovibrational states which are depleted by dissociation. This indirectly affects other physical processes that are coupled to the internal energy of the gas, for instance, dissociation, which has strong favoring for higher vibrational energy. Therefore, accurate rate and energy transfer modeling for recombination reactions is crucial for predictive simulations.

Recombination models used in computational fluid dynamics (CFD) solvers are obtained via the principle of detailed balance, requiring \textit{equilibrium} dissociation rates (typically those from Park model\cite{park1988two,park1993review, park1989assessment} ). The Park model uses an effective temperature ($\sqrt{T T_v}$) in Arrhenius rates, where $T$ is translational temperature of the gas and $T_v$ is an additional `vibrational' temperature to couple dissociation rates to the vibrational energy of the gas. The parameters in the Park model are empirically adjusted to fit experimental results \cite{byron1966shock,appleton1968shock,hanson1972shock}, which have large uncertainties (about an order of magnitude for nitrogen dissociation rates). For estimating recombination rates using the principle of detailed balance, equilibrium dissociation rates are obtained by setting $T=T_v$ in the Park model. However, this does not necessarily correspond to equilibrium, because shock-heated dissociating gas is inherently in nonequilibrium \cite{valentini2016dynamics,valentini2015N4,grover2019jtht,grover2019direct,panesi2013rovibrational,jaffe2018comparison}, and therefore experimental measurements, upon which the Park model is based, may correspond to an arbitrary nonequilibrum state. More specifically, the nonequilibrium state may be a quasi-steady state (QSS), where high rovibrational energy states are depleted relative to the corresponding Boltzmann distribution. This could result in under-predicted recombination rates, since dissociation rates corresponding to QSS are lower than  corresponding equilibrium rates.

Recently, accurate potential energy surfaces (PESs) have been developed for the purpose of studying air chemistry relevant to hypersonic flows, for instance N$_2$-N$_2$ and N–N$_2$ collisions  \cite{paukku2013global,paukku2014erratum}, O$_2$–O$_2$ \cite{paukku2017potential} and O-O$_2$ \cite{O2OTruhlar} collisions, N$_2$-O$_2$ collisions \cite{varga2016potential}, and N$_2$–O collisions \cite{lin2016global}.
Using these PESs, \textit{ab initio} methods such as direct molecular simulations (DMS) \cite{valentini2016dynamics,grover2019direct}, master-equation analysis \cite{panesi2014pre,macdonald2018_QCT, andrienko2018vibrational,magin2012coarse} and quasi-classical trajectory calculations (QCT) \cite{chaudhry2018qct,voelkel2017multitemperature}  have quantified the coupling of ro-vibrational energy to the state-specific dissociation rates of air species. 
Based on the obtained \textit{ab initio} data, a nonequilibrium dissociation model has been derived using first principles by the authors in Refs.~\cite{singh2019nonboltzmann,singh2019consistentI,singh2019consistentII,singh2020aiaanon,singh2020aiaaconsistent,Singhpnas}. The dissociation model is analytically consistent between the kinetic-scale and continuum scale \cite{singh2019consistentI}.  The continuum model incorporates the non-Boltzmann vibrational energy distribution model\cite{singh2019nonboltzmann} for dissociating gas and is shown to reproduce \textit{ab initio} data at continuum scale \cite{singh2019consistentII}. The model can be implemented in large-scale CFD solvers and direct simulation Monte Carlo (DSMC) solvers~\cite{singh2019consistentI,singh2019consistentII}. Other recent efforts  of developing a continuum nonequilibrium dissociation model based on recent \textit{ab initio} data can be found in Refs.~\cite{andrienko2015high,kustova2016advanced,chaudhry2020implementation,chaudhry2020vehicle}.
In this work, we extend the modeling efforts to include recombination reactions. 

In this article, an analytically consistent recombination model, using microscopic reversibility and the principle of detailed balance is developed. First, an analytical model for the probability of recombining into a specific rotational and vibrational energy state is developed using the state-specific dissociation rate model \cite{singh2019consistentI}. Kinetic rate of recombination is shown to have strong favoring for high vibrational energy. This is also true for rotational energy, however, the probability of recombining to form quasi-bound molecules reduces for higher internal energy. A model for the average vibrational energy of molecules in recombination is also derived and its connection to average vibrational energy of dissociating molecules is highlighted. Using the principle of detailed balance, an analytical recombination rate model, consistent with the dissociation rate model, is derived.  

As a final modeling step, an extension of the non-Boltzmann distribution model for dissociating gas to include the re-population of high-energy states due to recombination reactions is derived. The model extension is developed using the surprisal \cite{levine1978information,levine2009molecular,levine1971collision} analysis distribution model for dissociation. The derivation uses only microscopic reversibility, master-equation, and does not require any additional adjustable parameters. Incorporating the extension due to recombination in the generalized non-Boltzmann distribution model proposed by the authors in Ref.~\cite{singh2019nonboltzmann}, now recovers the Boltzmann distribution under equilibrium. Finally, the recombination model is then analyzed using isothermal relaxation simulations and results are compared with the Park two temperature model. 

\section{Theory and Background}
Consider rovibrational relaxation, dissociation and recombination reactions in an ensemble of gas consisting of diatomic molecules AB colliding with partner C. Collisional process considered in this work include the following:
 \begin{equation}
 \begin{split}
     AB (j,v) + C \xrightarrow{k_{jv-j'v'}}  AB(j',v')+C \\
     AB (j,v) + C \xrightarrow{k_{jv-d}}  A+B+C \\
     A+B+C  \xrightarrow{k_{r-jv}}  AB (j,v) + C
     \end{split}
 \end{equation}
 Here $(j,v)$ represents the rovibrational quantum state of the molecule and $C$ can be either a diatom or atom.  $k_{jv-d}$ is the dissociation rate constant of molecules from state $(j,v)$.
 $k_{r-jv}$ is the recombination rate constant of two atoms recombining in to a state $(j,v)$.  An evolution equation for the rovibrational energy populations, $[AB(i)]$ at a given translational temperature ($T$), can be written as:
   \begin{equation}
   \begin{split}
    \frac{ d\ [AB(i)] }{dt} =  \sum_{i'\neq i} k_{i'-i}  [AB(i')][C] 
    - \left ( \sum_{i'\neq i} k_{i-i'} \right)[C] [AB(i)] \\ -  k_{i-d} [C] [AB(i)] +  k_{r-i} [C] [A] [B] 
     \label{master_eqn_recomb11}
     \end{split}
 \end{equation}
 Here, $k_{i-i'} $ is the rate constant for transitioning from state $i (\equiv (j,v)) $ to $i' (\equiv (j',v'))$ during a collision with partner species $C$, $d$ denotes the dissociated state and $r-i$ denotes recombination to a state $i$. $k_{i-i'} (\equiv k_{i-i'} (T))$ depends on $T$ but for notational brevity, we have dropped the dependence. In principle one may employ master equation and solve evolution equations for the ro-vibrational population in each ($j,v$) state. However solving the full set of master equations requires computation of large numbers of transition rates (for instance $10^{15}$ for N$_2$-N$_2$ collisions) and therefore a coarse-grained description is necessary. A possible reduced order model is to bin states into groups  \cite{macdonald2018_QCT,macdonald2018construction_DMS} and track evolution of population within the groups. 
 
A more coarse grained approach is called the `multi-temperature framework', commonly employed in continuum-scale simulations. Here, evolution equations for macroscopic quantities, more specifically species concentration ($[AB]$), average vibrational energy ($\langle \epsilon_v \rangle$), and average rotational energy ($\langle \epsilon_{rot} \rangle$), are solved. Time evolution equations  for these quantities, under suitable approximations, can be obtained as: 
 \begin{equation}
    \frac{ d [AB]}{dt} = -k_{AB-C} [AB][C]+\bm{k_{r_{AB-C}}} [A][B][C]
     \label{Rate_eqn_general_recomb}
 \end{equation}

\begin{equation}
\begin{split}
  \frac{ d \langle \epsilon_v \rangle}{dt} = \frac{\langle \epsilon_v^* \rangle-\langle  \epsilon_v \rangle}{\tau_{\text{mix,v}}} - k_{AB-C} [C] (\langle \epsilon_v^d \rangle - \langle \epsilon_v \rangle) \\ + \bm{k_{r_{AB-C}}} \cfrac{[A][B][C]}{[AB]} (\bm{\langle \epsilon_v^{rec} \rangle}  - \langle \epsilon_v \rangle)
    \end{split}
     \label{LandauTeller_modified_general_recomb}
 \end{equation}
 
 \begin{equation}
 \begin{split}
    \frac{ d \langle \epsilon_{rot} \rangle}{dt} = \frac{\langle \epsilon_{rot}^* \rangle-\langle  \epsilon_{rot} \rangle}{\tau_{\text{mix,j}}} - k_{AB-C} [C] (\langle \epsilon_{rot}^d \rangle - \langle \epsilon_{rot} \rangle) \\ + \bm{k_{r_{AB-C}}} \cfrac{[A][B][C]}{[AB]} (\bm{\langle \epsilon_{rot}^{rec} \rangle}  - \langle \epsilon_{rot} \rangle)
    \end{split}
     \label{Jeans_modified_general_recomb}
 \end{equation}
 where $\langle \epsilon_{...}^* \rangle$ are corresponding average energies at  equilibrium. $k_{AB-C}$ is the dissociation rate constant, $\langle \epsilon_v^d \rangle$ is the average vibrational energy of dissociating molecules, and $\langle \epsilon_{rot}^d \rangle$ is the average rotational energy of dissociating molecules. 
 In the Landau-Teller (first term in the RHS of Eq.~\ref{LandauTeller_modified_general_recomb}) and Jeans equation (first term in the RHS of Eq.~\ref{Jeans_modified_general_recomb}) terms, $\tau_{\text{mix,v}}$ is the mixture vibrational relaxation time constant and $\tau_{\text{mix,rot}}$ is the mixture rotational relaxation time constant given by \cite{Lee1984}:
\begin{equation}
    \tau_{mix,i} = \cfrac{[AB] +[C]}{ \cfrac{[AB]}{\tau_{AB-AB,i}}+\cfrac{[C]}{\tau_{AB-C, i}}}
\end{equation}
 where $\tau_{AB-C, i}$ is the relaxation time constant due to collision of $AB$ and $C$ and $i$ refers to either rotation or vibration.  $k_{AB-C}$ is related to the state-specific dissociation rate constant ($k_{jv-d}$):
\begin{equation}
\begin{split}
    k_{AB-C} =  \sum _{v=0}^{v_{\max}} \sum _{j=0}^{j_{\max}(v)}  k_{jv-d} f(j,v)
    \end{split}
\end{equation}
 Similarly, $\langle \epsilon_v^d \rangle$ and $\langle \epsilon_{rot}^d \rangle$ are related to state-specific quantities by the following mathematical moment equations:
 \begin{equation}
\begin{split}
       \langle \epsilon_v^d \rangle = \cfrac{\sum _{v=0}^{v_{\max}} \sum _{j=0}^{j_{\max}(v)}  \epsilon_{int}(0,v) k_{jv-d} f(j,v)}{\sum _{v=0}^{v_{\max}} \sum _{j=0}^{j_{\max}(v)}  k_{jv-d} f(j,v)}
    \\
      \langle \epsilon_{rot}^d \rangle = \cfrac{\sum _{v=0}^{v_{\max}} \sum _{j=0}^{j_{\max}(v)}  (\epsilon_{int}(j,v) -\epsilon_{int}(0,v)) k_{jv-d} f(j,v)}{\sum _{v=0}^{v_{\max}} \sum _{j=0}^{j_{\max}(v)}  k_{jv-d} f(j,v)}
    \end{split}
\end{equation}
 where $\epsilon_{int}(0,v)$ is the vibrational energy corresponding to a state $(j,v)$, obtained using the vibration prioritized framework\cite{jaffe1987}, commonly used in the recent literature \cite{Singhpnas,macdonald2018_QCT,grover2019jtht}. The framework provides an approximate way to separate otherwise coupled rotational and vibrational parts of internal energy. Using this framework, rotational energy for the state $(j,v)$ is approximated as $\epsilon_{int}(j,v)-\epsilon_{int}(0,v)$. Further discussion of Eqs.~\ref{Rate_eqn_general_recomb}--\ref{Jeans_modified_general_recomb} and the links to master equation (Eq.~\ref{master_eqn_recomb11}) can be found in Ref.~\cite{singh2019consistentI}.

  Expressions for $k_{AB-C}$, $\langle \epsilon_v^d \rangle$ and $\langle \epsilon_{rot}^d \rangle$ were recently developed in Ref.~\cite{singh2019consistentI} from the state-specific rates consistent with \textit{ab initio} data. The expressions for these quantities were derived using generalized non-Boltzmann distributions ($f(j,v)$) developed in Ref.~\cite{singh2019nonboltzmann}.
    Detailed comparison of predictions using Eqs.~\ref{Rate_eqn_general_recomb}--\ref{Jeans_modified_general_recomb} with Direct Molecular Simulation (equivalent to full master equation) results have been carried out in Ref.~\cite{singh2019consistentII}. However, these studies do not model recombination processes.
    
    Specifically, recombination terms (written in bold face in  Eqs.~\ref{Rate_eqn_general_recomb}--\ref{Jeans_modified_general_recomb}) have not been considered in the earlier work \cite{singh2019consistentI,singh2019consistentII}. The focus of this article, therefore, is to derive the recombination rate constant ($k_{r_{AB-C}}$), as well as the  average vibrational ($\langle \epsilon_{v}^{rec} \rangle$) and average rotational energy ($\langle \epsilon_{rot}^{rec} \rangle$) of molecules formed in recombination reactions.

In terms of the organisation of the article, Sec.~III presents the governing equations required for the derivation of the kinetic recombination rate and the average vibrational and rotational energy of the molecules formed in recombination reactions. Section IV develops the analytical expressions for the kinetic recombination rate, continuum scale recombination rate, and average vibrational energy of the recombining molecules. 
Section V presents the analysis of the kinetic recombination rates. 
Section VI presents a generalized non-Boltzmann internal energy distribution model ($f(j,v)$) that includes re-population of $(j,v)$ states due to recombination.
Section VII presents results and discussion of isothermal relaxation simulations carried out using the consistent kinetic-continuum dissociation and recombination models. 
Finally, the summary and conclusions of the article are presented in Sec.~VII.

\section{Kinetic Framework to Continuum Framework}
In this section we present the framework that will enable us to derive the recombination model in Section IV.
\subsection{Recombination Rate}
 The microscopic reversibility relation can be used to relate a state-specific rate ($k_{r-jv}$) of recombining into a quantum state ($j,v$) to the state-specific dissociation rate ($k_{jv-d}$) from the state ($j,v$),  in the following manner \cite{tolman1979principles,lewis1925new}:
\begin{equation}
    \cfrac{k_{r-jv}}{k_{jv-d}} = \cfrac{g_{jv} Q_{AB}(T)}{\left[g_{A} g_{B} Q_{A}(T)Q_{B}(T)\right]} \exp\left[ \cfrac{\Delta \epsilon_{f_{AB}} -\epsilon_{int}(j,v)}{k_B T}\right]
    \label{jv_recomb_rate_111}
\end{equation}
 where $g_{jv}$ is the degeneracy of energy level $(j,v)$, $g_A$ and $g_B$ are the degeneracies (nuclear and electronic spin) of the atoms, $\Delta \epsilon_{f_{AB}}$ is the formation energy, $k_B$ is the Boltzmann constant, and $Q_{...}(T)$ denotes translational energy partition functions defined as:
 \begin{equation}
 \begin{split}
      \Qn_{AB}(T) = \left[\cfrac{2 \pi k_B m_{AB} T}{h_p^2}\right]^{3/2}, \\
      \Qn_A(T) = \left[\cfrac{2 \pi k_B m_{A} T}{h_p^2}\right]^{3/2}, \\
      \Qn_B(T) = \left[\cfrac{2 \pi k_B m_{B} T}{h_p^2}\right]^{3/2},
      \label{trans_partition_function}
 \end{split}
 \end{equation}
Following from Eq.~\ref{jv_recomb_rate_111}, the state-specific recombination rate can be written as follows:
\begin{equation}
\begin{split}
   k_{r-jv}  = k_{jv-d} \cfrac{g_{jv} \Qn_{AB}(T)}{\left[g_{A} g_{B} Q_{A}(T)Q_{B}(T)\right]} \exp\left[ \cfrac{\Delta \epsilon_{f_{AB}} -\epsilon_{int}(j,v)}{k_B T}\right]
   \\
     = \cfrac{ \Qn_{AB}(T)}{\left[g_{A} g_{B} Q_{A}(T)Q_{B}(T)\right]} \exp\left[ \cfrac{\Delta \epsilon_{f_{AB}} }{k_B T}\right] g_{jv} k_{jv-d} \exp\left[ -\cfrac{\epsilon_{int}(j,v)}{k_B T}\right]
     \label{k_rjv}
\end{split}   
\end{equation}
where $h_p$ is Planck's constant, $m_{AB}$ is the mass of molecule AB, and $m_C$ is the mass of partner $C$. The continuum-level recombination rate can then be obtained by summing $k_{r-jv}$ over all $(j,v)$ pairs as:
\begin{equation}
\begin{split}
   k_r & = \sum _{v=0}^{v_{\max}}  \sum _{j=0}^{j_{\max}(v)} k_{r-jv} 
   \\
   & = \cfrac{ \Qn_{AB}(T)}{\left[g_{A} g_{B} Q_{A}(T)Q_{B}(T)\right]} \exp\left[ \cfrac{\Delta \epsilon_{f_{AB}} }{k_B T}\right] \\ & \times 
 \sum _{v=0}^{v_{\max}}  
   \sum _{j=0}^{j_{\max}(v)} g_{jv} k_{jv-d} \exp\left[ -\cfrac{\epsilon_{int}(j,v)}{k_B T}\right]
   \\
  & =  \cfrac{ \Qn_{AB}(T) Z(T,T) }{\left[g_{A} g_{B} Q_{A}(T)Q_{B}(T)\right]} \exp\left[ \cfrac{\Delta \epsilon_{f_{AB}} }{k_B T}\right] \\
  & \times \sum _{v=0}^{v_{\max}} 
  \sum _{j=0}^{j_{\max}(v)}  k_{jv-d} \cfrac{g_{jv} \exp\left[ -\cfrac{\epsilon_{int}(j,v)}{k_B T}\right]}{Z(T,T)}
   \\
  & = \cfrac{ \Qn_{AB}(T) Z(T,T) }{\left[g_{A} g_{B} Q_{A}(T)Q_{B}(T)\right]} \exp\left[ \cfrac{\Delta \epsilon_{f_{AB}} }{k_B T}\right] \sum _{v=0}^{v_{\max}}  \sum _{j=0}^{j_{\max}(v)}  k_{jv-d} f_0(j,v) 
   \\
 &  =\cfrac{ \Qn_{AB}(T) Z(T,T) }{\left[g_{A} g_{B} Q_{A}(T)Q_{B}(T)\right]} \exp\left[ \cfrac{\Delta \epsilon_{f_{AB}} }{k_B T}\right] k_d^*(T)
     \label{k_rec_general}
\end{split}   
\end{equation}
Here it is important to note that  $k_d^*$ is the dissociation rate constant \textit{at equilibrium}. As seen in Eq.~\ref{k_rec_general}, $f_0(j,v)$ represents an equilibrium internal energy distribution and, therefore, the sum of state-specific dissociation rates taken over the underlying internal energy population yields:
\begin{equation}
    k_d^* =  \sum _{v=0}^{v_{\max}} \sum _{j=0}^{j_{\max}(v)}  k_{jv-d} f_0(j,v). 
\end{equation}
Finally note that $Z(T,T)$ is the partition function \textit{at equilibrium} for internal energy,   defined as:
\begin{equation}
    Z(T,T) = \sum _{v=0}^{v_{\max}}  \sum _{j=0}^{j_{\max}(v)} g_{jv} \exp\left[ -\cfrac{\epsilon_{int}(j,v)}{k_B T}\right]
\end{equation}
In terms of the equilibrium constant $(K_C)$, Eq.~\ref{k_rec_general} can then be written as:
\begin{equation}
        k_r = \cfrac{k_d^*(T)}{K_C}  
        \label{detailed_balance}
\end{equation}
where $K_c$ is
 \begin{equation}
    \cfrac{1}{K_c} = \cfrac{[AB]^*}{[A]^*[B]^*} =  \cfrac{ \Qn_{AB}(T) Z(T,T) }{\left[g_{A} g_{B} Q_{A}(T)Q_{B}(T)\right]} \exp\left[ \cfrac{\Delta \epsilon_{f_{AB}} }{k_B T}\right]
     \label{equilibrium_constant}
 \end{equation}
Therefore in using microscopic reversibility, it is clear that, recombination rates are only dependent on the translational temperature of the gas and are independent of whether the gas is in equilibrium or not. This has an important implication for continuum recombination modeling. Specifically, recombination rates based on detailed balance (Eq.~\ref{detailed_balance}) require dissociation rates \textit{at equilibrium} ($  k_d^* $). Currently, dissociation rates used in CFD models (i.e the Park models) are empirically fit to experimental data that most likely corresponds to nonequilibrium (more specifically, Quasi-Steady State, QSS, described later in Sec.~VI) conditions. When these nonequilibrium dissociation rates are used to obtain recombination rates using detailed balance (as they currently are in CFD models), this could lead to inaccuracy. Since the new dissociation model is analytical and can easily be evaluated for nonequilibrium conditions and equilibrium conditions, the purpose of this article is to derive a more consistent recombination rate model. 
Before deriving the recombination rate, we first look at the average internal energy of recombining molecules.

\subsection{Average vibrational and rotational energy of molecules formed via recombination}
The average vibrational energy of molecules, $\langle \epsilon_{v}^{rec} \rangle$, formed in the recombination reaction in an ensemble of gas is given by:
\begin{equation}
    \langle \epsilon_{v}^{rec} \rangle = \cfrac{\sum _{v=0}^{v_{\max}}  \sum _{j=0}^{j_{\max}(v)} \epsilon_{int}(0,v) k_{r-jv}}{\sum _{v=0}^{v_{\max}}  \sum _{j=0}^{j_{\max}(v)} k_{r-jv}}
    \label{ev_recomb_eqn}
\end{equation}
Interestingly, it is straightforward to show that the average vibrational energy of molecules formed via recombination, $\langle \epsilon_v^{rec} \rangle(T)$, is same as the average vibrational energy of dissociating molecules $\langle \epsilon_v^d \rangle$ at \textit{equilibrium}. Let us consider the average vibrational energy of dissociating molecules, 
\begin{equation}
    \langle \epsilon_{v}^{d} \rangle = \cfrac{\sum _{v=0}^{v_{\max}}  \sum _{j=0}^{j_{\max}(v)} \epsilon_{int}(0,v) k_{jv-d} f(j,v)}{\sum _{v=0}^{v_{\max}}  \sum _{j=0}^{j_{\max}(v)} k_{jv-d} f(j,v)} 
    \label{ev_avg_diss_eqn}
\end{equation}
where $f(j,v)$ is the general (not necessarily Boltzmann) internal energy distribution of the molecules in the gas.  Let us use the relation in Eq.~\ref{k_rjv}, to express $k_{jv-d}$ as 
\begin{equation}
    k_{jv-d} = \chi(T) \cfrac{k_{r-jv} }{g_{jv}}\exp\left[\cfrac{\epsilon_{int}(j,v)}{k_B T} \right]
    \label{k_djv_simple}
\end{equation}
where $\chi(T)$ contains dependence on $T$ and other variables, whose explicit form is omitted for brevity. Now we insert the expression for $k_{jv-d}$ from Eq.~\ref{k_djv_simple} into Eq.~\ref{ev_avg_diss_eqn}.
\begin{equation}
    \langle \epsilon_{v}^{d} \rangle = \cfrac{\sum _{v=0}^{v_{\max}}  \sum _{j=0}^{j_{\max}(v)} \epsilon_{int}(0,v) \cfrac{k_{r-jv} }{g_{jv}}\exp\left[\cfrac{\epsilon_{int}(j,v)}{k_B T} \right] f(j,v)}{\sum _{v=0}^{v_{\max}}  \sum _{j=0}^{j_{\max}(v)} \cfrac{k_{r-jv} }{g_{jv}}\exp\left[\cfrac{\epsilon_{int}(j,v)}{k_B T} \right] f(j,v)}. 
    \label{ev_avg_diss_eqn_11}
\end{equation}
Next, consider the case when the gas is in equilibrium, we have
\begin{equation}
    f_0(j,v) = \cfrac{g_{jv}\exp\left[-\cfrac{\epsilon_{int}(j,v)}{k_B T}\right]}{\sum _{v=0}^{v_{\max}}  \sum _{j=0}^{j_{\max}(v)} g_{jv} \exp\left[\cfrac{-\epsilon_{int}(j,v)}{k_B T}\right]}
    \label{boltz_distro_11}
\end{equation}
Inserting this Boltzmann expression for the distribution ($f_0(j,v)$) in Eq.~\ref{ev_avg_diss_eqn}, we obtain  an expression for the average energy of dissociating molecules at equilibrium,
\begin{equation}
\begin{split}
    \langle \epsilon_{v}^{d} \rangle^* = \cfrac{\sum _{v=0}^{v_{\max}}  \sum _{j=0}^{j_{\max}(v)} \epsilon_{int}(0,v) k_{r-jv} }{\sum _{v=0}^{v_{\max}}  \sum _{j=0}^{j_{\max}(v)}k_{r-jv} },
    \end{split}
    \label{ev_avg_diss_recomb_eqn}
\end{equation}
which is the definition of average vibrational energy for recombining molecules (Eq.~\ref{ev_recomb_eqn}). Therefore, the average vibrational energy of recombining molecules is equal to the  average vibrational energy of dissociating molecules at equilibrium ($\langle \epsilon_{v}^{rec} \rangle  = \langle \epsilon_{v}^{d} \rangle^*$).
\begin{equation}
      \langle \epsilon_{v}^{rec} \rangle  = \langle \epsilon_{v}^{d} \rangle^*
    \label{avg_recomb_vib}
\end{equation}
Equation~\ref{avg_recomb_vib} has also been derived in Ref.~\cite{olejniczak1995vibrational}. For the average rotational energy of dissociating molecules, $\langle \epsilon_{rot}^d \rangle $, we proposed in Ref.\cite{singh2019consistentII} that $\langle \epsilon_{rot}^d \rangle  = \epsilon_d - \langle \epsilon_{v}^{d} \rangle $ . In the view of the above analysis, we propose the analogous model for average rotational energy in recombination: 
\begin{equation}
      \langle \epsilon_{rot}^{rec} \rangle  = \epsilon_d - \langle \epsilon_{v}^{d} \rangle^*
    \label{avg_recomb_rot}
\end{equation}
At this stage, the continuum recombination model has been derived. Specifically, since the dissociation model is analytical, the expressions for $ \langle \epsilon_{v}^{d} \rangle^*$ and $k_d^*(T)$ can be inserted in Eqs.~\ref{k_rec_general}, \ref{avg_recomb_vib} and \ref{avg_recomb_rot}, and the required recombination rate constant ($k_{r_{AB-C}}$), as well as the average vibrational ($\langle \epsilon_{v}^{rec} \rangle$) and average rotational energy ($\langle \epsilon_{rot}^{rec} \rangle$) of recombined molecules can be directly obtained. These quantities can then be used in the continuum equations (Eqs.~\ref{Rate_eqn_general_recomb}--\ref{Jeans_modified_general_recomb}). 

In the next section, we present the full expressions for continuum recombination quantities, $k_{r_{AB-C}}, \, \langle \epsilon_{v}^{rec} \rangle$ and $\langle \epsilon_{rot}^{rec} \rangle$, and the state-specific recombination rates, primarily for two reasons. First, state-specific recombination rates provide insight into which states (vibrational and rotational) are preferred when atoms recombine. Second, state-specific recombination rates can be  used in the DSMC method using the procedure described in Ref.~\cite{gimelshein2017dsmc}.

\section{Derivation of $k_{r-jv}$, $k_r$, and $\langle\epsilon_v^{rec} \rangle$}
In this section, we derive the expressions for $k_{r-jv}$, $k_r$, and $\langle\epsilon_v^{rec} \rangle$ for nitrogen gas. To estimate the expression for $ k_r$, we need the expression for the state-specific dissociation rate ($k_{jv-d}$) in Eq.~\ref{k_rec_general}. The model for $k_{jv-d}$ can be obtained from a state-specific dissociation probability/cross-section averaged over a Maxwell-Boltzmann distribution of relative translational energy. An analytical model for dissociation cross-sections consistent with \textit{ab-initio} data was proposed by the authors in Ref.~\cite{singh2019consistentI}. Using the state-specific dissociation model, the expression for $k_{jv-d}$ is derived in the appendix (Eq.~\ref{Rate_Derive4}).  The expression for  $k_{jv-d}$ is reproduced here for clarity: 
 \begin{equation}
\begin{split}
    k_{jv-d}(T)  & =
   \frac{1}{S} \left(\frac{8 k_B T}{\pi \mu_C}\right)^{1/2} \pi b_{\max}^2  C_1 \exp\left[- \frac{\epsilon_d}{k_B T} \right]   \left(\frac{k_B T}{\epsilon_d}\right)^{\alpha-1}  
    \\ & 
    \times \Gamma[1+\alpha]
    \exp\left[\beta \frac{\epsilon_{rot}^{\text{eff}}}{\epsilon_d}\right] \exp\left[\gamma \frac{\epsilon_{v}}{\epsilon_d}\right] \exp\left[\delta \frac{|\epsilon_{int}-\epsilon_d|}{\epsilon_d}\right]
    \\   &
    \times \exp\left[-\frac{\theta_{CB}\epsilon_{rot}}{k_B T} \right]
  \exp\left[\frac{\epsilon_{int}}{k_B T} \right] 
   \end{split}
  \label{kjv_d_rate}
\end{equation}
where $S$ is the symmetry factor, $b_{\mx}$ is the maximum impact parameter, $\epsilon_d$ is the ground state dissociation energy, $\epsilon_{rot}^{\ef} = \epsilon_{rot}-\theta_{CB} \epsilon_{rot} $, $\epsilon_v$ is the vibrational energy and $\epsilon_{int} =\epsilon_{rot}+\epsilon_v$ is the internal energy. $\alpha$, $\beta$, $\gamma$ and $\delta$ capture the dependence of state-specific dissociation probability on translational energy, rotational energy, vibrational energy, and internal energy respectively. $\theta_{CB} \epsilon_{rot}$ is the increase in effective dissociation energy due to the centrifugal barrier. For complete details of this expression (Eq.~25), the reader is referred to Section IV of Ref.~\cite{singh2019consistentI}.

The state-specific recombination rate ($k_{r-jv}$) can be obtained by inserting the  expression for $k_{jv-d}(T)$ from Eq.~\ref{kjv_d_rate} in Eq.~\ref{k_rjv}. The full expression for $k_{r-jv}$ is:
\begin{equation}
\begin{split}
   k_{r-jv}   = &
      \cfrac{ \Qn_{AB}(T)}{\left[g_{A} g_{B} Q_{A}(T)Q_{B}(T) \right]}  \left\{ \frac{1}{S} \left(\frac{8 k_B T}{\pi \mu_C}\right)^{1/2} \pi b_{\max}^2  C_1  \right. 
       \\  & \left.
       \times \exp\left[- \frac{\epsilon_d}{k_B T} \right]  
     \left(\frac{k_B T}{\epsilon_d}\right)^{\alpha-1} \Gamma[1+\alpha]\right\} 
     \exp\left[ \cfrac{\Delta \epsilon_{f_{NN}} }{k_B T}\right] 
       \\
     &
    \times \exp\left[\beta \frac{\epsilon_{rot}^{\text{eff}}}{\epsilon_d}\right] 
     \exp\left[\gamma \frac{\epsilon_{v}}{\epsilon_d}\right]
         \exp\left[\delta \frac{|\epsilon_{int}-\epsilon_d|}{\epsilon_d}\right]
         \\
     & \times \exp\left[-\frac{\theta_{CB}\epsilon_{rot}}{k_B T} \right] \hspace{1.5in}
     \label{k_rjv_model}
\end{split}   
\end{equation}
The above expression in Eq.~\ref{k_rjv_model} for the state-specific recombination rate can be directly used in the DSMC method, since it is common to compute a cell-averaged translational temperature (T)\cite{boyd2017nonequilibrium}. Another approach for DSMC is to convert the expression for $k_{r-jv}$ into a probability that depends on the relative translational energy of the three-bodies undergoing collision by following the procedure in Ref.~\cite{gimelshein2017dsmc}, for example.

The overall recombination rate, as given in Eq.~\ref{k_rec_general} can be obtained by summing $k_{r-jv}$ from the model given in Eq.~\ref{k_rjv_model} over all $(j,v)$ states. The expression for the continuum recombination rate is:
\begin{equation}
\begin{split}
   k_r  = & \sum _{v=0}^{v_{\max}}  \sum _{j=0}^{j_{\max}(v)} k_{r-jv} =
      \cfrac{ \Qn_{AB}(T)}{\left[g_{A} g_{B} Q_{A}(T)Q_{B}(T) \right]} \left\{ \frac{1}{S} \left(\frac{8 k_B T}{\pi \mu_C}\right)^{1/2}   \right. \\ 
      & \times \left.
      \pi b_{\max}^2  C_1 \exp\left[- \frac{\epsilon_d}{k_B T} \right] 
      \left(\frac{k_B T}{\epsilon_d}\right)^{\alpha-1} \Gamma[1+\alpha]\right\} 
      \\
      &
     \times         \exp\left[ \cfrac{\Delta \epsilon_{f_{NN}} }{k_B T}\right] 
     \sum _{v=0}^{v_{\max}}  \sum _{j=0}^{j_{\max}(v)} (2j+1) \exp\left[\beta \frac{\epsilon_{rot}^{\text{eff}}}{\epsilon_d}\right]
      \exp\left[\gamma \frac{\epsilon_{v}}{\epsilon_d}\right] 
      \\
      &
    \times \exp\left[\delta \frac{|\epsilon_{int}-\epsilon_d|}{\epsilon_d}\right] \exp\left[-\frac{\theta_{CB}\epsilon_{rot}}{k_B T} \right] 
     \label{k_rec}
\end{split}   
\end{equation}
 The above summation requires specifying $\epsilon_v$ as function of vibrational quantum number (v) and rotational energy as function of rotational quantum number ($j$).  All procedures as laid out in Ref.~\cite{singh2019consistentI} for the dissociation rate are exactly followed here. To summarize, the modified simple harmonic oscillator model is employed, where instead of a unique characteristic temperature, multiple (in this case three) characteristic temperatures are used. 
 For rotational energy, the rigid rotor assumption is used.
These modified simple harmonic oscillator and rigid rotor models approximate \textit{ab-initio} energies accurately \cite{singh2019consistentI}. Using these models, analytically obtained dissociation rate constants have been shown to be in excellent agreement with QCT calculations in Refs.\cite{singh2019consistentI,singh2019consistentII}. 
 Further details of these approximations can be found in Sec.~IV A of Ref.~\cite{singh2019consistentI}.
 
By evaluating the sum in Eq.~\ref{k_rec} using these approximations, and substituting $\Delta \epsilon_{f_{NN}}=\epsilon_d$, the following expression for the recombination rate is obtained:
 \begin{equation}
\begin{split}
   k_r(T)& = 
   \\
   & \underbrace{A  T^\eta \cfrac{ \Qn_{AB}(T)}{\left[g_{A} g_{B} Q_{A}(T)Q_{B}(T)\right]}}_{\text{Prefactor}}   *\left[ \Hn_r(\epsilon_d,0,1) +  \Hn_r(\epsilon_d^{\mx},\epsilon_d,2) \right]
  \end{split}
  \label{Rate_Final_Boltz}
\end{equation}

\begin{equation*}
\eta =\alpha-\frac{1}{2}; \hspace{0.15in} A =  \frac{1}{S}  \left(\frac{8 k_B }{\pi \mu_C}\right)^{1/2} \pi b_{\max}^2  C_1 \Gamma[1+\alpha] \left(\frac{k_B }{\epsilon_d}\right)^{\alpha-1}
\end{equation*}

\begin{equation}
\begin{split}
    & \Hn_r(\epsilon_i,\epsilon_j,n) =\exp[(-1)^{n-1} \delta ]
    \\ 
     & \times \frac{\exp\left[ \epsilon_i \zeta_{rot,r}\right]  \gn (\zeta_{vr,r}) - \exp\left[ \epsilon_j\zeta_{rot,r}\right] \gn (\zeta_{v}-\epsilon_j \zeta_{rot,r}/\epsilon_d )}{k_B \theta_{rot} \zeta_{rot,r}}  ;
    \label{HBoltz_gen_rec}~,
\end{split}
\end{equation}

where,
\begin{equation*}
\begin{split}
\zeta_{rot,r}  = \frac{\beta-\theta_{CB}+(-1)^n\delta}{\epsilon_d} -\frac{\theta_{CB}}{k_B T};\hspace{0.10 in}\\
\label{delta_defs_rec}
\end{split}
\end{equation*}
\begin{equation*}
\begin{split}
\zeta_{v,r}  = \frac{\gamma+(-1)^n\delta}{\epsilon_d} \hspace{0.5in} \zeta_{vr,r} =\zeta_{v}-\zeta_{rot}
\label{delta_defs_rec}
\end{split}
\end{equation*}
and $\gn (...)$ (Eq.~\ref{gn_Boltz}) is given in the appendix. Here, $H_r(\epsilon_d,0,1)$ multiplied by the `Prefactor' gives the recombination rate into bound states and $\Hn_r(\epsilon_d^{\mx},\epsilon_d,2)$ times `Prefactor' gives the recombination rate of forming quasi-bound molecules. Note that aside from the `Prefactor', these expressions ($H (...)$) are identical to the dissociation model expressions corresponding to \textit{equilibrium} (Boltzmann) internal energy distribution functions (see Eq.~25 in Ref.~\cite{singh2019consistentI}).  Since many of the dissociation terms would already be computed, implementation of this recombination model in a CFD code would be straightforward. 

Next we derive the expression for the average vibrational energy of molecules, $\langle \epsilon_{v}^{rec} \rangle$, formed in recombination using Eq.~\ref{ev_recomb_eqn}. Inserting the expression for $k_{r-jv}$ from Eq.~\ref{k_rjv_model}  into Eq.~\ref{ev_recomb_eqn}, the expression for the average vibrational energy of molecules ($\langle \epsilon_v^{rec} \rangle(T)$) formed in the recombination reaction is: 
\begin{equation}
    \langle \epsilon_v^{rec} \rangle(T) =  \frac{\Phi_r(\epsilon_d,0,1)+\Phi_r(\epsilon_d^{\mx},\epsilon_d,2)}{\Hn_r(\epsilon_d,0,1)+\Hn_r(\epsilon_d^{\mx},\epsilon_d,2)}
    \label{edrecomb_gen_sum_Boltz}
\end{equation}

\begin{equation*}
\begin{split}
    & \Phi_r(\epsilon_i,\epsilon_j,n) = \exp[(-1)^{n-1}\delta] 
   \\ &\times \frac{\exp\left[ \epsilon_i \zeta_{rot,r}\right]  \gn' (\zeta_{vr,r}) - \exp\left[ \epsilon_j \zeta_{rot,r}\right] \gn' (\zeta_{v,r}-\epsilon_j \zeta_{rot,r}/\epsilon_d )}{k_B \theta_{rot} \zeta_{rot,r}} ;
    \label{Hnb_Boltz_recomb}
\end{split}
\end{equation*}
Similar to the recombination rate expression, the expression for the average vibrational energy of recombined molecules is identical to the dissociation expression corresponding to equilibrium (Boltzmann) distribution functions (see Eq.~41 in Ref.\cite{singh2019consistentI}). For completeness, we present  Eq.~\ref{hatgderivativesum} for the expression of $\gn'(...)$ in the appendix.

\section{Key physical mechanisms related to $k_{r-jv}$, $k_r$, and $\langle\epsilon_v^{rec} \rangle$ }
  In this section, we analyze the  probability of atoms recombining into a given internal energy state. First, we consider the recombination probability as a function of internal energy at a constant temperature ($T$). Then, we analyze the effect of $T$ on the state-specific recombination probability.
  
 \subsection{Recombination probability to a state, $\epsilon_{int}(j,v)$, at a given temperature}
\begin{figure}
       \vspace{-0.00in}
  \subfigure[]
  {
  \includegraphics[width=3.50in,trim={0.20cm 0.1cm 0.15cm 1.75cm}, clip]{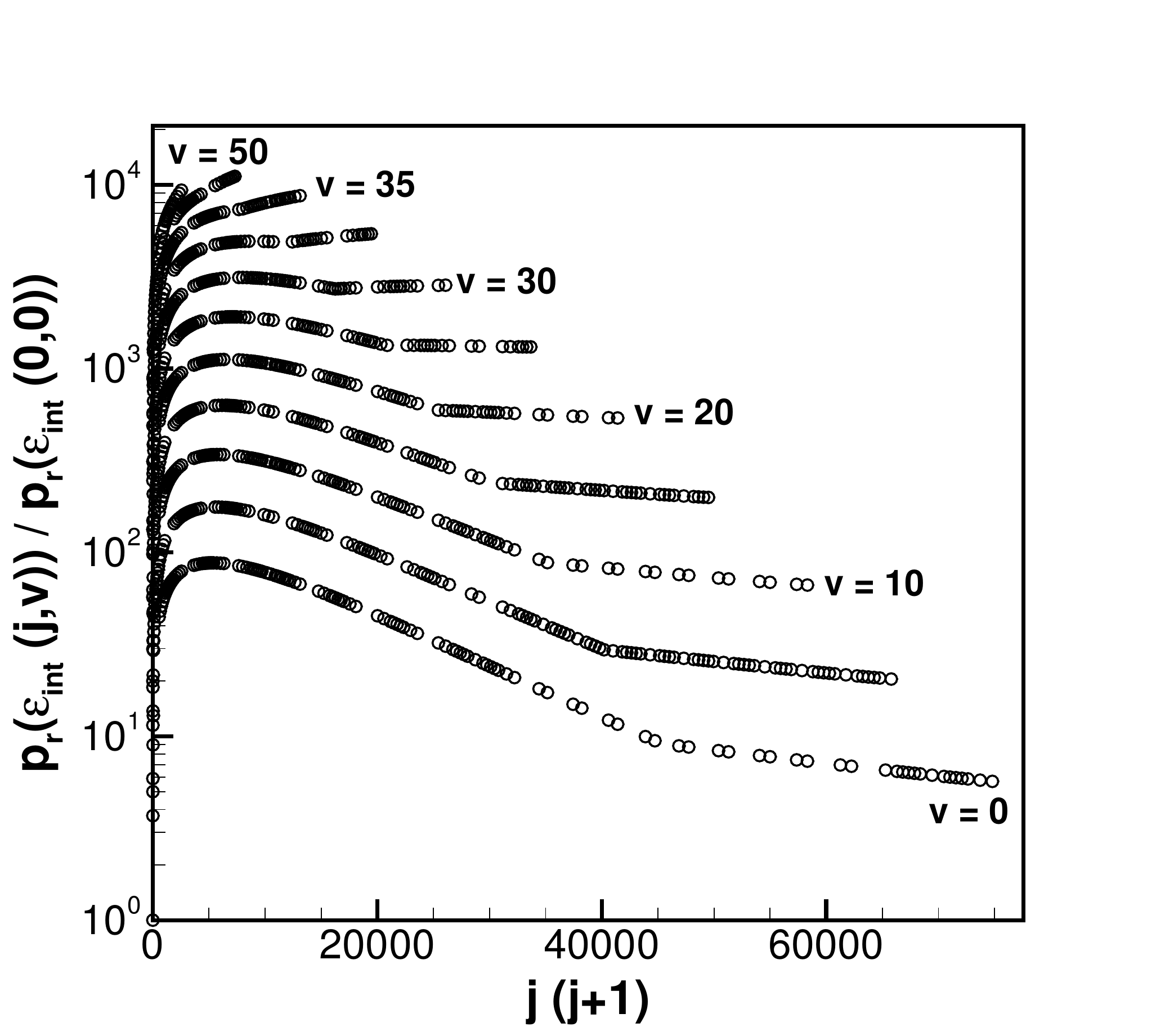}
   \label{probability_state_j_log}
   } 
          \vspace{-0.00in}
  \subfigure[]
  {
    \includegraphics[width=3.50in,trim={0.20cm 0.1cm 0.15cm 1.75cm}, clip]{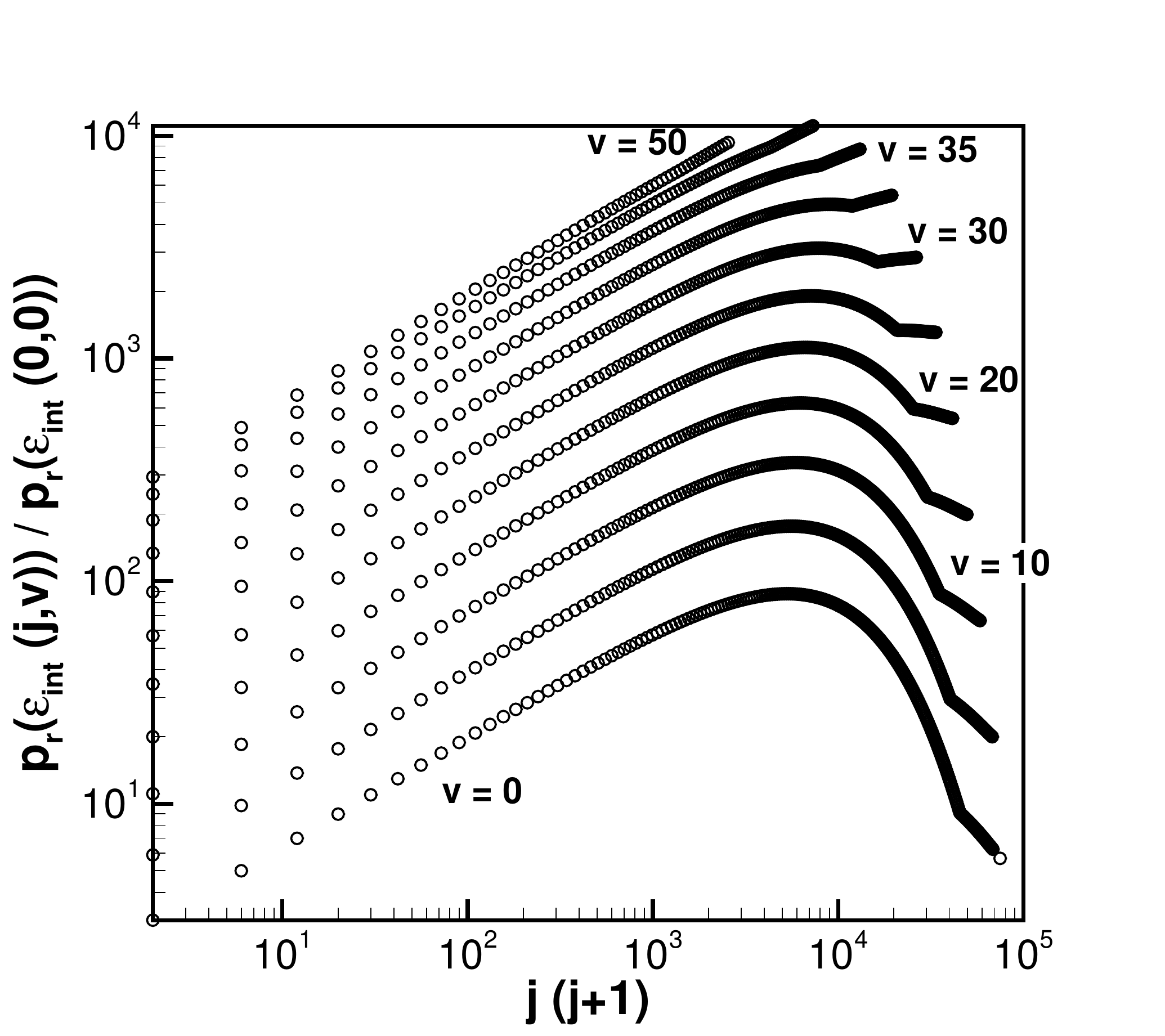}
   \label{probability_state_j}
   } 
    \caption{(a) Probability of recombining to an internal energy state  as function of $j(j+1)$ at different $v$ (b) Results from (a) but plotted on a log scale. T = 5, 000 K is used and the probability is normalized by the ground internal energy state probability. } 
   \label{probability_state_recomb_3D_2D}
\end{figure}

\begin{figure}
         \subfigure[]
  {
    \includegraphics[width=3.50in,trim={0.20cm 0.1cm 0.15cm 1.75cm}, clip]{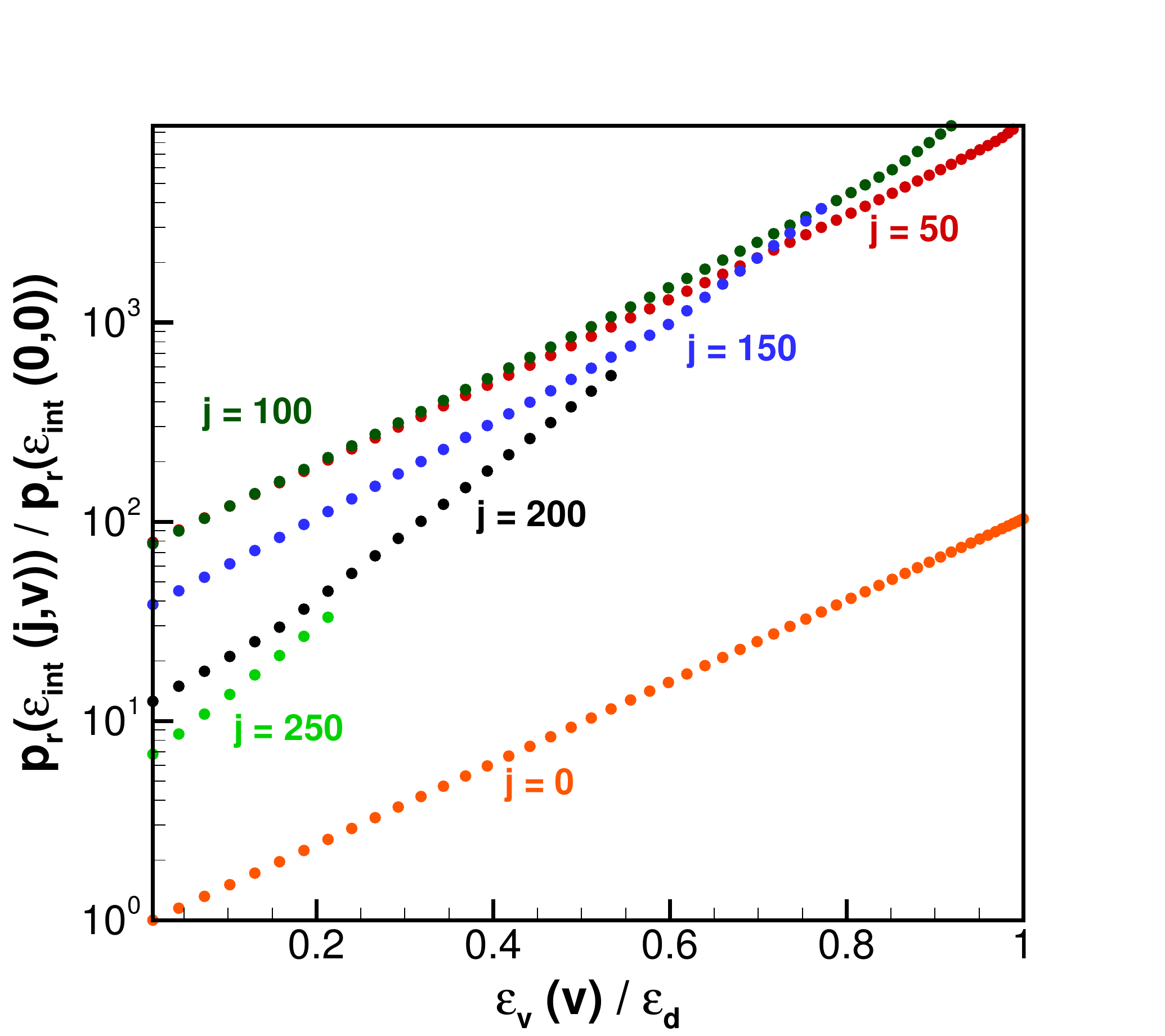}
   \label{probability_state_v}
   } 
          \vspace{-0.0000in}
        \subfigure[]
  {
    \includegraphics[width=3.50in,trim={0.20cm 0.1cm 0.15cm 1.75cm}, clip]{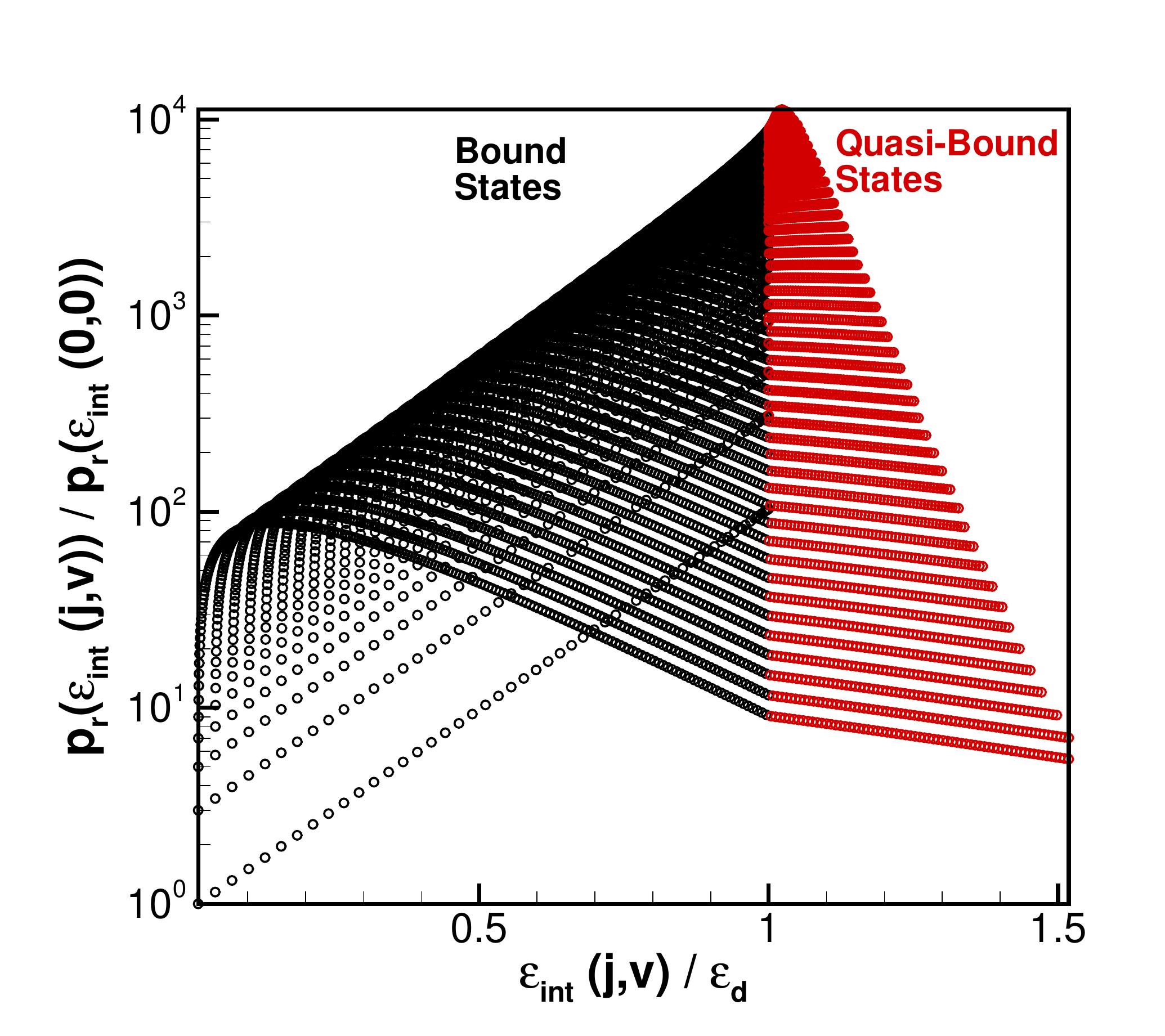}
   \label{eint_prob_relative_2d}
   } 
    \caption{(a) Probability of recombining to an internal energy state as a function of $v$ at different $j$. (For brevity only few $v$ with a gap of 5 are plotted, out of which only a few are mentioned). (b) Probability of recombining to an internal energy as a function of $\epsilon_{int}(j,v)$. T = 5, 000 K is used and the probability is normalized by the ground internal energy state probability. } 
   \label{probability_state_recomb_3D_2D_2}
\end{figure}

 First, we consider the relative probability of recombining into a given internal energy state. 
 While the expression for the state-specific rate was derived in Eq.~\ref{k_rjv_model}, the relative probability allows us to analyze the favoring of the state in which the recombining atoms end up as a molecule. 
 Mathematically, the expression for the relative probability ($p(\epsilon_{int}(j,v))/p(\epsilon_{int}(0,0))$) using Eq.~\ref{k_rjv_model} can be written as:
 \begin{equation}
 \begin{split}
    & \cfrac{p(\epsilon_{int}(j,v))}{p(\epsilon_{int}(0,0))}=\cfrac{k_{r-jv}}{k_{r-jv}|_{(j=0,v=0)}} \hspace{3.0in}\\
     &= \cfrac{(2j+1) \exp\left[\beta \cfrac{\epsilon_{rot}^{\text{eff}}}{\epsilon_d}\right] \exp\left[\gamma \cfrac{\epsilon_{v}}{\epsilon_d}\right] \exp\left[\delta \cfrac{|\epsilon_{int}-\epsilon_d|}{\epsilon_d}\right]
     \exp\left[-\cfrac{\theta_{CB}\epsilon_{rot}}{k_B T} \right]}
      { \exp\left[\gamma \cfrac{\epsilon_{0}}{\epsilon_d}\right] \exp\left[\delta \cfrac{|\epsilon_{0}-\epsilon_d|}{\epsilon_d}\right]
     }
     \label{relative_prob_int}
     \end{split}
 \end{equation}
 where $\epsilon_{0}$ is the zero-point energy of the molecule. We plot this relative probability in Figs.~\ref{probability_state_recomb_3D_2D} and \ref{probability_state_recomb_3D_2D_2} as a function of $v$ and $j(j+1)$\footnote{$\epsilon_{v} \propto v $ and $\epsilon_{rot} \propto j(j+1)$} for all allowed ro-vibrational ($j,v$) states.
 Figures \ref{probability_state_j_log} and ~\ref{probability_state_j} shows that the probability to recombine into a state with low vibrational energy (equivalently $v$) is low. This trend is not true for recombining probability to a rotational state, where at a given $v$, the recombining probability is low for both low  and high rotational energies (equivalently $j(j+1)$), and takes a maximum for some $j$ in between. This specific variation of the recombination probability can be seen in Figs.~\ref{probability_state_j_log} and \ref{probability_state_j}. In Fig.~\ref{probability_state_v}, the variation of the probability  with normalized vibrational energy ($\epsilon_v(v)/\epsilon_d$) for different values of $j$ is shown. At each $j$, the probability increases with $\epsilon_v(v)$. On the other hand, for a given $v$, the probability first increases and then decreases as $j$ increases as shown in Fig.~\ref{probability_state_v}. This is due to the effect of degeneracy and centrifugal barrier as explained in the earlier paragraph. This occurs because the degeneracy with $j$ is $(2j+1)$, therefore there are more ways of observing higher $j$ states. But for higher $j$, the centrifugal barrier reduces the probability of recombining (terms with $\beta$ and $\theta_{CB}$ in Eq.~\ref{relative_prob_int}).    The same recombination probability for all ro-vibrational states but now as function of normalized $\epsilon_{int}$ is shown in Fig.~\ref{eint_prob_relative_2d}. A diatomic molecule can have the same internal energy for different vibrational and rotational states, therefore multiple values of the probability exist at a given internal energy. Fig.~\ref{eint_prob_relative_2d}, shows that the mostly likely internal energy for a diatomic molecule to form is close to the dissociation energy ($\epsilon_d$). In fact, this can be directly deduced from  detailed balance and the fact that the average internal energy of dissociating molecules is $\epsilon_d$ \cite{bender2015improved}.

 In summary, for two states with the same internal energy, the state of lower rotational energy is preferred for recombination. However, higher degeneracy associated with higher rotational energy may increase the overall probability of observing molecules in  higher rotational energy states. For fixed rotational energy, the probability of recombining is higher in a state of higher vibrational energy.

  \subsection{Temperature dependence on the state-specific recombination probability}\label{sec_t_depend_recomb}
\begin{figure}
\centering 
   \subfigure[\textcolor{red}{}]
  {
    \includegraphics[width=3.75in]{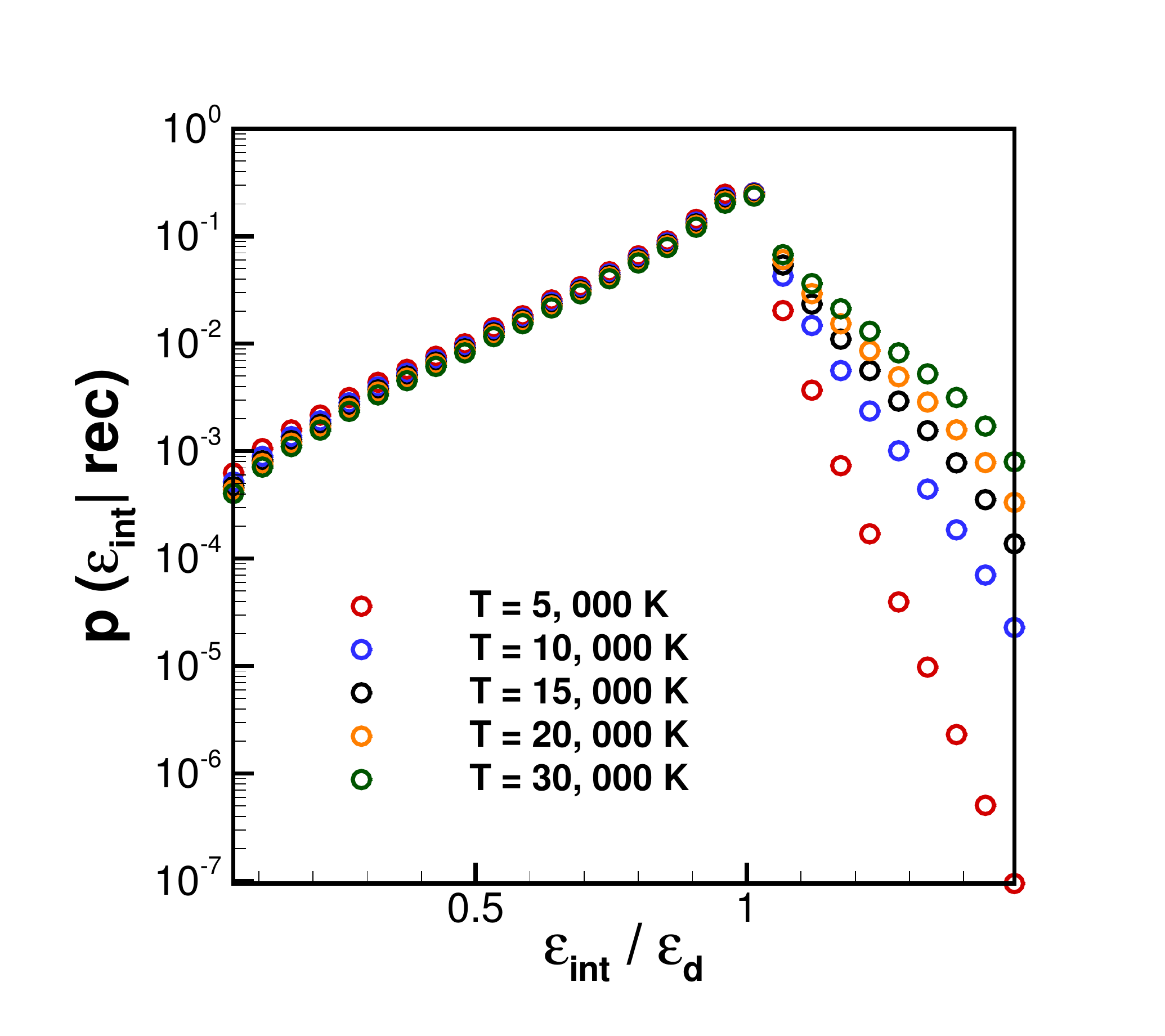}
   \label{eint_prob_relative}
   } 
  \subfigure[]
  {
    \includegraphics[width=3.50in]{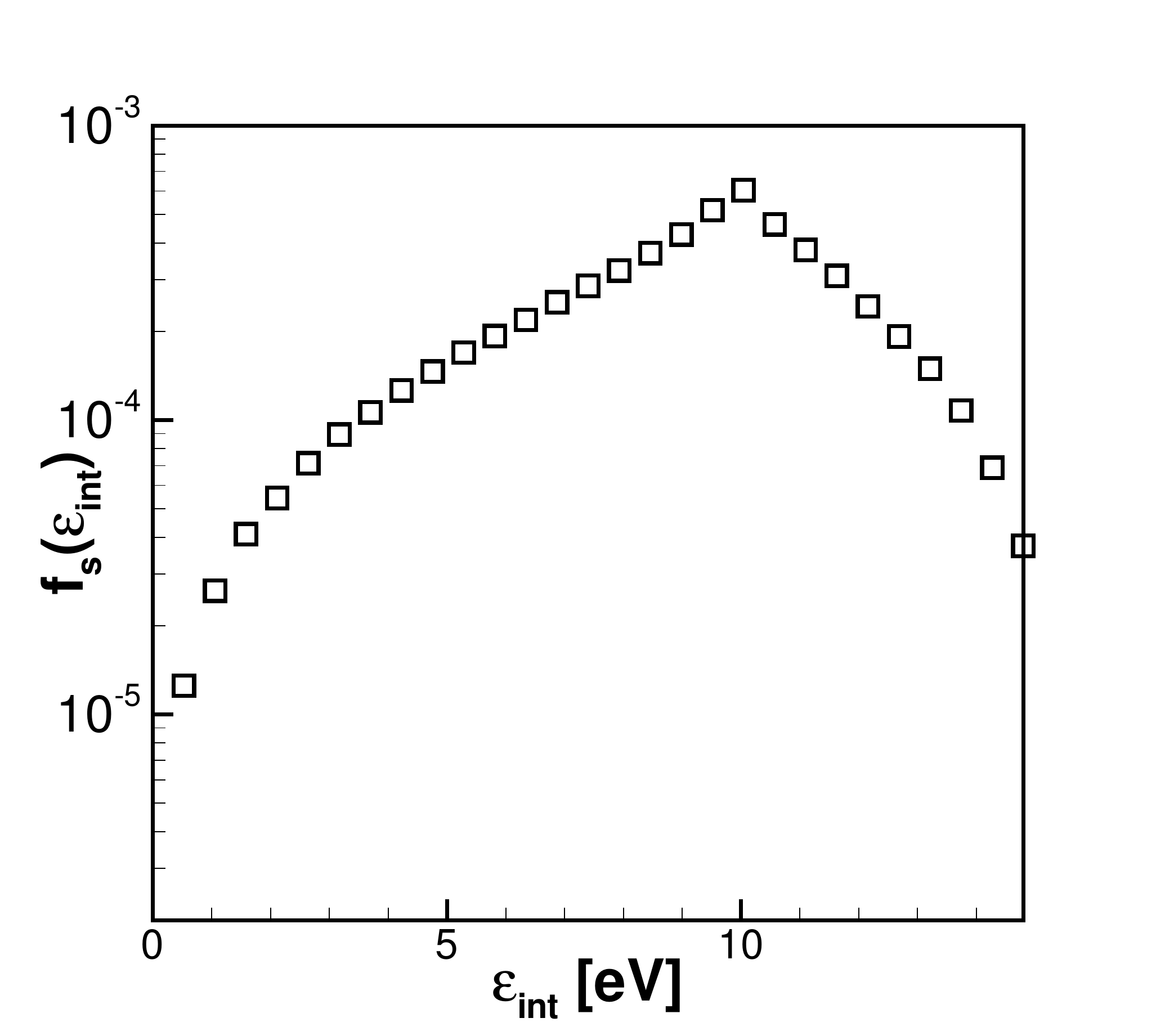}
   \label{freq_eint_fig}
   }  
    \caption{(a) Probability of recombining to an internal energy state (b) Frequency (Eq.~\ref{freq_eint}) of observing an internal energy $\epsilon_{int}$ in a diatomic molecule.} 
\end{figure}

 In this subsection, we consider the recombination probability at different temperatures as a function of internal, rotational, and vibrational energies. Figure~\ref{eint_prob_relative} shows the normalized recombination probability at five different translational temperatures, ranging from $T= 5,000$ K to $T= 30,000$ K as a function of internal energy. Note that in Fig.~\ref{eint_prob_relative}, the probabilities of recombination at allowed ro-vibrational quantum states are binned in internal energy groups. The probability of recombining to a bound state ($\epsilon_{int} < 9.91$ eV) with higher internal energy is higher, and is weakly dependent on the temperature. In contrast, the probability of recombining to a quasi bound state decreases as internal energy is increased. The quasi-bound recombining probability has a relatively stronger dependence on the temperature. In order to understand these trends, we need to consider the distribution of diatomic internal energy into the allowed  rovibrational quantum states and the role of the centrifugal barrier term, which results in the observed temperature dependence of the recombination probability. 

 Let us first consider the distribution of internal energy among ro-vibrational quantum states. A rovibrational quantum state is defined by rotational ($j$) and vibrational ($v$) quantum number and the degeneracy associated with $j$ is $(2j+1)$. As mentioned earlier, different ro-vibrational quantum states can have the same internal energy ($\epsilon_{int}(j,v)$). If the internal energy is binned, then the probability with which an internal energy state can exist in a diatomic molecule can be written as $f_s(\epsilon_{int})$,
 \begin{equation}
     f_s(\epsilon_{int}) = \cfrac{ \sum _{j',v'}  \left( 2 j' +1 \right)I_{\Delta \epsilon_{int}} (j',v') }{\sum _{\epsilon_{int}} \sum _{j',v'} \left( 2 j' +1 \right)I_{\Delta \epsilon_{int}} (j',v')}
     \label{freq_eint}
 \end{equation}
 where $I_{\Delta \epsilon_{int}} (j',v') =1\  \forall \{j',v'\}$ such that $ \epsilon_{int}(j',v') \in  [\epsilon_{int}-\Delta \epsilon_{int}/2, \epsilon_{int}+\Delta \epsilon_{int} /2] $ and $0$ otherwise. The variation of this quantity ($f_s(\epsilon_{int})$) with $\epsilon_{int}$ is shown in Fig.~\ref{freq_eint_fig}, which is also peaked around $\epsilon_d$. If hypothetically, the probability of recombination to form a molecule with a given internal energy is assumed independent of the magnitude of the internal energy,  then the overall probability distribution of the recombined molecules will be identical to $f_s(\epsilon_{int})$. Therefore, to first order, the general trend of the relative probability (i.e. increase with internal energy for bound state and decrease with internal energy for quasi-bound state in Fig.~\ref{eint_prob_relative} resembles the trend of $f_s(\epsilon_{int})$) in  Fig.~\ref{freq_eint_fig}. Physically, this just means that if there are more ways of observing a given internal energy in a diatomic molecule, the probability that atoms will recombine in that state will be higher.

Sections II-V focused on the derivation of the analytical recombination model using microscopic reversibility and state-specific dissociation rate expressions. Together with the dissociation rate model developed in Ref.~\cite{singh2019consistentI}, the recombination model provides closure to Eqs.~\ref{Rate_eqn_general_recomb}-\ref{Jeans_modified_general_recomb}, which can be applied to model chemical kinetics coupled to internal energy relaxation. However, before testing the combined dissociation and recombination model, there is one more important aspect to consider.  The dissociation model \cite{singh2019consistentI} is based on non-Boltzmann distributions in which high energy states are depleted. Therefore, the dissociation model must be extended to include the effects of recombination reactions, which re-populate the depleted high energy states. This extension of the dissociation model to include recombination reactions is considered next in Section VI.

\section{Non-Boltzmann Distribution Modeling Including Recombination Reactions}
In this section, an extension of the non-Boltzmann distribution model \cite{singh2019nonboltzmann} to include the effect of recombination is derived. The  model extension is developed using the surprisal \cite{levine1978information,levine2009molecular,levine1971collision} analysis based QSS distribution model \cite{Singhpnas}  undergoing \textit{only} dissociation. The derivation employs only microscopic reversibility \cite{tolman1979principles,lewis1925new}, master-equation, and does not require any additional adjustable parameters.
Before deriving the effect of recombination on non-Boltzmann distributions, for the sake of completeness, first a brief review of the depleted QSS distribution model is presented. 

\subsection{Review: Non-Boltzmann Distribution Modeling for a Dissociating Gas}
In prior work \cite{singh2019nonboltzmann,Singhpnas}, the proposed model for non-Boltzmann vibrational energy distributions, $f^{NB}(v)$, behind strong shocks is given by:
\begin{equation}
     f^{NB}(v) = \cfrac{\tilde{f}(v;T_v)+\Lambda f^d(v;T)}{1+\Lambda} 
     \label{fnb_overall}
\end{equation}
 where  $\tilde{f}(v;T_v)$ and $f^d(v;T)$ are defined as follows:
 \begin{equation}
 \begin{split}
     f^d(v;T) = \cfrac{\exp\left[-\cfrac{\epsilon_v(v)}{k_B T}-\lambda_{1,v} \cfrac{\langle \epsilon_t \rangle}{\epsilon_d} v\right]}{ \sum_v \exp\left[-\cfrac{\epsilon_v(v)}{k_B T}-\lambda_{1,v} \cfrac{\langle \epsilon_t \rangle}{\epsilon_d} v\right]}
     \label{depleted_distros1}
     \end{split}
 \end{equation}
  \begin{equation}
 \begin{split}
       \tilde{f}(v;T_v) = \cfrac{\exp\left[-\cfrac{\Delta_{\epsilon} v}{k_B T_v}-\lambda_{1,v} \cfrac{\langle \epsilon_t \rangle}{\epsilon_d} v\right]}{ \sum\limits_{v}^{v_{\mx}}  \exp\left[-\cfrac{\Delta_{\epsilon} v}{k_B T_v}-\lambda_{1,v} \cfrac{\langle \epsilon_t \rangle}{\epsilon_d} v\right]}
       \label{ftv_term}
        \end{split}
 \end{equation}
Here, the parameter $\Lambda$ ensures that $ \sum_v \epsilon_v(v) f^{NB}(v)  = \langle \epsilon_v \rangle$. $T_v$ is called vibrational temperature, such that a Boltzmann distribution at $T_v$ recovers the average vibrational energy of the gas. $\Delta_{\epsilon} = \epsilon_v(1)-\epsilon_v(0)$ and $\lambda_{1,v}$ is a free parameter chosen to match the depletion observed in Direct Molecular Simulation (DMS) calculations. Singh and Schwartzentruber \cite{Singhpnas} developed Eq.~\ref{depleted_distros1} using a surprisal framework to model the depletion of high energy states due to QSS dissociation.  Eq.~\ref{depleted_distros1} is  in-fact a Boltzmann distribution corrected for depletion of the high energy tail using the parameter $\lambda_{1,v}$. In Eq.~\ref{ftv_term}, the low energy states which have nearly constant energy spacing ($\Delta_{\epsilon}$) are populated according to a Boltzmann distribution, such as in the simple harmonic oscillator assumption, and the high energy states are depleted due to dissociation.

\subsection{Extension to a General Non-Equilibrium Distribution Model Including  Recombination}
 In order to develop a general non-equilibrium distribution model, we analyze the system using master equation. More specifically, we analyze evolution of the vibrational energy distribution function. An evolution equation (similar to Eq.~\ref{master_eqn_recomb11}) for the vibrational energy population $[AB(v)]$ at a given translational temperature ($T$), can be written as:
   \begin{equation}
   \begin{split}
    \frac{ d\ [AB(v)] }{dt} =  \sum_{v'\neq v} k_{v'-v}  [AB(v')][C] 
    - \left ( \sum_{v'\neq v} k_{v-v'} \right)[C] [AB(v)] \\ -  k_{v-d} [C] [AB(v)] +  k_{r-v} [C] [A] [B] 
     \label{master_eqn_recomb1}
     \end{split}
 \end{equation}
 Here, $k_{v-v'} $ is the rate constant for transitioning from state $v$ to $v'$ during a collision with partner $C$, $d$ denotes the dissociated state and $r-v$ denotes recombination to a state $v$. $k_{v-v'} (\equiv k_{i-i'} (T,\langle \epsilon_{rot} \rangle))$ depends on $T$ and $\langle \epsilon_{rot} \rangle$ but for notational brevity, we have dropped the dependence. Note that, in Eq.~\ref{master_eqn_recomb1}, the rate constants are  averaged over the rovibrational states of the colliding partner when it is a diatom. We transform Eq.~\ref{master_eqn_recomb1} to obtain an evolution equation for $f(v) (= [AB(v)]/[AB])$ as:
 \begin{equation}
   \begin{split}
       \frac{ d f(v)  }{dt} =  \sum_{v'\neq v} k_{v'-v}  f(v')[C] 
    - \left ( \sum_{v'\neq v} k_{v-v'} \right)[C] f(v)  
   \\ -   k_{v-d} [C] f(v)
      +  k_{r-v} [C] \cfrac{[A] [B]}{[AB]}  \\ -f(v) \left(-k_d [C] +k_{rec} [C] \cfrac{[A][B]}{[AB]}\right) 
     \label{master_eqn_recomb3}
     \end{split}
 \end{equation}
 where 
 \begin{equation}
     k_d = \sum_v  k_{v-d} f(v) \hspace{0.5in} k_{rec} = \sum_v k_{r-v}
     \label{rates_d_rec_eqn}
 \end{equation}
 where $k_d$ is the overall dissociation rate coefficient and $k_{rec}$ is the overall recombination rate coefficient. Equation ~\ref{master_eqn_recomb1} has been summed over all $v$'s to obtain $d \ [AB] /dt $ in Eq.~\ref{master_eqn_recomb3}.
 It is nearly impossible to develop an exact analytical solution of Eq.~\ref{master_eqn_recomb3}, without making overly simplified assumptions which are inaccurate at high temperatures \cite{rubin1956relaxation, rubin1956relaxationprob,bazley1958studies,montroll1957application}. We therefore are interested in finding an approximate solution of this equation by considering this equation in two limits: dissociation dominated and recombination dominated.

   \subsubsection{Dissociation dominated gas}
 Let us first consider Eq.~\ref{master_eqn_recomb3}, when the gas is not undergoing any appreciable recombination reactions ($k_{r-v} = k_{rec} = 0$):
  \begin{equation}
   \begin{split}
       \frac{ d f(v)  }{dt} =  \sum_{v'\neq v} k_{v'-v}  f(v')[C] 
    - \left ( \sum_{v'\neq v} k_{v-v'} \right)[C] f(v)  
    \\ 
    -   [C] f(v)\left( k_{v-d}-k_d  \right)
     \label{master_eqn_recomb4}
     \end{split}
 \end{equation}
As stated in Eq.~\ref{depleted_distros1}, an approximate solution for $f(v)$ in the QSS phase has been developed by the authors \cite{Singhpnas}.  
 \begin{equation}
 \begin{split}
     f^d(v;T) = \cfrac{f_0(v)  \exp\left[ \hat{\delta}_v v\right]}{\sum_v f_0(v) \exp\left[ \hat{\delta}_v v\right] } 
     \\
     \hat{\delta}_v = -\lambda_{1,v} \cfrac{\langle \epsilon_t \rangle}{\epsilon_d}
     \label{fqss_eqn_recomb}
     \end{split}
 \end{equation}
 where $f_0(v)$ is Boltzmann distribution, and $\lambda_{1,v}$ is the parameter obtained from DMS results. 
Let us analyze QSS by inserting Eq.~\ref{fqss_eqn_recomb} in Eq.~\ref{master_eqn_recomb4}
   \begin{equation}
   \begin{split}
       \frac{ d f(v)  }{dt} =  \sum_{v'\neq v} k_{v'-v}  f(v')[C] 
    - \left ( \sum_{v'\neq v} k_{v-v'} \right)[C] f(v)  \hspace{2.00 in}
    \\ 
    -   f(v) [C] \left(k_{v-d}  -\sum_v k_{v'-d} f(v') \right) = 0 \hspace{2.00 in}
    \\
    \implies
    \sum_{v'\neq v} k_{v'-v} f_0(v')  \exp\left[ \hat{\delta}_v v'\right]  
    - \left ( \sum_{v'\neq v} k_{v-v'} \right) f_0(v)  \exp\left[ \hat{\delta}_v v\right]  \hspace{2.00 in}
    \\  
    -   f_0(v)  \exp\left[ \hat{\delta}_v v\right]  \left(k_{v-d}  -\sum_{v'} k_{v'-d} f_0(v')  \exp\left[ \hat{\delta}_v v'\right]   \right) =0 \hspace{2.00 in}
     \label{master_eqn_recomb5}
     \end{split}
 \end{equation}
 Based on the magnitude of $\lambda_{1,v} =0.08$, we can approximate the exponential in Eq.~\ref{master_eqn_recomb5} using series\footnote{At very high temperature and for higher quantum state this approximation may be inaccurate.}
    \begin{equation}
   \begin{split}
  \sum_{v'\neq v} k_{v'-v} f_0(v')  \left[1+ \hat{\delta}_v v'\right]  
    - \left ( \sum_{v'\neq v} k_{v-v'} \right) f_0(v)  \left[1+\hat{\delta}_v v\right]  \\  
    -   f_0(v)  \exp\left[ 1+\hat{\delta}_v v\right] \left(k_{v-d}  - k_d^* -  \sum_{v'} k_{v'-d} \hat{\delta}_v v'  \right) =0
     \label{master_eqn_recomb6}
     \end{split}
 \end{equation}
 We next transform $k_{v'-v}$ using microscopic reversibility ($k_{v'-v} =k_{v-v'} f_0(v)/f_0(v')$) and rearrange terms in Eq.~\ref{master_eqn_recomb6} to obtain:
     \begin{equation}
   \begin{split}
  \sum_{v'\neq v} k_{v-v'} f_0(v)    
    - \left ( \sum_{v'\neq v} k_{v-v'} \right) f_0(v)    \\  
      \sum_{v'\neq v} k_{v-v'} f_0(v)  \left[\hat{\delta}_v v'\right]  
    - \left ( \sum_{v'\neq v} k_{v-v'} \right) f_0(v)  \left[\hat{\delta}_v v\right]
    \\
    -   f_0(v)  \left[ 1+\hat{\delta}_v v\right]  \left(k_{v-d}   - k_d^* -  \sum_{v'} k_{v'-d} \hat{\delta}_v v'   \right) =0
     \label{master_eqn_recomb7}
     \end{split}
 \end{equation}
 Removing the first two terms in Eq.~\ref{master_eqn_recomb7} as they are the negative of one another, and normalizing the entire equation by $f_0(v)$, we obtain:
 \begin{equation}
   \begin{split}
    \sum_{v'\neq v} k_{v-v'}   \left[\hat{\delta}_v v'\right] 
     - \left ( \sum_{v'\neq v} k_{v-v'} \right)  \left[\hat{\delta}_v v\right]
     \\
        -    \left[ 1+\hat{\delta}_v v\right]  \left(k_{v-d}  - k_d^* -  \sum_{v'} k_{v'-d} \hat{\delta}_v v'  \right) =0
     \label{master_eqn_fvr_rec_diss_dom}
     \end{split}
 \end{equation}
 We further ignore $\hat{\delta}$ terms in relation to unity:
   \begin{equation}
   \begin{split}
    \sum_{v'\neq v} k_{v-v'}   \left[\hat{\delta}_v v'\right] 
     - \left ( \sum_{v'\neq v} k_{v-v'} \right)  \left[\hat{\delta}_v v\right]
        -      \left(k_{v-d}  - k_d^*  \right) =0
     \label{master_eqn_fvr_rec_diss_dom}
     \end{split}
 \end{equation}
 Equation \ref{master_eqn_fvr_rec_diss_dom} can be seen as an equation for the parameter $\hat{\delta}_v$, which depends on $T$ alone. A solution requires state-specific transition rates and dissociation rates.  The objective of deriving this equation here is to compare it to the analogous equation for a recombination dominated gas.

 \subsubsection{Recombination dominated gas}
 In this subsection, we consider ro-vibrational relaxation where recombination reactions are dominant. For this, we can neglect the dissociation terms from Eq.~\ref{master_eqn_recomb3} to obtain ($k_{v-d} = k_d = 0$): 
 \begin{equation}
   \begin{split}
       \frac{ d f(v)  }{dt} =  \sum_{v'\neq v} k_{v'-v}  f(v')[C] 
    - \left ( \sum_{v'\neq v} k_{v-v'} \right)[C] f(v)  \hspace{1.75 in} \\ 
        +  k_{r-v} [C] \cfrac{[A] [B]}{[AB]}  -f(v) k_{rec} [C] \cfrac{[A][B]}{[AB]}\hspace{1.75 in}
     \label{master_eqn_recomb8}
     \end{split}
 \end{equation}
We first express recombination rates in terms of the equilibrium constant and dissociation rates using the principle of microscopic reversibility \cite{tolman1979principles,lewis1925new}, which is :
 \begin{equation}
    k_{r-v} =  k_{v-d}  \exp\left[-\cfrac{\epsilon_{v}(v)}{k_B T} \right] \cfrac{1}{\chi(T)}  =  k_{v-d} f_0(v) \cfrac{1}{K_C}  
    \label{micro_rever_state}
\end{equation}
where $\chi (T)$ depends on partition functions and 
 \begin{equation}
   K_C = \cfrac{\chi(T)}{Z_v(T) Z_{rot} (T)}. 
    \label{k_equilib_recomb}
\end{equation}
Inserting Eqs.~\ref{micro_rever_state} and \ref{k_equilib_recomb} in Eq.~\ref{master_eqn_recomb8} gives

 \begin{equation}
   \begin{split}
       \frac{ d f(v)  }{dt} =  \sum_{v'\neq v} k_{v'-v}  f(v')[C] 
    - \left ( \sum_{v'\neq v} k_{v-v'} \right)[C] f(v)  
    \\
        +   [C] \cfrac{[A] [B]}{[AB]} \cfrac{1}{K_C}   \left(  k_{v-d} f_0(v)    - f(v)  k_d^* \right)\hspace{0.75 in}
     \label{master_eqn_recomb9}
     \end{split}
 \end{equation}
 We hypothesize the steady state solution of  Eq.~\ref{master_eqn_recomb9}, $f^r(v;T)$ is similar to the QSS distribution for the dissociation case (considered in the previous subsection): 
  \begin{equation}
 \begin{split}
     f^r(v;T)  = \cfrac{f_0(v)  \exp\left[ \hat{\delta}_v^r v\right]}{\sum_v f_0(v) \exp\left[ \hat{\delta}^r_v v\right] } 
     \label{recomb_delta_distro}
     \end{split}
 \end{equation}
  where $\hat{\delta}_v^r$ is unknown. Inserting Eq.~\ref{recomb_delta_distro} in Eq.~\ref{master_eqn_recomb9}, and following similar steps taken in deriving Eq.~\ref{master_eqn_recomb7}, yields
  \begin{equation}
   \begin{split}
       \sum_{v'\neq v} k_{v-v'}   \left[\hat{\delta}_v^r v'\right] 
     - \left ( \sum_{v'\neq v} k_{v-v'} \right)  \left[\hat{\delta}_v^r v\right]
     \\
             +   \cfrac{[A] [B]}{[AB]} \cfrac{1}{K_C}  \left(  k_{v-d}     - k_d^* - \hat{\delta}_v^r k_d^*\right) = 0 
     \label{master_eqn_fvr_34}
     \end{split}
 \end{equation}
 Ignoring the $\hat{\delta}_v^r$ term in relation to unity, and rearranging terms gives
    \begin{equation}
   \begin{split}
       \sum_{v'\neq v} k_{v-v'}   \left[\hat{\delta}_v^r K_C \cfrac{[AB]}{[A] [B]} v'\right] 
     - \left ( \sum_{v'\neq v} k_{v-v'} \right)  \left[\hat{\delta}_v^r K_C \cfrac{[AB]}{[A] [B]} v\right]
     \\
             +     \left(  k_{v-d}     - k_d^* \right) = 0 \hspace{1.75 in}
     \label{recomb_dom_final_1}
     \end{split}
 \end{equation}
Equation~\ref{recomb_dom_final_1} can be seen as an evolution equation for the parameter $\hat{\delta}_v^r$, which should depend on $T$ and the concentration of different species. But because $\hat{\delta}_v$ is an approximate solution of Eq.~\ref{master_eqn_fvr_rec_diss_dom}, then 
 \begin{equation}
 \begin{split}
 \hat{\delta}_v^r K_C \cfrac{[AB]}{[A] [B]} = - \hat{\delta}_v \\
     \implies \hat{\delta}_v^r  =  - \cfrac{[A] [B]}{[AB]} \cfrac{1}{K_C}  \hat{\delta}_v
     \label{delta_r_v}
      \end{split}
 \end{equation}
 is the corresponding approximate solution of Eq.~\ref{recomb_dom_final_1}. This result is physical on two grounds (a) the opposite sign of $\hat{\delta}_v^r$ and $\hat{\delta}_v$ terms corresponds to the overpopulation and depletion of high-vibrational states due to recombination and dissociation respectively, (b) at equilibrium ($\frac{[A]^* [B]^*}{[AB]^*} \frac{1}{K_C} = 1$, where $^*$ denotes equilibrium compositions), and the extent of overpopulation due to recombination is balanced by the extent of depletion due to dissociation. 
 Note that even when the thermal state of the gas may be close to equilibrium ($T \simeq T_v$), the underlying distributions may still be depleted due to dissociation or overpopulated due to recombination. It is only when chemical equilibrium is reached, that the gas obeys the corresponding equilibrium Boltzmann distribution as expected.

 \subsection{Non-Equilibrium Vibrational Energy Distribution Model for Dissociating and Recombining Gas at High Temperature}
 The full non-equilibrium vibrational energy distribution model can now be formulated using Eqs.~\ref{fnb_overall} and \ref{delta_r_v} as follows:
 \begin{equation}
     f^{NB}(v) = \cfrac{\tilde{f}(v;T_v)+\Lambda f^{r,d}(v;T)}{1+\Lambda} 
     \label{fnb_overall_diss_recomb}
\end{equation}
 where $f^{r,d}(v;T)$ and  $\tilde{f}(v;T_v,T_0)$  are defined as follows:
 \begin{equation}
 \begin{split}
     f^{r,d}(v;T) = \cfrac{\exp\left[-\cfrac{\epsilon_v(v)}{k_B T}-\lambda_{1,v} \cfrac{\langle \epsilon_t \rangle}{\epsilon_d} \left(1 - \cfrac{[A] [B]}{[AB]} \cfrac{1}{K_C}\right)v\right]}{ \sum_v \exp\left[-\cfrac{\epsilon_v(v)}{k_B T}-\lambda_{1,v} \cfrac{\langle \epsilon_t \rangle}{\epsilon_d}\left(1 - \cfrac{[A] [B]}{[AB]} \cfrac{1}{K_C}\right) v\right]}
     \label{depleted_repopulated_distros}
     \end{split}
 \end{equation}
  \begin{equation}
 \begin{split}
       \tilde{f}(v;T_v) = \cfrac{\exp\left[-\cfrac{\Delta_{\epsilon} v}{k_B T_v}-\lambda_{1,v} \cfrac{\langle \epsilon_t \rangle}{\epsilon_d} \left(1 - \cfrac{[A] [B]}{[AB]} \cfrac{1}{K_C}\right)v\right]}{ \sum\limits_{v}^{v_{\mx}}  \exp\left[-\cfrac{\Delta_{\epsilon} v}{k_B T_v}-\lambda_{1,v} \cfrac{\langle \epsilon_t \rangle}{\epsilon_d}\left(1 - \cfrac{[A] [B]}{[AB]} \cfrac{1}{K_C}\right) v\right]}
       \label{ftv_term_depleted_repopulated}
        \end{split}
 \end{equation}
 where the QSS distribution (Eq.~\ref{depleted_distros1}) in Eq.~\ref{fnb_overall_diss_recomb} now includes the re-population effects in Eq.~\ref{depleted_repopulated_distros}. The distribution goes to a Boltzmann distribution when the gas reaches equilibrium ($\Lambda \rightarrow \infty$), as desired. It is good to recall at this stage that that in the derivation, we assumed that the re-population term due to recombination is small. While this holds for most conditions of interest, if during a simulation the deviation from equilibrium state is arbitrarily large ($\frac{[A]^* [B]^*}{[AB]^*} \frac{1}{K_C} >> 1$), practical implementation may require limiting the value of the re-population term. 
  
 \begin{figure}
\centering 
    \subfigure[DMS and Model (no recombination)]
 {
 \includegraphics[width=3.5in]{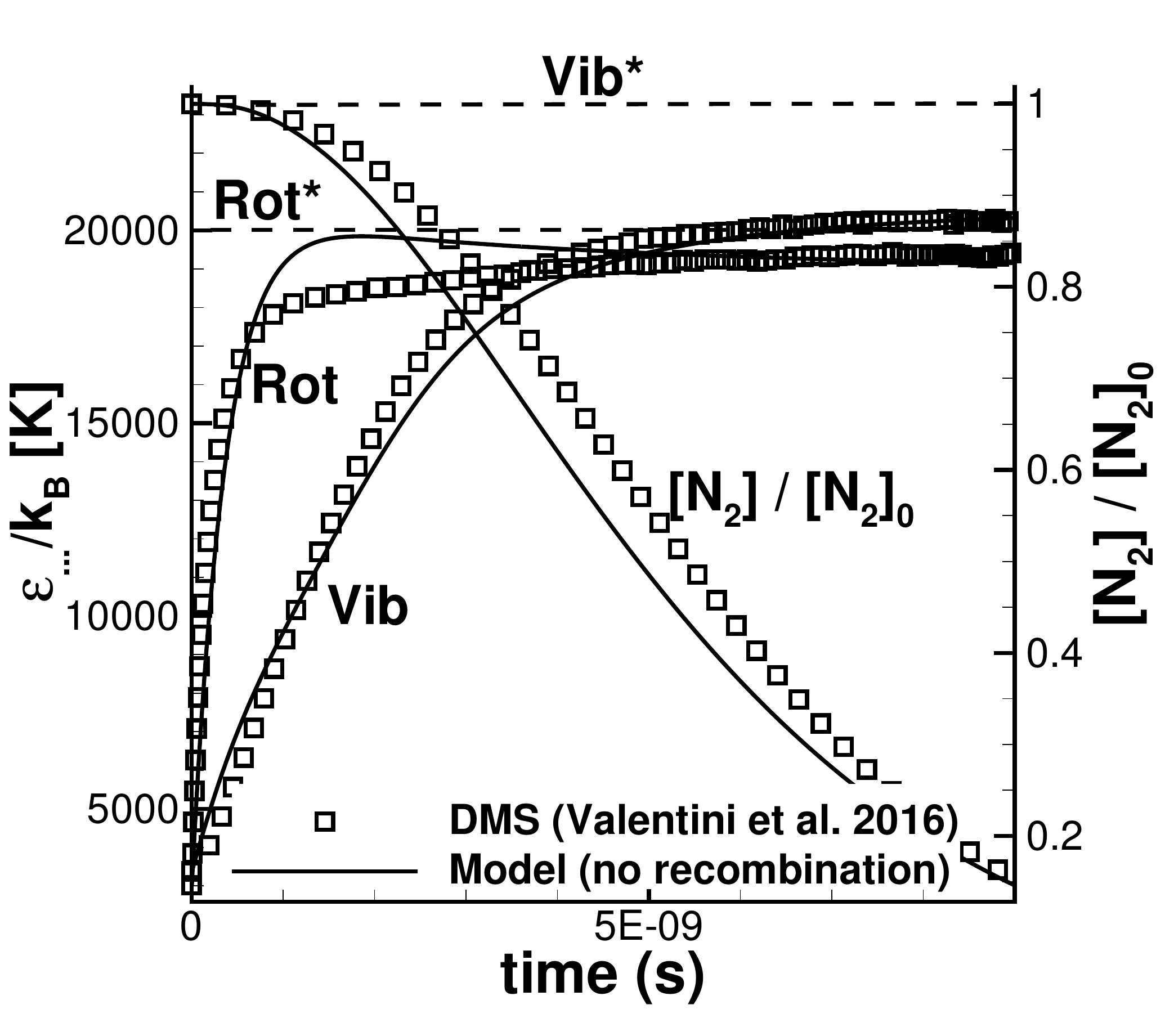}
 \label{DMS_model_no_recomb}
 }
     \subfigure[Model results]
 {
    \includegraphics[width=3.5in]{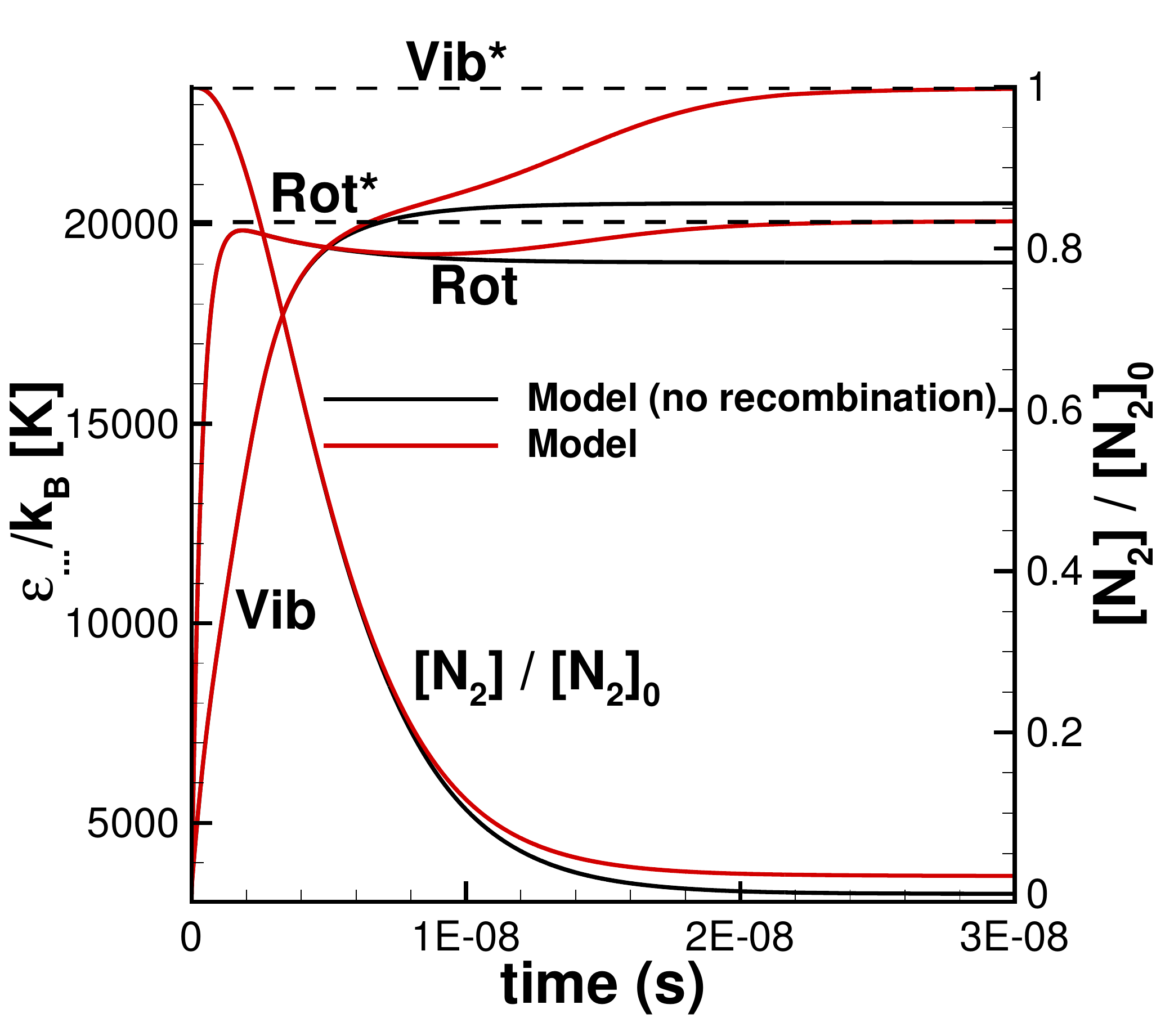}
   \label{recomb_and_recomb}
 }  
  \caption{(a) Isothermal ro-vibrational relaxation of nitrogen at $T=20, 000$ K. Image is taken from Ref.~\cite{singh2019consistentII} (Fig. 7)  (b) Model results including recombination. DMS results are taken from Ref.~\cite{valentini2016dynamics}.} 
   \label{zerod_20K_N3_Full_recomb}
\end{figure}

\section{Results and Discussion} 
 In order to assess the quantitative behavior of the recombination model, we perform simulations for ro-vibrational relaxation of nitrogen gas under isothermal conditions. The dissociation model has been shown to agree with Direct Molecular Simulation (DMS) calculations in Ref.~\cite{singh2019consistentII}, where recombination reactions are not included. In this section, we first test the recombination model for isothermal relaxation and then compare to predictions using the Park model. 
 
 
 \subsection{Isothermal relaxation }
 For the sake of completeness and a better understanding of the role of recombination, we simulate ro-vibrational relaxation of nitrogen with an initial density of $1.28$ kg/m$^3$. We reproduce such a calculation from our earlier work \cite{singh2019consistentII}  in Fig.~\ref{DMS_model_no_recomb}. For the DMS calculation, at each time step, the  translational energy of the molecules are  re-initialized, from a Maxwell-Boltzmann distribution at a temperature of $T=20, 000$ K to maintain isothermal condition. Initial internal energy of the molecules is sampled from a Boltzmann distribution at $T_v =T_{rot} = 3, 000 $ K. As time progresses, the gas starts to excite, and the average rotational and vibrational energies excite to steady values. The average rotational energy excites at a faster rate than the average vibrational energy. The steady state value of the average energies are different from the corresponding equilibrium values denoted by $\langle \epsilon_v \rangle^* \equiv \text{Vib}^*$ and $\langle \epsilon_{rot}\rangle^* \equiv \text{Rot}^*$ \footnote{The differences i.e. $\langle \epsilon_v \rangle^* \neq \langle \epsilon_{rot} \rangle^* \neq k_B T$ are the consequence of the diatomic energy variation with quantum state and the chosen vibration-prioritized \cite{jaffe1987}  framework of separating internal energy into vibration and rotational mode.}. The steady state is not the equilibrium state, and is referred to as quasi-steady state (QSS).  The system reaches QSS because the dissociation, shown by the decrease in the concentration of nitrogen molecules, balances the excitation process due to inelastic collisions. This means that the removal of the average internal energy from the system due to dissociation is balanced by the addition due to excitation. The atoms do not recombine in the current DMS calculations~\footnote{The focus of DMS so far has been to study the dissociation dominated conditions, observed in the regions immediately behind the shock waves. The addition of recombination in DMS for expanding flow studies is an ongoing effort.}~, therefore the system remains in QSS and does not reach equilibrium. 
 
Consider the same simulation (as shown in Fig,~\ref{DMS_model_no_recomb}) again, but now including the analytical recombination model for modeling recombination reactions. The results for the simulations are shown in Fig.~\ref{recomb_and_recomb}. The system in this case reaches the equilibrium state, where the internal energies reach their equilibrium values. The early phase of excitation is dominated by dissociation, and recombination remains negligible until the system reaches the QSS phase (time$~10^{-8}$s). After the system has reached QSS, recombination of atoms increases the average vibrational and rotational energies of the system and the system progresses towards equilibrium. 

   \begin{figure}
\centering 
  \subfigure[Zoomed in view of steady state part of \ref{recomb_and_recomb}]
  {
  \includegraphics[width=3.5in]{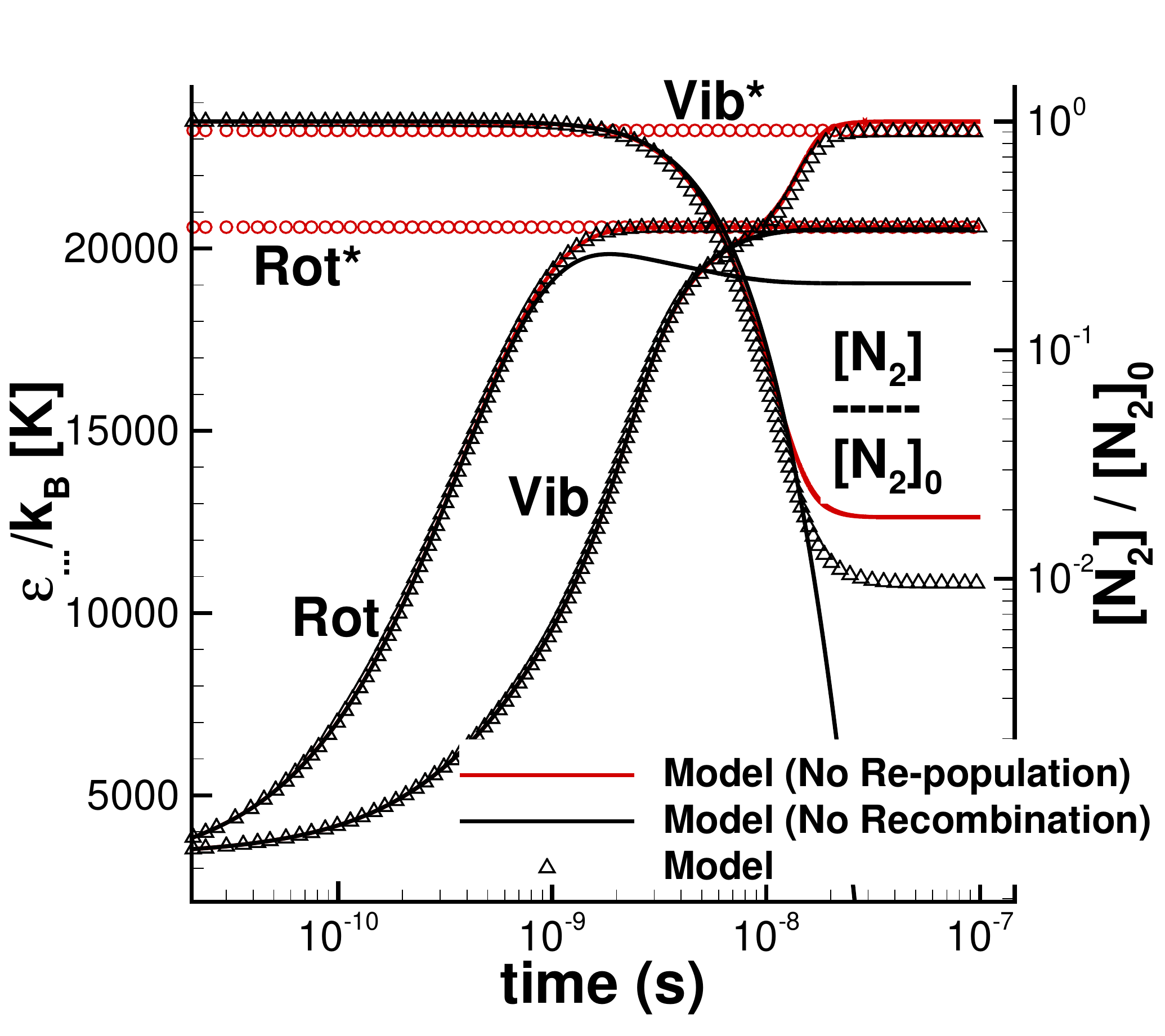}
   \label{recomb_zoomed_steady_State_repop_11}
   } 
   \subfigure[]
     {
  \includegraphics[width=3.5in]{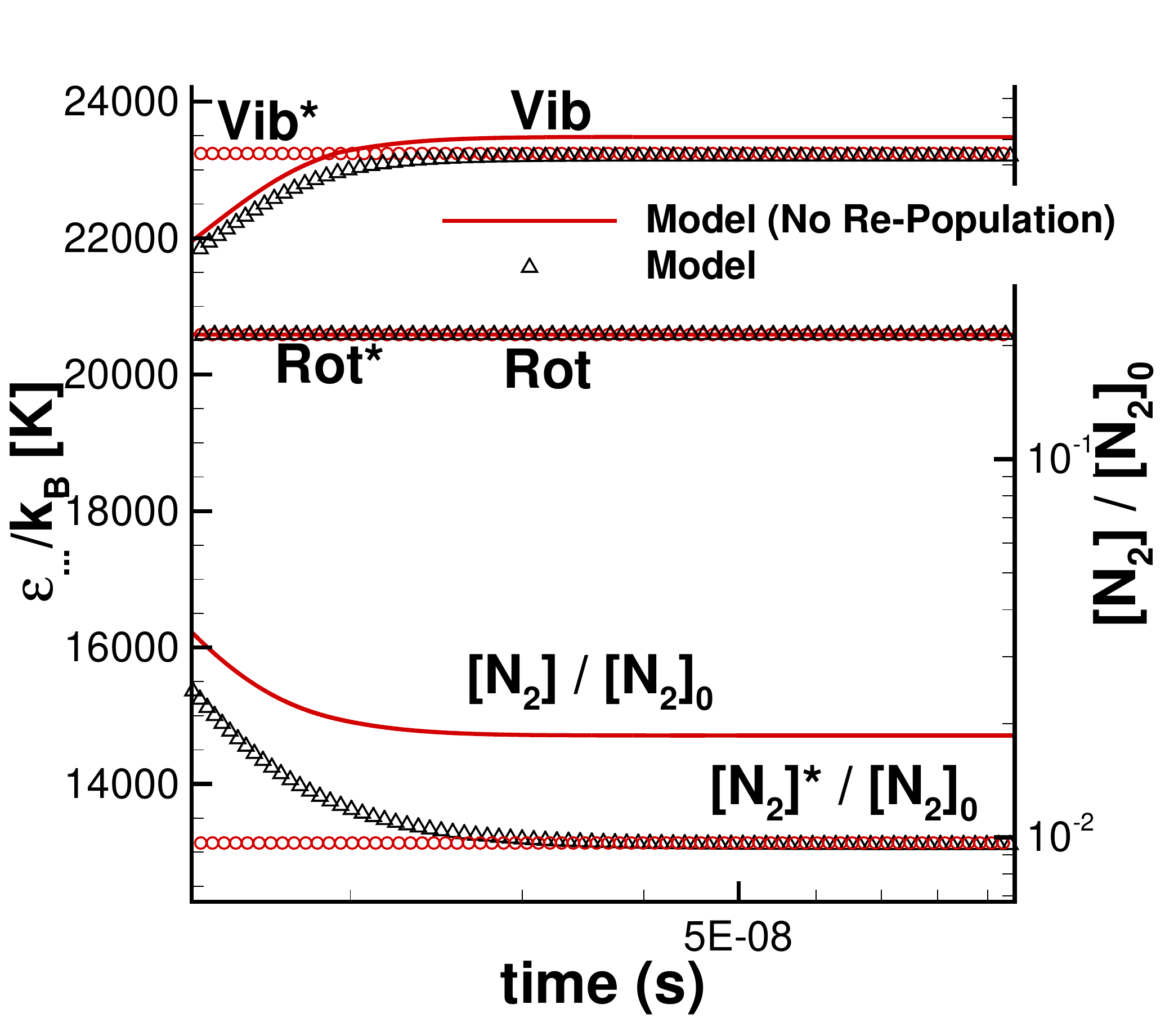}
   \label{recomb_zoomed_steady_State_repop_12}
   }  
  \caption{(a) Isothermal ro-vibrational relaxation of nitrogen at $T=20, 000$K. Model results with no recombination, recombination and re-population due to recombination reactions. (b) Zoomed in view of steady state region of (a).  DMS results are taken from Ref.~\cite{valentini2016dynamics}. `$^*$' refers to equilibrium values, shown by red circles. } 
   \label{zerod_20K_N3_Full_recomb_repop}
\end{figure}

  \subsection{Effect of recombination on non-Boltzmann distributions and dissociation}
  In this section, the effect of extension to the non-Boltzmann distribution model derived in Section VI is shown. In order to illustrate the effect of re-population, the simulation in Fig.~\ref{zerod_20K_N3_Full_recomb} is considered again in Fig.~\ref{zerod_20K_N3_Full_recomb_repop}, where  model results \textit{without} re-population terms are also included. 
As shown in the zoomed-in views of the steady state in Fig.~\ref{recomb_zoomed_steady_State_repop_12}, the system with no-repopulation term does not reach equilibrium.  In the steady state where re-population effects are absent, the dissociation rate is  based on the depleted distributions (Eq.~\ref{depleted_distros1}). Therefore, less overall dissociation relative to recombination results in the average vibrational energy being higher than the equilibrium value. The deviation from equilibrium is more apparent in the concentration where due to a lower dissociation rate, the steady state concentration of nitrogen molecules is higher than the corresponding equilibrium value. When re-population of high energy states due recombination is included as depicted by `Model', the concentration as well as average ro-vibrational energies reach the corresponding equilibrium values.

 \begin{figure}
\centering 
  {
 \includegraphics[width=3.5in]{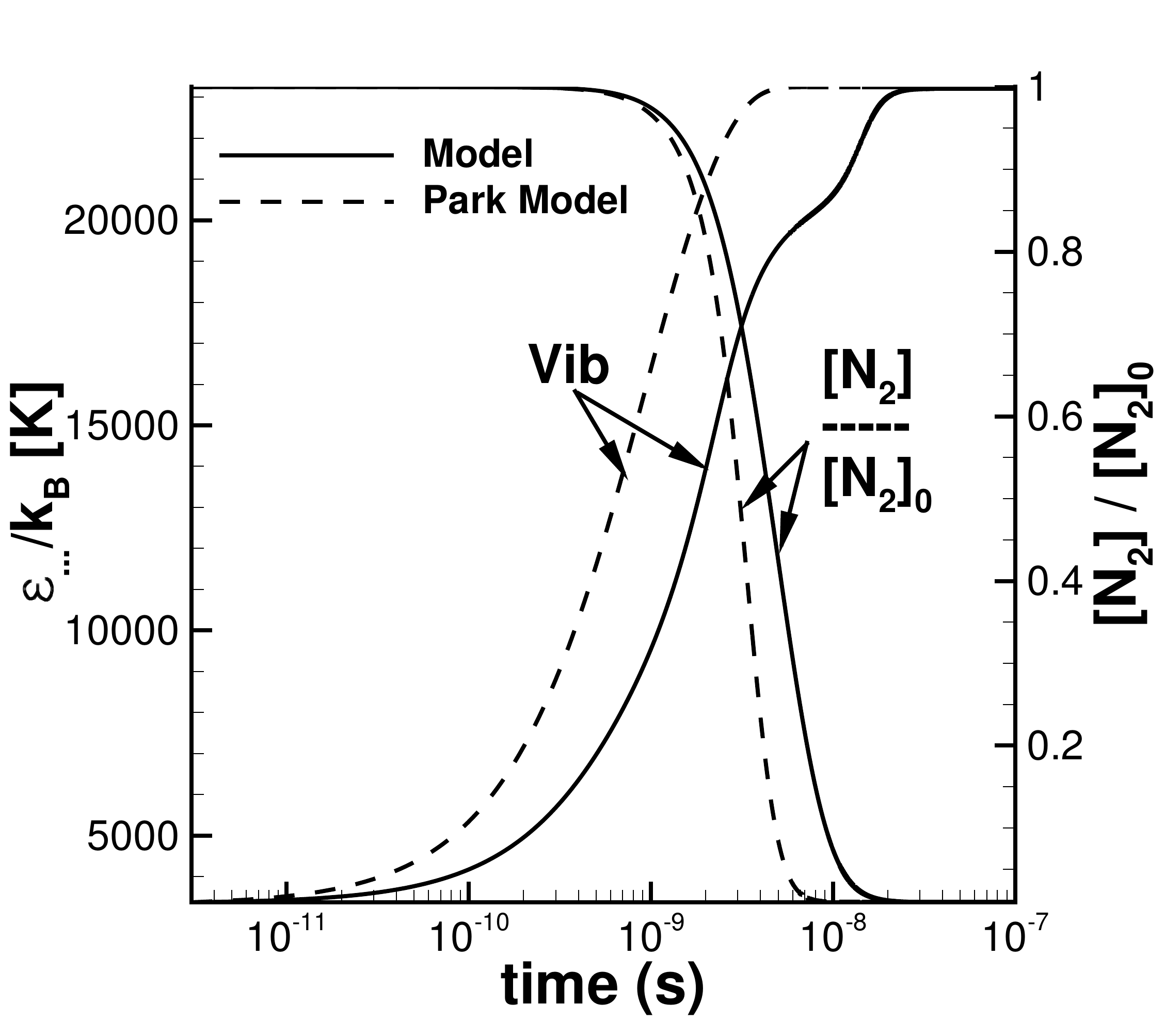}
   \label{recomb_zoomed_steady_State}
   }  
  \caption{Isothermal ro-vibrational relaxation of nitrogen at $T=20, 000$K. Comparison of Park model with \textit{ab-initio} model.} 
   \label{zerod_20K_N3_Full_recomb_Park}
\end{figure}
 \begin{figure}
\centering 
  \subfigure[Average vibrational energy]
  {
 \includegraphics[width=3.5in]{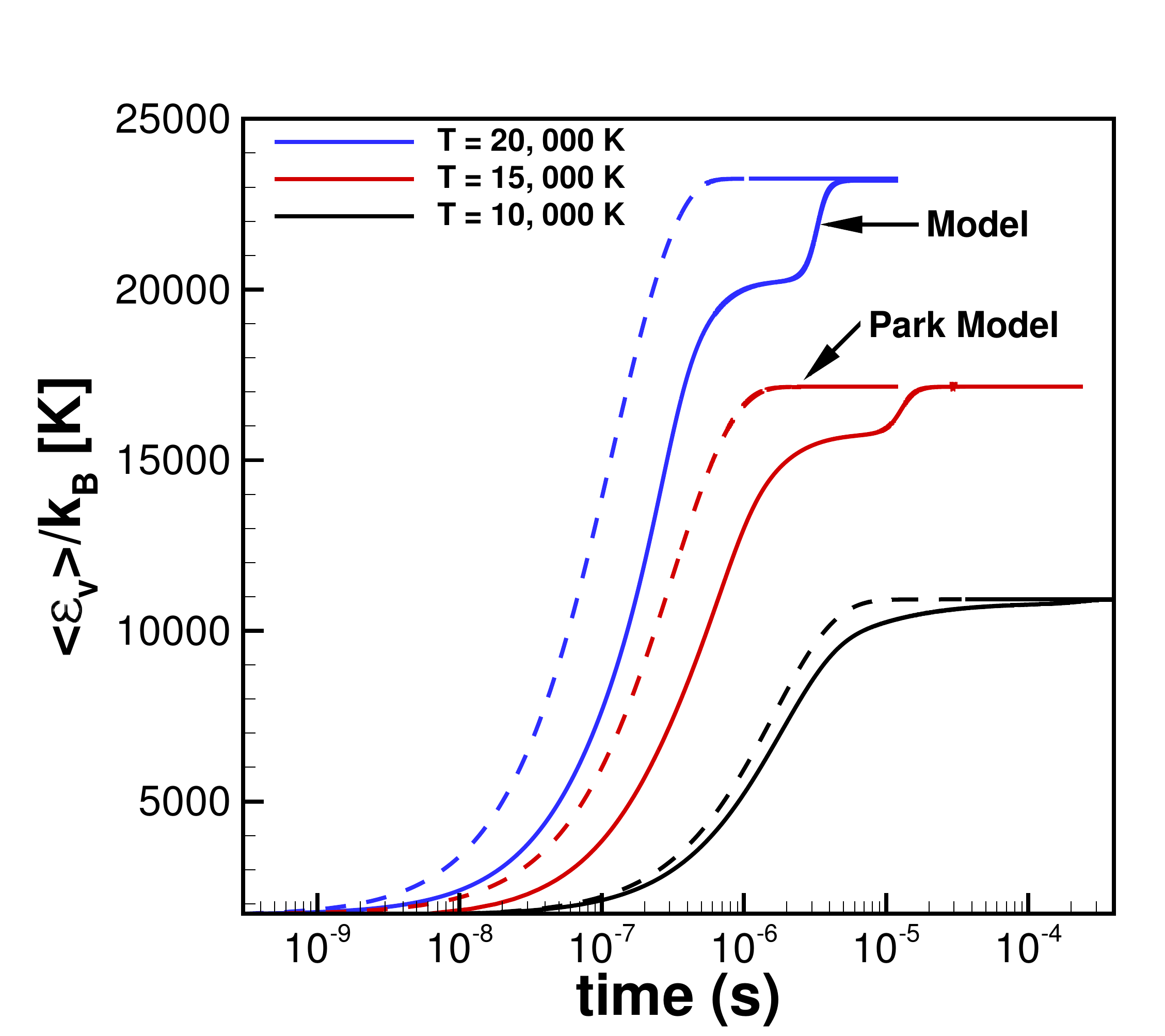}
   \label{recomb_zoomed_steady_State}
   }  
     \subfigure[Concentration of nitrogen]
     {
 \includegraphics[width=3.5in]{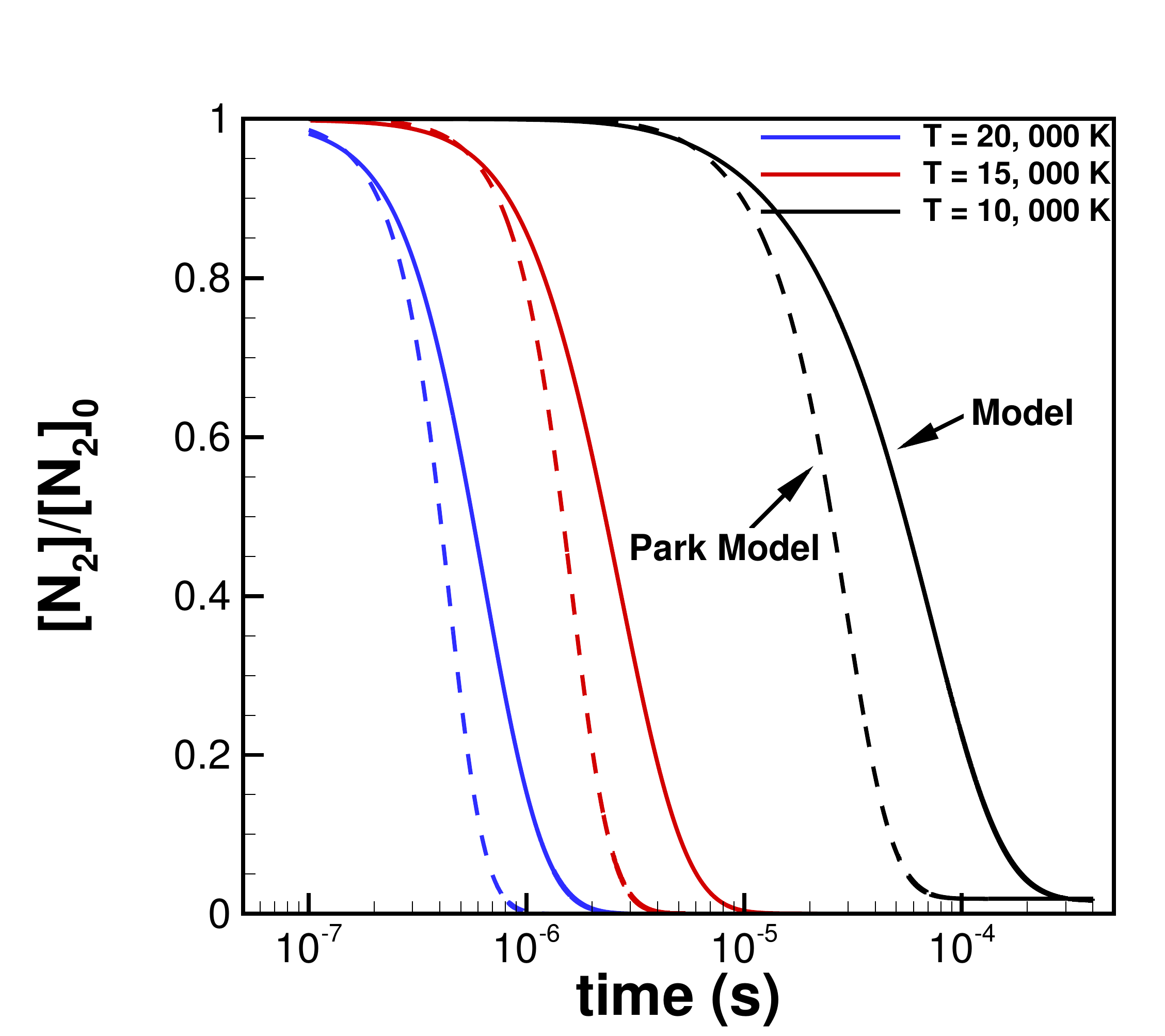}
   \label{recomb_zoomed_steady_State}
   }  
  \caption{Comparison of the model (shown as solid lines) with Park model (shown as dashed lines) for compression type flows. For all the simulations, the initial condition has pure nitrogen molecules $0.01$ kg/m$^3$ with ro-vibrational energy at $T_v = T_{rot} = 300$ K. } 
   \label{compression_zerod_N3_Full_recomb_Park_overall}
\end{figure}

 \subsection{Model Comparison to the Park Model}
Next, the results obtained from the Park model are compared with the analytical model in Fig.~\ref{zerod_20K_N3_Full_recomb_Park}. 
In the case of the Park model, the average vibrational energy of recombined atoms is set equal to the average vibrational energy of the molecules in the gas. The recombination rates in the Park model are obtained using the principle of detailed balance (Eq.~\ref{detailed_balance}). 
 The Park model significantly over-predicts the rate of dissociation. Because the resulting recombination rates are higher (higher dissociation rates), the system reaches chemical equilibrium faster compared to the results of the new model.  Therefore, the Park model may predict a gas to be in equilibrium at a point in the flow, when in fact the gas should still be in nonequilibrium. More comparative simulations at conditions representative of post-shock conditions showing higher dissociation and faster approach to equilibrium are shown in Fig.~\ref{compression_zerod_N3_Full_recomb_Park_overall}.

\begin{figure}
\centering 
  {
 \includegraphics[width=3.5in]{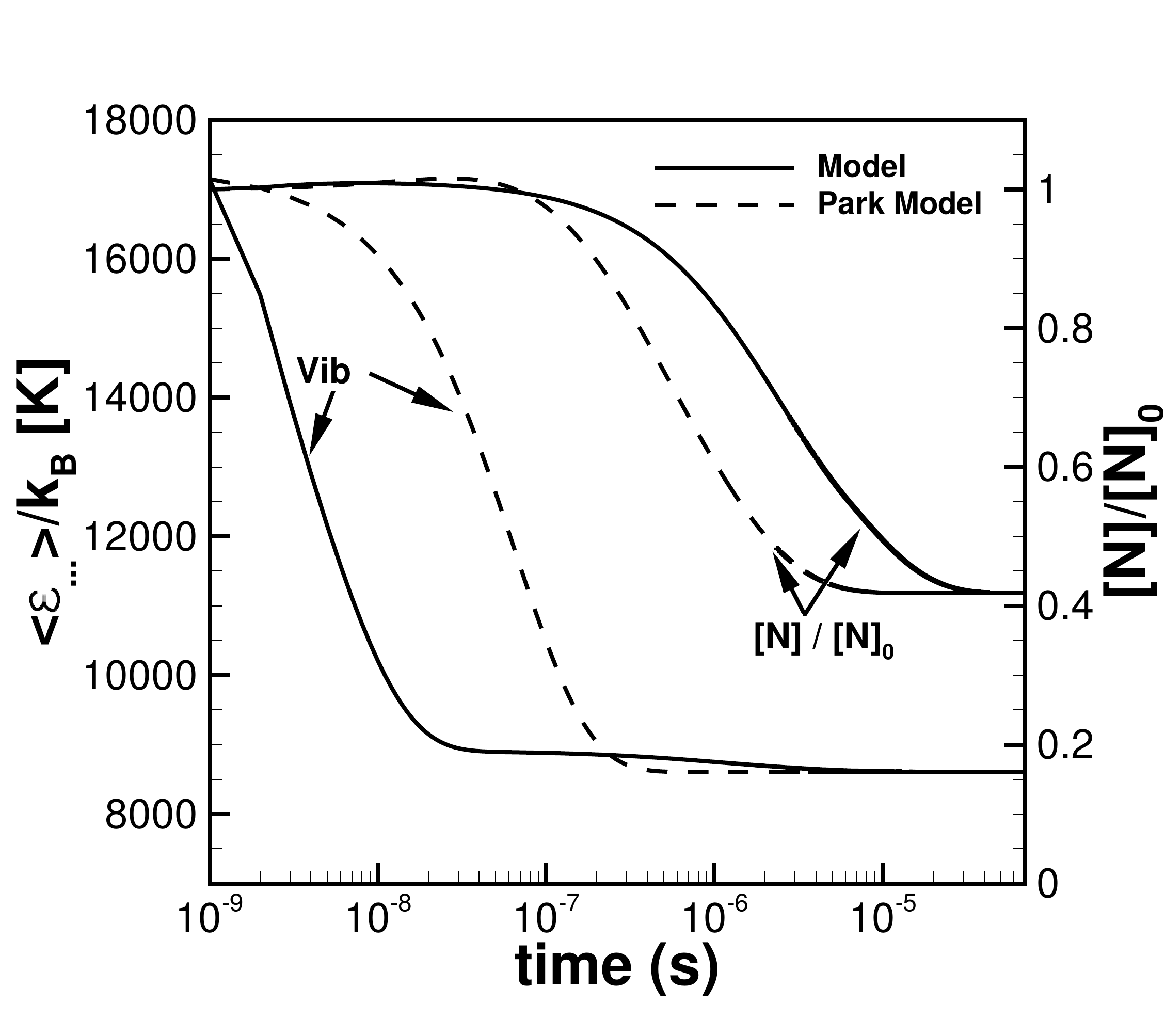}
   \label{recomb_zoomed_steady_State}
   }  
  \caption{Isothermal ro-vibrational relaxation of nitrogen at $T=T_{rot}=8, 000$K. Comparison of Park model with \textit{ab-initio} model. For this simulation, the initial density of atomic nitrogen is $0.2$ kg/m$^3$ and molecular nitrogen is $0.1$ kg/m$^3$, and $T_{v} = 15, 000$ K.  } 
   \label{Expanding_Case_Full_recomb_Park}
\end{figure}

In the last set of comparisons between the \textit{ab initio} model and the Park model, we consider a simulation at a condition representative of expanding flow i.e $T_v>T$ in Fig.~\ref{Expanding_Case_Full_recomb_Park}. In the \textit{ab initial} model, due to the lower vibration time constant from N$_2$-N interactions (see Fig.~5 in Ref~\cite{valentini2016dynamics}), the average vibrational energy relaxes faster in the \textit{ab initio} model compared to the Park model. The rapid decrease in the concentration of atomic nitrogen shows that recombination rate in the Park model is significantly faster than in the \textit{ab initio} model.

Because the recombination rates in the case of the Park model are based on the principle of detailed balance, the system reaches the equilibrium state. Recall that the dissociation rate coefficients (in the case of Park model) for $T=T_v$ are based on the fit to experiments, which have large uncertainties. Even in the case where accurate measurements can be made for the dissociation rate coefficients, the experimental rates may include non-Boltzmann (QSS) distributions effects. It is also worthwhile to consider that a vibrational temperature $T_v$ approximately equals to $T$ does not necessarily mean the gas is in equilibrium\footnote{The distinction between thermal and chemical equilibrium is not meaningful when the chemical processes are coupled to thermal evolution.}. $T_v$ extracted from the population of low lying vibrational states does not contain any information of the high-energy tail of the distribution, which affects and is affected by, the dissociation process. Therefore, merely setting $T=T_v$ in the case of Park's model does not recover equilibrium rates, which are needed in the principle of detailed balance to infer recombination rates.

\section{Summary and Conclusions}
In this article, a state-specific recombination probability model (Eq.~\ref{k_rjv_model})  is derived using the principle of microscopic reversibility and state-specific dissociation rates (derived in Ref.~\cite{singh2019consistentI}). The state-specific recombination probability is shown to have high favoring for higher vibrational energy states as expected. The role of the centrifugal barrier in limiting the probability into higher rotational energy states is also derived and analyzed. The state-specific model can be used in DSMC for recombination, which is consistent with the dissociation model developed in earlier work (see Eq.~10 of Ref.~\cite{singh2019consistentI}). 

 Using the state-specific recombination probability, a continuum-level recombination model is derived analytically in Eq.~\ref{detailed_balance}.  The  expression (Eqs.~\ref{avg_recomb_vib} and \ref{avg_recomb_rot}) for the average internal energy of molecules formed in a recombination reaction, consistent with the state specific recombination rates, is also developed. Additionally, the vibrational energy of the formed molecules in recombination is  mathematically shown to be equal to the average vibrational energy of dissociating molecules \textit{at equilibrium}. 

As a final modeling step, an extension to the generalized non-equilibrium vibrational energy distribution model is developed in Eq.~\ref{fnb_overall_diss_recomb} to now include recombination effects. The derivation uses the principle of microscopic reversibility, the QSS distribution model for a dissociating gas, and master equation with no new adjustable parameters.  The model reproduces the depletion of high-energy states due to dissociation and re-population of high energy states due to recombination. The extended  nonequilibrium distribution model is incorporated in the continuum dissociation rate model. 

Using the \textit{ab-initio} recombination model, isothermal relaxation simulations are performed and it is shown that once the re-population effects due to recombination reactions are taken into account in the continuum dissociation model, the system is driven out of a \textit{near} equilibrium stationary state  into the equilibrium state. The complete \textit{ab initio} continuum dissociation model is given by Eqs.~\ref{rate_full_nb} and \ref{edv_gen_sum_NB} which uses the extended non-Boltzmann vibrational energy distribution model  including recombination from Eq.~\ref{fnb_overall_diss_recomb}. Similarly, the continuum recombination model is given by Eqs.~\ref{detailed_balance}, \ref{avg_recomb_vib} and \ref{avg_recomb_rot}.

Isothermal relaxation simulations are also performed using the Park model, where the recombination rates are derived using the  principle of detailed balance. It is pointed out that as high-temperature dissociation is inherently a nonequilibrium process, using experimental dissociation rate data to calculate recombination rates (using the principle of detailed balance) may be inaccurate because \textit{equilibrium} dissociation rates are required according to the principle of microscopic reversibility. In isothermal relaxation simulations, it is shown that faster dissociation in the case of the Park model results in faster recombination and therefore faster approach to equilibrium, compared to the \textit{ab-initio} model. The implication of this is that an ensemble of high-temperature gas otherwise in nonequilibrium may in fact be predicted to be in (or near) equilibrium if Park's model is used. 

\section*{Acknowledgement}
This work was supported by Air Force Office of Scientific Research Grants  FA9550-16-1-0161 and  FA9550-19-1-0219 and was also partially supported
by Air Force Research Laboratory Grant FA9453-17-2-0081. Narendra Singh was partially supported by Doctoral Dissertation Fellowship at the University of Minnesota. 

\appendix

\begin{widetext}
\section{Park model \cite{park1989assessment} parameters} \label{Existing_CFD}

\begin{enumerate}
  \item  For the dissociation rate constant, Park's two temperature model \cite{park1989assessment,park1993review} is used as follows:
\begin{equation}
    k^d_{N_2-X} = C_x T_{\text{eff}}^\eta \exp\left[-\cfrac{\theta_d}{T_{\text{eff}}} \right]
\end{equation}
where $C_{N_2} = 0.01162 $ cm$^3$molecule$^{-1}$s$^{-1}$, $C_{N} = 0.0498$ cm$^3$molecule$^{-1}$s$^{-1}$, $\eta = -1.6$, $\theta_d =113, 200$ K and $T_{\text{eff}} = \sqrt{T_v T}$.
\item Time constants used in the existing CFD formulation  are based on Millikan and White experimental fits \cite{millikan1963systematics} along with the Park high temperature correction \cite{park1993review}: 
\begin{equation}
    p \tau_{v,X} = \exp \left[a_m \left(T^{-1/3}-b_m\right)-18.42 \right]
\end{equation}
where $ p \tau_{v,X}$ is in atm-s, $p$ is pressure, $a_{N_2} = 221, a_{N} = 180, b_{N_2} = 0.0290, b_{N} =0.0262$.
\item  The average energy of dissociating molecules, $\langle \epsilon_v^d \rangle$ is set as $\langle \epsilon_v \rangle$.
\item  The average vibrational energy of recombining atoms, $\langle \epsilon_v^r \rangle$ is set as $\langle \epsilon_v \rangle$.
\end{enumerate}

\section{DSMC Probability Model and Derivation of State Specific Dissociation Model}
The state-specific dissociation rate constant, $k_{jv-d}(T)$ is obtained using the state-specific dissociation probability ($p(d|\epsilon_{rel}, \epsilon_{rot}, \epsilon_{v}) $) and a Maxwell-Boltzmann distribution as:
 \begin{equation*}
\begin{split}
  k_{jv-d}(T) =  \hspace{5.0in}
     \end{split}
  \label{Rate_Derive1}
\end{equation*}
 \begin{equation}
\begin{split}
  \frac{1}{S} \left(\frac{8 k_B T}{\pi \mu_C}\right)^{1/2} \sum _{v=0}^{v_{\max}}  \sum _{j=0}^{j_{\max}} \int _0^\infty p(d|\epsilon_{rel}, \epsilon_{rot}, \epsilon_{v})   \pi b_{\max}^2  \left(\frac{\epsilon_{rel}}{k_B T}\right) \exp\left[- \frac{\epsilon_{rel}}{k_B T} \right] d \left(\frac{\epsilon_{rel}}{k_B T}\right) 
  \\
  =\frac{1}{S} \left(\frac{8 k_B T}{\pi \mu_C}\right)^{1/2} \pi b_{\max}^2  C_1 \sum _{v=0}^{v_{\max}}  \sum _{j=0}^{j_{\max}}\exp\left[\beta \frac{\epsilon_{rot}^{\text{eff}}}{\epsilon_d}\right] \exp\left[\gamma \frac{\epsilon_{v}}{\epsilon_d}\right] \exp\left[\delta \frac{|\epsilon_{int}-\epsilon_d|}{\epsilon_d}\right] \\
  \times \int _{\epsilon_d^{\ef}-\epsilon_{int}}^\infty  \left[\frac{(\epsilon_{rel}+\epsilon_{int}-\epsilon_d^{\ef})^{\alpha}}{\epsilon_{rel}}\right] 
 \left(\frac{1}{\epsilon_d}\right)^{\alpha-1}    \left(\frac{\epsilon_{rel}}{k_B T}\right) \exp\left[- \frac{\epsilon_{rel}}{k_B T} \right] d \left(\frac{\epsilon_{rel}}{k_B T}\right) 
  \end{split}
  \label{Rate_Derive2}
\end{equation}
With substitution of $(\epsilon_{rel}+\epsilon_{int}-\epsilon_d^{\ef})/(k_B T) = x$, one can reduce the above expression to the following:
\begin{equation}
\begin{split}
    k_{jv-d}(T)  =\frac{1}{S} \left(\frac{8 k_B T}{\pi \mu_C}\right)^{1/2} \pi b_{\max}^2  C_1 \exp\left[- \frac{\epsilon_d}{k_B T} \right]   \left(\frac{k_B T}{\epsilon_d}\right)^{\alpha-1} \int _{0}^\infty  x^{\alpha}
   \exp\left(- x  \right) d x \times \\
    \left\{ \sum _{v=0}^{v_{\max}}  \sum _{j=0}^{j_{\max}} \exp\left[\beta \frac{\epsilon_{rot}^{\text{eff}}}{\epsilon_d}\right] \exp\left[\gamma \frac{\epsilon_{v}}{\epsilon_d}\right] \exp\left[\delta \frac{|\epsilon_{int}-\epsilon_d|}{\epsilon_d}\right]
  \exp\left[\frac{\epsilon_{int}}{k_B T} \right]  \exp\left[-\frac{\theta_{CB}\epsilon_{rot}}{k_B T} \right]
  \right\}
   \end{split}
  \label{Rate_Derive3}
\end{equation}
Solving the integral above ($\alpha>-1$ is required for convergence, which is the case):
 \begin{equation}
\begin{split}
   & k_{jv-d}(T)  =\frac{1}{S} \left(\frac{8 k_B T}{\pi \mu_C}\right)^{1/2} \pi b_{\max}^2  C_1 \exp\left[- \frac{\epsilon_d}{k_B T} \right]   \left(\frac{k_B T}{\epsilon_d}\right)^{\alpha-1}   \\
  & \times \Gamma[1+\alpha] 
   \left\{ \sum _{v=0}^{v_{\max}}  \sum _{j=0}^{j_{\max}} \exp\left[\beta \frac{\epsilon_{rot}^{\text{eff}}}{\epsilon_d}\right] \exp\left[\gamma \frac{\epsilon_{v}}{\epsilon_d}\right] \exp\left[\delta \frac{|\epsilon_{int}-\epsilon_d|}{\epsilon_d}\right] \right. 
   \\ &\times \left.
  \exp\left[\frac{\epsilon_{int}}{k_B T} \right]
  \exp\left[-\frac{\theta_{CB}\epsilon_{rot}}{k_B T} \right]
 \right\}
   \end{split}
  \label{Rate_Derive4}
\end{equation}
where the quantity in curly brackets can be expressed as contribution from bound and quasi-bound molecules. Eqs.~\ref{Rate_Derive1}--\ref{Rate_Derive4} are also presented in the appendix of Ref.~\cite{singh2019consistentI}, where details of the state-specific reaction probability are given. The equations are reproduced here for completeness. 

\section{Ab Initio Continuum Model Equations} \label{Abinitio_Model}
The inputs required for the zero-dimensional continuum relaxation calculations (Eqs.\ref{Rate_eqn_general_recomb}--\ref{equilibrium_constant}) are presented in this appendix; namely, the rate constant ($ k^d_{N_2-X}$ ), and the average energy of the dissociating molecules ($\langle \epsilon_v^d \rangle$) are presented in this appendix.

\subsection{Dissociation Rate Constant}
The full model expression for the new non-equilibrium model was derived in Sec.~D of Ref.~\cite{singh2019consistentI}, and is listed again here for completeness:
\begin{equation}
     k^{NB} = \cfrac{  \tilde{k}(T,T_{rot},T_v; T_0) +\Lambda \hat{k}(T,T_{rot},T) }  {1 + \Lambda } ~.
     \label{rate_full_nb}
\end{equation}
Here, $\tilde{k}(T,T_{rot},T_v; T_0) $ represents the contribution predominantly from the low energy states, and $\hat{k}(T,T_{rot},T)$ represents the contribution from the high energy tail of the distribution. $\Lambda$ controls the relative importance of each term capturing the overpopulation and depletion effects. 
\subsubsection{Contribution from the low-energy states $ \tilde{k}(T,T_{rot},T_v; T_0)$}
The expression for $\tilde{k}(T,T_{rot},T_v; T_0)$ is:
\begin{equation}
\begin{split}
 \tilde{k}(T,T_{rot},T_v; T_0) 
  = A T^{\eta} \exp\left[- \frac{\epsilon_d}{k_B T} \right]    *\left[\tilde{\Hn}(\epsilon_d,0,1) +  \tilde{\Hn}(\epsilon_d^{\mx},\epsilon_d,2) \right]
  \end{split}
  \label{Rate_Final__derive}
\end{equation}

\begin{equation*}
\eta =\alpha-\frac{1}{2}; \hspace{0.35in} A= \frac{1}{S}  \left(\frac{8 k_B }{\pi \mu_C}\right)^{1/2} \pi b_{\max}^2  C_1 \Gamma[1+\alpha] \left(\frac{k_B }{\epsilon_d}\right)^{\alpha-1}
\end{equation*}

\begin{equation}
\begin{split}
    \tilde{\Hn}(x,y,n) = \hspace{3.0in} \\ \frac{\exp[(-1)^{n-1}\delta]}{ \tilde{\Zn}(T_v,T_0,T_{rot})} \frac{\exp\left[ x \hat{\zeta}_{rot} \right]  \tilde{\gn} (\hat{\zeta}_{vr}) - \exp\left[ y\hat{\zeta}_{rot} \right] \tilde{\gn} (\zeta_{v}-y\hat{\zeta}_{rot}/\epsilon_d )}{k_B \theta_{rot} \hat{\zeta}_{rot}}  ;
    \label{Ztvtr_rate__derive}
\end{split}
\end{equation}

where
\begin{equation*}
\begin{split}
\hat{\zeta}_{rot}  = -\frac{1}{k_B T_{rot}}+\frac{1}{k_B T}
+\frac{\beta-\theta_{CB}+(-1)^n\delta}{\epsilon_d} -\frac{\theta_{CB}}{k_B T}+\frac{\hat{\delta}_{rot} }{\theta_{rot} k_B};\hspace{0.10 in}\\
\label{deltarot_defs_QSS}
\end{split}
\end{equation*}
\begin{equation*}
\begin{split}
\zeta_{v}  =  +\frac{1}{k_B T_0} +\frac{1}{k_B T}  +\frac{\gamma+(-1)^n\delta}{\epsilon_d} ; \hspace{0.5in} \hat{\zeta}_{vr} = \zeta_{v}-\hat{\zeta}_{rot}
\label{deltav_defs_QSS}
\end{split}
\end{equation*}

\begin{equation*}
     \begin{split}
         \Delta_{\epsilon} = \epsilon_v(1)-\epsilon_v(0) = k_B \theta_v^I
     \end{split}
 \end{equation*}

 \begin{equation}
\begin{split}
    \tilde{\Zn}(T_v,T_0,T_{rot}) = \cfrac{T_{rot}}{\theta_{rot} -\cfrac{ T_{rot}}{\theta_{rot}k_B}\hat{\delta}_{rot} }
    \left \{ \tilde{\gn} \left(+\frac{1}{ k_B T_0}\right)  - \exp\left[-\cfrac{\epsilon_d^{\mx}}{k_B T_{rot}} +\epsilon_d^{\mx} \cfrac{\hat{\delta}_{rot}}{k_B \theta_{rot}} \right]  \times \tilde{\gn} \left(+\frac{1}{ k_B T_0}+ \cfrac{1}{k_B T_{rot}} -\cfrac{\hat{\delta}_{rot}}{k_B \theta_{rot}}\right)\right \} ;
    \label{Ztvtr_qss_derive}
\end{split}
\end{equation}
 
\begin{equation}
\begin{split}
    \hat{\delta}_v &=  - \lambda_{1,v} \frac{  3 k_B T }{2\epsilon_d}\left(1 - \cfrac{[A] [B]}{[AB]} \cfrac{1}{K_C}\right) \\
    \hspace{0.15in}  \hat{\delta}_{rot} &=  - \lambda_{1,j} \frac{  3 k_B T }{2\epsilon_d}\left(1 - \cfrac{[A] [B]}{[AB]} \cfrac{1}{K_C}\right) 
        \hspace{0.25in} 
        \label{deltas_repop}
    \end{split}
\end{equation}
where the expression for $\tilde{g}$ is:,
\begin{equation*}
  \tilde{\gn}(x)= \sum_m \tilde{\gn}_m(x)
\end{equation*}

\begin{equation*}
    \tilde{\gn}_m(x)=  \exp\left[x E_{m^-}+ \tilde{\delta}_v m^-\right]\frac{1-\exp[ (m^+-m^-) (x \theta_v^m\ k_B+ \tilde{\delta}_v) ]}{1-\exp[x\  \theta_v^m\ k_B+ \tilde{\delta}_v ]} 
\end{equation*}

\begin{equation*}
   \tilde{\delta}_v =  \hat{\delta}_v  -\cfrac{\Delta_{\epsilon} }{k_B T_v}-\cfrac{\Delta_{\epsilon} }{k_B T_0} \hspace{0.25in} 
\end{equation*}
and where $\hat{\delta}_v$ and $\hat{\delta}_{rot}$ accounts for the depletion in the population due to dissociation.

The expression for the derivative of $\tilde{\gn}(x)$ denoted as $\tilde{g}'(x)$ is:
\begin{equation*}
   \tilde{\gn}'(x)= \sum_m \tilde{\gn}'_m(x)
\end{equation*}
\begin{equation*}
    \tilde{\gn}'_m(x)= \frac{\partial \tilde{\gn}_m}{\partial x} = \tilde{\gn}_m\frac{\partial \log \tilde{\gn}_m}{\partial x} 
\end{equation*}

\begin{equation*}
\begin{split}
   \tilde{\gn}'_m(x)=\tilde{\gn}_m(x)\left\{ E_{m^-} - \frac{(m^+-m^-)  \theta_v^m\ k_B \exp[ (m^+-m^-) (x \theta_v^m\ k_B +\tilde{\delta}_v )]}{1-\exp[ (m^+-m^-)( x\theta_v^m\ k_B +\tilde{\delta}_v)]} 
   + \frac{ \theta_v^m\ k_B \exp[x\  \theta_v^m\ k_B +\tilde{\delta}_v ]}{1-\exp[x\  \theta_v^m\ k_B +\tilde{\delta}_v  ]} \right \} 
\end{split}
\end{equation*}

\subsubsection{ Contribution from the high-energy states, $\hat{k}(T, T_{rot},T_{v})$ }
The expression for $\hat{k}(T, T_{rot},T_{v})$ is:
\begin{equation}
\begin{split}
   \hat{k}(T, T_{rot},T_{v}) 
  = A T^{\eta} \exp\left[- \frac{\epsilon_d}{k_B T} \right]    *\left[\hat{\Hn}(\epsilon_d,0,1) +  \hat{\Hn}(\epsilon_d^{\mx},\epsilon_d,2) \right]
  \end{split}
  \label{Rate_Final_QSS}
\end{equation}
\begin{equation*}
\eta =\alpha-\frac{1}{2}; \hspace{0.35in} A= \frac{1}{S}  \left(\frac{8 k_B }{\pi \mu_C}\right)^{1/2} \pi b_{\max}^2  C_1 \Gamma[1+\alpha] \left(\frac{k_B }{\epsilon_d}\right)^{\alpha-1}
\end{equation*}
\begin{equation}
\begin{split}
    \hat{\Hn}(\epsilon_i,\epsilon_j,n) = \hspace{3.0in} \\ \frac{\exp[(-1)^{n-1}\delta]}{ \hat{\Zn}(T_{rot},T_v)} \frac{\exp\left[ \epsilon_i \hat{\zeta}_{rot} \right]  \hat{\gn} (\hat{\zeta}_{vr}) - \exp\left[ \epsilon_j \hat{\zeta}_{rot} \right] \hat{\gn} (\zeta_{v}-\epsilon_j \hat{\zeta}_{rot}/\epsilon_d )}{k_B \theta_{rot} \hat{\zeta}_{rot}}~,
    \label{HTvtr_rate_QSS}
\end{split}
\end{equation}
where,
\begin{equation*}
\begin{split}
\hat{\zeta}_{rot}  = -\frac{1}{k_B T_{rot}}+\frac{1}{k_B T}
+\frac{\beta-\theta_{CB}+(-1)^n\delta}{\epsilon_d} -\frac{\theta_{CB}}{k_B T}+\frac{\hat{\delta}_{rot} }{\theta_{rot} k_B};\hspace{0.10 in}\\
\label{deltarot_defs_QSS}
\end{split}
\end{equation*}
\begin{equation*}
\begin{split}
\zeta_{v}  =  -\frac{1}{k_B T_v} +\frac{1}{k_B T}  +\frac{\gamma+(-1)^n\delta}{\epsilon_d} ; \hspace{0.5in} \hat{\zeta}_{vr} = \zeta_{v}-\hat{\zeta}_{rot}
\label{deltav_defs_QSS}
\end{split}
\end{equation*}

\begin{equation}
\begin{split}
    \hat{\Zn}(T_{rot},T_{v}) = \cfrac{T_{rot}}{\theta_{rot} -\cfrac{ T_{rot}}{\theta_{rot}k_B}\hat{\delta}_{rot} }
    \left \{ \hat{\gn} \left(-\frac{1}{ k_B T_v}\right)  - \exp\left[-\cfrac{\epsilon_d^{\mx}}{k_B T_{rot}} +\epsilon_d^{\mx} \cfrac{\hat{\delta}_{rot}}{k_B \theta_{rot}} \right]  \times \hat{\gn} \left(-\frac{1}{ k_B T_v}+ \cfrac{1}{k_B T_{rot}} -\cfrac{\hat{\delta}_{rot}}{k_B \theta_{rot}}\right)\right \} ;
    \\
    \label{Ztvtr_qss}
\end{split}
\end{equation}

\begin{equation*}
    \hat{\delta}_v =  - \lambda_{1,v} \frac{  3 k_B T }{2\epsilon_d}  \hspace{0.25in}  \hat{\delta}_{rot}=  - \lambda_{1,j} \frac{  3 k_B T }{2\epsilon_d} \hspace{0.25in} ~,
\end{equation*}
where the expression for $\hat{g}$ is:,
\begin{equation*}
  \hat{\gn}(x)= \sum_m \hat{\gn}_m(x)
\end{equation*}
\begin{equation*}
    \hat{\gn}_m(x)=  \exp\left[x E_{m^-}+ \hat{\delta}_v m^-\right]\frac{1-\exp[ (m^+-m^-) (x \theta_v^m\ k_B+ \hat{\delta}_v) ]}{1-\exp[x\  \theta_v^m\ k_B+ \hat{\delta}_v ]} 
\end{equation*}

We note that rate, $\hat{k}(T, T_{rot},T_{v})$ alone could be used to model nonequilibrium because the overpopulation effect, embedded in the full expression in Eq.~\ref{rate_full_nb} does not play a significant role in the overall dissociation trend as shown in Ref.~\cite{singh2019consistentII}.

\subsection{ Average vibrational ($\langle \epsilon^{d}_v \rangle$) and rotational ($\langle \epsilon^{d}_{rot} \rangle$) energy of dissociating molecules} {\label{avgedvrotappend}}
The average vibrational energy of dissociated molecules is calculated in the manner analogous to the rate constant :
\begin{equation}
   \langle \epsilon^{d}_v \rangle ^{NB}(T,T_{rot},T_{v}) =  \cfrac{\langle \tilde{\epsilon}^{d}_v \rangle(T,T_{rot},T_v; T_0) + \langle \hat{\epsilon}^{d}_v\rangle(T,T_{rot},T_{v}) \Lambda  k_r}{1+\Lambda  k_r}
    \label{edv_gen_sum_NB}
\end{equation}
where 
\begin{equation}
    k_r =\cfrac{\hat{k}(T,T_{rot},T) }{\tilde{k}(T,T_{rot},T_v; T_0)}
\end{equation}
and where the expression for $\tilde{\epsilon}^{d}_v(T,T_{rot},T_v; T_0)$ is:

\begin{equation}
    \langle \tilde{\epsilon}^{d}_v\rangle(T,T_{rot},T_{v}) =  \frac{\tilde{\Phi}(\epsilon_d,0,1)+\tilde{\Phi}(\epsilon_d^{\mx},\epsilon_d,2)}{\tilde{\Hn}(\epsilon_d,0,1)+\tilde{\Hn}(\epsilon_d^{\mx},\epsilon_d,2)}
    \label{edv_gen_sum_QSS}
\end{equation}
The functions in Eq.~\ref{edv_gen_sum_QSS} are given by:
\begin{equation*}
\begin{split}
    & \tilde{\Phi}(\epsilon_i,\epsilon_j,n) =
    \\ 
    & \cfrac{\exp[(-1)^{n-1}\delta]}{ \tilde{\Zn}(T_v,T_0,T_{rot})} \frac{\exp\left[ \epsilon_i \hat{\zeta}_{rot}\right]  \tilde{\gn}' (\hat{\zeta}_{vr}) - \exp\left[ \epsilon_j \hat{\zeta}_{rot}\right] \tilde{\gn}' (\zeta_{v}-\epsilon_j \hat{\zeta}_{rot}/\epsilon_d )}{k_B \theta_{rot}\hat{\zeta}_{rot}} ;
    \label{Hnb_QSS}
\end{split}
\end{equation*}
\begin{equation*}
\begin{split}
    & \tilde{\Hn}(\epsilon_i,\epsilon_j,n) =
    \\
    & \cfrac{\exp[(-1)^{n-1}\delta]}{ \tilde{\Zn}(T_v,T_0,T_{rot})}  \frac{\exp\left[ \epsilon_i \hat{\zeta}_{rot}\right]  \tilde{\gn} (\hat{\zeta}_{vr}) - \exp\left[ \epsilon_j \hat{\zeta}_{rot}\right] \tilde{\gn} (\zeta_{v}-\epsilon_j \zeta_{rot}/\epsilon_d )}{k_B \theta_{rot}\hat{\zeta}_{rot}} ;
    \\
    \\
    \label{Ztvtr_QSS}
\end{split}
\end{equation*}

and the expression for $\langle \hat{\epsilon}^{d}_v\rangle(T,T_{rot},T_{v})$ is:
\begin{equation}
    \langle \hat{\epsilon}^{d}_v\rangle(T,T_{rot},T_{v}) =  \frac{\hat{\Phi}(\epsilon_d,0,1)+\hat{\Phi}(\epsilon_d^{\mx},\epsilon_d,2)}{\hat{\Hn}(\epsilon_d,0,1)+\hat{\Hn}(\epsilon_d^{\mx},\epsilon_d,2)}
    \label{avg_edv_QSS_CF2}
\end{equation}
where
\begin{equation}
\begin{split}
    \hat{\Phi}(\epsilon_i,\epsilon_j,n) = \hspace{3.0in} \\
    \cfrac{\exp[(-1)^{n-1}\delta]}{\hat{\Zn}(T_{rot},T_{v})} \frac{\exp\left[ \epsilon_i \hat{\zeta}_{rot}\right]  \hat{\gn}' (\hat{\zeta}_{vr}) - \exp\left[ \epsilon_j \hat{\zeta}_{rot}\right] \hat{\gn}' (\zeta_{v}-\epsilon_j \hat{\zeta}_{rot}/\epsilon_d )}{k_B \theta_{rot} \hat{\zeta}_{rot}} ;
    \label{Hnb_QSS_11}
\end{split}
\end{equation}
\begin{equation}
\begin{split}
    \hat{\Hn}(\epsilon_i,\epsilon_j,n) =  \hspace{3.0in} \\  \cfrac{\exp[(-1)^{n-1}\delta]}{\hat{\Zn}(T_{rot},T_{v})}  \frac{\exp\left[ \epsilon_i \hat{\zeta}_{rot}\right]  \hat{\gn} (\hat{\zeta}_{vr}) - \exp\left[ \epsilon_j \hat{\zeta}_{rot}\right] \hat{\gn} (\zeta_{v}-\epsilon_j \zeta_{rot}/\epsilon_d )}{k_B \theta_{rot} \hat{\zeta}_{rot}} ;
    \label{Ztvtr_QSS_12}
\end{split}
\end{equation}
and where the derivatives of $\hat{\gn}$ can be expressed in the following manner:
\begin{equation}
   \hat{\gn}'(x)= \sum_m \hat{\gn}'_m(x)
   \label{hatgderivativesum}
\end{equation}
\begin{equation}
    \hat{\gn}'_m(x)= \frac{\partial \hat{\gn}_m}{\partial x} = \hat{\gn}_m\frac{\partial \log \hat{\gn}_m}{\partial x} 
    \label{hatgderivative_m}
\end{equation}

\begin{equation}
\begin{split}
   \hat{\gn}'_m(x)=\hat{\gn}_m(x)\left\{ E_{m^-} - \frac{(m^+-m^-)  \theta_v^m\ k_B \exp[ (m^+-m^-) ( x \theta_v^m\ k_B +\hat{\delta}_v )]}{1-\exp[ (m^+-m^-)( x \theta_v^m\ k_B +\hat{\delta}_v)]} 
   + \frac{ \theta_v^m\ k_B \exp[x\  \theta_v^m\ k_B +\hat{\delta}_v ]}{1-\exp[x\  \theta_v^m\ k_B +\hat{\delta}_v  ]} \right \} 
\end{split}
\label{hatgderivative_CF2}
\end{equation}

As discussed in the article, we propose that the following simple approximation is accurate for the average rotational energy of dissociated molecules:
\begin{equation}
   \langle \epsilon^{d}_{rot} \rangle ^{NB}(T,T_{rot},T_{v}) =  \epsilon_d -  \langle \epsilon^{d}_{v} \rangle ^{NB}(T,T_{rot},T_{v})
    \label{avg_edrot_gen}
\end{equation}
The above proposition is also based on the finding in Ref.\cite{bender2015improved}, where the average internal energy of dissociating molecules is
approximately $\epsilon_d$ for the considered range of conditions.

\subsection{Calculation of the parameter $\Lambda$}
As seen in Eq.~B1,  $\Lambda$ controls the relative importance from both the overpopulation phase and the depleted QSS phase. As derived in the appendix of Ref.~\cite{singh2019consistentI}(see Eqs. D1-D4 of Ref.~\cite{singh2019consistentI}),  $\Lambda$  is given by:
\begin{equation}
    \Lambda = \cfrac{\langle \epsilon_v \rangle - \langle \tilde{\epsilon_v} \rangle(T_v,T_0,T_{rot})}{ \langle \hat{\epsilon}_v \rangle (T_{rot},T_{v}) -\langle \epsilon_v\rangle }
\end{equation}

The parameter $\Lambda$ requires two quantities $\langle \tilde{\epsilon_v} \rangle$ and $ \langle \hat{\epsilon}_v \rangle$, which are mathematically described as:

\begin{equation}
\begin{split}
    \langle \tilde{\epsilon_v} \rangle (T_v,T_0,T_{rot}) = \cfrac{1}{\tilde{Z}(T_v,T_0,T_{rot})}\cfrac{T_{rot}}{\left[\theta_{rot} -\cfrac{ T_{rot}}{\theta_{rot}k_B}\hat{\delta}_{rot} \right]} \left \{ -\tilde{\gn} '\left(+\frac{1}{ k_B T_0}\right)  + \exp\left[-\cfrac{\epsilon_d^{\mx}}{k_B T_{rot}} +\epsilon_d^{\mx} \cfrac{\hat{\delta}_{rot}}{k_B \theta_{rot}} \right]  
    \right. \\
    \left.
    \times \tilde{\gn}' \left(+\frac{1}{ k_B T_0}+ \cfrac{1}{k_B T_{rot}} -\cfrac{\hat{\delta}_{rot}}{k_B \theta_{rot}}\right)\right \} 
    \end{split}
\end{equation}

\begin{equation}
\begin{split}
    \langle \hat{\epsilon_v} \rangle (T_{rot},T_{v}) = \cfrac{1}{\hat{\Zn}(T,T)}\cfrac{T_{rot}}{\left[ \theta_{rot} -\cfrac{ T_{rot}}{ \theta_{rot}k_B}\hat{\delta}_{rot} \right] }
    \left \{- \hat{\gn}' \left(-\frac{1}{ k_B T_v}\right)  + \exp\left[-\cfrac{\epsilon_d^{\mx}}{k_B T_{rot}} +\epsilon_d^{\mx} \cfrac{\hat{\delta}_{rot}}{k_B \theta_{rot}} \right] \times \hat{\gn}' \left(-\frac{1}{ k_B T_v}+ \cfrac{1}{k_B T_{rot}} -\cfrac{\hat{\delta}_{rot}}{k_B \theta_{rot}}\right)\right \} 
    \end{split}
\end{equation}



\subsection{Other expressions required by the \textit{ab initio} model}
The expression for $\gn$, required for the recombination rate is:
\begin{equation}
   \gn(x)= \sum_m \gn_m(x)
   \label{gn_Boltz}
\end{equation}
\begin{equation}
    \gn_m(x)=  \exp\left[x E_{m^-}\right]\frac{1-\exp[x (m^+-m^-) \theta_v^m\ k_B ]}{1-\exp[x\  \theta_v^m\ k_B ]} 
\end{equation}
The derivative of the $\gn$ function, required in the estimation of average vibrational energy of molecule formed in recombination, is:
\begin{equation*}
   \gn(x)'= \sum_m \gn_m(x)'
\end{equation*}
\begin{equation*}
    \gn_m(x)'= \frac{\partial \gn_m}{\partial x} = \gn_m\frac{\partial \log \gn_m}{\partial x} 
\end{equation*}
\begin{equation*}
\begin{split}
   \gn_m'(x)=\gn_m(x)\left\{ E_{m^-} - \frac{(m^+-m^-) \theta_v^m\ k_B \exp[x (m^+-m^-) \theta_v^m\ k_B ]}{1-\exp[x (m^+-m^-) \theta_v^m\ k_B ]} 
   + \frac{ \theta_v^m\ k_B \exp[x\  \theta_v^m\ k_B ]}{1-\exp[x\  \theta_v^m\ k_B ]} \right \} ~,
\end{split}
\end{equation*}
where $g_m'(0)\equiv \lim_{x \rightarrow 0} g_m'(x)$.

\subsection{Ab-initio Model Parameters for Nitrogen}
All of the parameters required for the model equations are listed in Table I. 

\begin{table}[H]
\centering 
\caption{\label{tab:table-name} Constants for approximation of \textit{ab initio} energies and parameters required in the model.}
\begin{tabular}{ | m{13em} | m{4.2cm}| }
\hline
Vibrational energy  & $\theta_v^{I}=3390$ K for $\vv \in [0,9) $,\\ 
(SHO) &  $\theta_v^{II}=0.75\  \theta_v^{I}$  for $\vv \in [9,31) $,  \\ 
 &  $\theta_v^{III}=0.45\  \theta_v^{I}$  for $\vv \in [31,55) $  \\ 
\hline
Rotational energy (Rigid Rotor) & $\theta_{rot} = 2.3$  K \\ 
\hline
Centrifugal barrier & $\theta_{CB} = 0.27$  \\ 
\hline
Diatomic energies & $\epsilon_d = 9.91 \text{eV} , \epsilon_d^{\max}=14.5 \text{eV}$  \\ 
\hline
Reaction probability & $C_{1} = 8.67\times 10^{-5}, \ \alpha =1.04 $, 
\\
                      &  $ \beta = 4.81, \ \gamma = 5.91,\ \delta = 1.20$  
\\ 
\hline
Non-Boltzmann distributions \cite{Singhpnas} & $\lambda_{1,v}=0.080, \lambda_{1,j}=4.33\times 10^{-5} $  \\ 
\hline
\end{tabular}
\label{parameters}
\end{table}

\end{widetext}
\bibliography{Recombination_Model_JCP}

\begin{thebibliography}{56}%
\makeatletter
\providecommand \@ifxundefined [1]{%
 \@ifx{#1\undefined}
}%
\providecommand \@ifnum [1]{%
 \ifnum #1\expandafter \@firstoftwo
 \else \expandafter \@secondoftwo
 \fi
}%
\providecommand \@ifx [1]{%
 \ifx #1\expandafter \@firstoftwo
 \else \expandafter \@secondoftwo
 \fi
}%
\providecommand \natexlab [1]{#1}%
\providecommand \enquote  [1]{``#1''}%
\providecommand \bibnamefont  [1]{#1}%
\providecommand \bibfnamefont [1]{#1}%
\providecommand \citenamefont [1]{#1}%
\providecommand \href@noop [0]{\@secondoftwo}%
\providecommand \href [0]{\begingroup \@sanitize@url \@href}%
\providecommand \@href[1]{\@@startlink{#1}\@@href}%
\providecommand \@@href[1]{\endgroup#1\@@endlink}%
\providecommand \@sanitize@url [0]{\catcode `\\12\catcode `\$12\catcode
  `\&12\catcode `\#12\catcode `\^12\catcode `\_12\catcode `\%12\relax}%
\providecommand \@@startlink[1]{}%
\providecommand \@@endlink[0]{}%
\providecommand \url  [0]{\begingroup\@sanitize@url \@url }%
\providecommand \@url [1]{\endgroup\@href {#1}{\urlprefix }}%
\providecommand \urlprefix  [0]{URL }%
\providecommand \Eprint [0]{\href }%
\providecommand \doibase [0]{http://dx.doi.org/}%
\providecommand \selectlanguage [0]{\@gobble}%
\providecommand \bibinfo  [0]{\@secondoftwo}%
\providecommand \bibfield  [0]{\@secondoftwo}%
\providecommand \translation [1]{[#1]}%
\providecommand \BibitemOpen [0]{}%
\providecommand \bibitemStop [0]{}%
\providecommand \bibitemNoStop [0]{.\EOS\space}%
\providecommand \EOS [0]{\spacefactor3000\relax}%
\providecommand \BibitemShut  [1]{\csname bibitem#1\endcsname}%
\let\auto@bib@innerbib\@empty
\bibitem [{\citenamefont {Singh}\ and\ \citenamefont
  {Schwartzentruber}(2020{\natexlab{a}})}]{singh2019consistentI}%
  \BibitemOpen
  \bibfield  {author} {\bibinfo {author} {\bibfnamefont {N.}~\bibnamefont
  {Singh}}\ and\ \bibinfo {author} {\bibfnamefont {T.}~\bibnamefont
  {Schwartzentruber}},\ }\bibfield  {title} {\enquote {\bibinfo {title}
  {Consistent kinetic-continuum dissociation model {I}: Kinetic formulation},}\
  }\href {\doibase 10.1063/1.5142752} {\bibfield  {journal} {\bibinfo
  {journal} {The Journal of Chemical Physics}\ } (\bibinfo {year}
  {2020}{\natexlab{a}}),\ 10.1063/1.5142752}\BibitemShut {NoStop}%
\bibitem [{\citenamefont {Park}(1988)}]{park1988two}%
  \BibitemOpen
  \bibfield  {author} {\bibinfo {author} {\bibfnamefont {C.}~\bibnamefont
  {Park}},\ }\bibfield  {title} {\enquote {\bibinfo {title} {Two-temperature
  interpretation of dissociation rate data for {N}2 and {O}2},}\ }in\
  \href@noop {} {\emph {\bibinfo {booktitle} {26th Aerospace Sciences
  Meeting}}}\ (\bibinfo {year} {1988})\ p.\ \bibinfo {pages} {458}\BibitemShut
  {NoStop}%
\bibitem [{\citenamefont {Park}(1993)}]{park1993review}%
  \BibitemOpen
  \bibfield  {author} {\bibinfo {author} {\bibfnamefont {C.}~\bibnamefont
  {Park}},\ }\bibfield  {title} {\enquote {\bibinfo {title} {Review of
  chemical-kinetic problems of future {NASA} missions. {I}-earth entries},}\
  }\href@noop {} {\bibfield  {journal} {\bibinfo  {journal} {Journal of
  Thermophysics and Heat transfer}\ }\textbf {\bibinfo {volume} {7}},\ \bibinfo
  {pages} {385--398} (\bibinfo {year} {1993})}\BibitemShut {NoStop}%
\bibitem [{\citenamefont {Park}(1989)}]{park1989assessment}%
  \BibitemOpen
  \bibfield  {author} {\bibinfo {author} {\bibfnamefont {C.}~\bibnamefont
  {Park}},\ }\bibfield  {title} {\enquote {\bibinfo {title} {Assessment of
  two-temperature kinetic model for ionizing air},}\ }\href@noop {} {\bibfield
  {journal} {\bibinfo  {journal} {Journal of Thermophysics and Heat Transfer}\
  }\textbf {\bibinfo {volume} {3}},\ \bibinfo {pages} {233--244} (\bibinfo
  {year} {1989})}\BibitemShut {NoStop}%
\bibitem [{\citenamefont {Byron}(1966)}]{byron1966shock}%
  \BibitemOpen
  \bibfield  {author} {\bibinfo {author} {\bibfnamefont {S.}~\bibnamefont
  {Byron}},\ }\bibfield  {title} {\enquote {\bibinfo {title} {Shock-tube
  measurement of the rate of dissociation of nitrogen},}\ }\href@noop {}
  {\bibfield  {journal} {\bibinfo  {journal} {The Journal of {C}hemical
  Physics}\ }\textbf {\bibinfo {volume} {44}},\ \bibinfo {pages} {1378--1388}
  (\bibinfo {year} {1966})}\BibitemShut {NoStop}%
\bibitem [{\citenamefont {Appleton}, \citenamefont {Steinberg},\ and\
  \citenamefont {Liquornik}(1968)}]{appleton1968shock}%
  \BibitemOpen
  \bibfield  {author} {\bibinfo {author} {\bibfnamefont {J.}~\bibnamefont
  {Appleton}}, \bibinfo {author} {\bibfnamefont {M.}~\bibnamefont {Steinberg}},
  \ and\ \bibinfo {author} {\bibfnamefont {D.}~\bibnamefont {Liquornik}},\
  }\bibfield  {title} {\enquote {\bibinfo {title} {Shock-tube study of nitrogen
  dissociation using vacuum-ultraviolet light absorption},}\ }\href@noop {}
  {\bibfield  {journal} {\bibinfo  {journal} {The Journal of {C}hemical
  Physics}\ }\textbf {\bibinfo {volume} {48}},\ \bibinfo {pages} {599--608}
  (\bibinfo {year} {1968})}\BibitemShut {NoStop}%
\bibitem [{\citenamefont {Hanson}\ and\ \citenamefont
  {Baganoff}(1972)}]{hanson1972shock}%
  \BibitemOpen
  \bibfield  {author} {\bibinfo {author} {\bibfnamefont {R.~K.}\ \bibnamefont
  {Hanson}}\ and\ \bibinfo {author} {\bibfnamefont {D.}~\bibnamefont
  {Baganoff}},\ }\bibfield  {title} {\enquote {\bibinfo {title} {Shock-tube
  study of nitrogen dissociation rates using pressure measurements},}\
  }\href@noop {} {\bibfield  {journal} {\bibinfo  {journal} {AIAA Journal}\
  }\textbf {\bibinfo {volume} {10}},\ \bibinfo {pages} {211--215} (\bibinfo
  {year} {1972})}\BibitemShut {NoStop}%
\bibitem [{\citenamefont {Valentini}\ \emph {et~al.}(2016)\citenamefont
  {Valentini}, \citenamefont {Schwartzentruber}, \citenamefont {Bender},\ and\
  \citenamefont {Candler}}]{valentini2016dynamics}%
  \BibitemOpen
  \bibfield  {author} {\bibinfo {author} {\bibfnamefont {P.}~\bibnamefont
  {Valentini}}, \bibinfo {author} {\bibfnamefont {T.~E.}\ \bibnamefont
  {Schwartzentruber}}, \bibinfo {author} {\bibfnamefont {J.~D.}\ \bibnamefont
  {Bender}}, \ and\ \bibinfo {author} {\bibfnamefont {G.~V.}\ \bibnamefont
  {Candler}},\ }\bibfield  {title} {\enquote {\bibinfo {title} {Dynamics of
  nitrogen dissociation from direct molecular simulation},}\ }\href@noop {}
  {\bibfield  {journal} {\bibinfo  {journal} {Physical Review Fluids}\ }\textbf
  {\bibinfo {volume} {1}},\ \bibinfo {pages} {043402} (\bibinfo {year}
  {2016})}\BibitemShut {NoStop}%
\bibitem [{\citenamefont {Valentini}\ \emph {et~al.}(2015)\citenamefont
  {Valentini}, \citenamefont {Schwartzentruber}, \citenamefont {Bender},
  \citenamefont {Nompelis},\ and\ \citenamefont {Candler}}]{valentini2015N4}%
  \BibitemOpen
  \bibfield  {author} {\bibinfo {author} {\bibfnamefont {P.}~\bibnamefont
  {Valentini}}, \bibinfo {author} {\bibfnamefont {T.~E.}\ \bibnamefont
  {Schwartzentruber}}, \bibinfo {author} {\bibfnamefont {J.~D.}\ \bibnamefont
  {Bender}}, \bibinfo {author} {\bibfnamefont {I.}~\bibnamefont {Nompelis}}, \
  and\ \bibinfo {author} {\bibfnamefont {G.~V.}\ \bibnamefont {Candler}},\
  }\bibfield  {title} {\enquote {\bibinfo {title} {Direct molecular simulation
  of nitrogen dissociation based on an ab initio potential energy surface},}\
  }\href@noop {} {\bibfield  {journal} {\bibinfo  {journal} {Physics of Fluids
  (1994-present)}\ }\textbf {\bibinfo {volume} {27}},\ \bibinfo {pages}
  {086102} (\bibinfo {year} {2015})}\BibitemShut {NoStop}%
\bibitem [{\citenamefont {Grover}\ \emph {et~al.}(2019)\citenamefont {Grover},
  \citenamefont {Schwartzentruber}, \citenamefont {Varga},\ and\ \citenamefont
  {Truhlar}}]{grover2019jtht}%
  \BibitemOpen
  \bibfield  {author} {\bibinfo {author} {\bibfnamefont {M.~S.}\ \bibnamefont
  {Grover}}, \bibinfo {author} {\bibfnamefont {T.~E.}\ \bibnamefont
  {Schwartzentruber}}, \bibinfo {author} {\bibfnamefont {Z.}~\bibnamefont
  {Varga}}, \ and\ \bibinfo {author} {\bibfnamefont {D.~G.}\ \bibnamefont
  {Truhlar}},\ }\bibfield  {title} {\enquote {\bibinfo {title} {Vibrational
  energy transfer and collision-induced dissociation in {O}+{O}2 collisions},}\
  }\href@noop {} {\bibfield  {journal} {\bibinfo  {journal} {Journal of
  Thermophysics and Heat Transfer}\ }\textbf {\bibinfo {volume} {33}},\
  \bibinfo {pages} {797--807} (\bibinfo {year} {2019})}\BibitemShut {NoStop}%
\bibitem [{\citenamefont {Grover}, \citenamefont {Torres},\ and\ \citenamefont
  {Schwartzentruber}(2019)}]{grover2019direct}%
  \BibitemOpen
  \bibfield  {author} {\bibinfo {author} {\bibfnamefont {M.~S.}\ \bibnamefont
  {Grover}}, \bibinfo {author} {\bibfnamefont {E.}~\bibnamefont {Torres}}, \
  and\ \bibinfo {author} {\bibfnamefont {T.~E.}\ \bibnamefont
  {Schwartzentruber}},\ }\bibfield  {title} {\enquote {\bibinfo {title} {Direct
  molecular simulation of internal energy relaxation and dissociation in
  oxygen},}\ }\href@noop {} {\bibfield  {journal} {\bibinfo  {journal} {Physics
  of Fluids}\ }\textbf {\bibinfo {volume} {31}},\ \bibinfo {pages} {076107}
  (\bibinfo {year} {2019})}\BibitemShut {NoStop}%
\bibitem [{\citenamefont {Panesi}\ \emph {et~al.}(2013)\citenamefont {Panesi},
  \citenamefont {Jaffe}, \citenamefont {Schwenke},\ and\ \citenamefont
  {Magin}}]{panesi2013rovibrational}%
  \BibitemOpen
  \bibfield  {author} {\bibinfo {author} {\bibfnamefont {M.}~\bibnamefont
  {Panesi}}, \bibinfo {author} {\bibfnamefont {R.~L.}\ \bibnamefont {Jaffe}},
  \bibinfo {author} {\bibfnamefont {D.~W.}\ \bibnamefont {Schwenke}}, \ and\
  \bibinfo {author} {\bibfnamefont {T.~E.}\ \bibnamefont {Magin}},\ }\bibfield
  {title} {\enquote {\bibinfo {title} {Rovibrational internal energy transfer
  and dissociation of {N}$_2$ ($^1\sum$g+)- {N} ($^4${S}$_u$) system in
  hypersonic flows},}\ }\href@noop {} {\bibfield  {journal} {\bibinfo
  {journal} {The Journal of {C}hemical physics}\ }\textbf {\bibinfo {volume}
  {138}},\ \bibinfo {pages} {044312} (\bibinfo {year} {2013})}\BibitemShut
  {NoStop}%
\bibitem [{\citenamefont {Jaffe}\ \emph {et~al.}(2018)\citenamefont {Jaffe},
  \citenamefont {Grover}, \citenamefont {Venturi}, \citenamefont {Schwenke},
  \citenamefont {Valentini}, \citenamefont {Schwartzentruber},\ and\
  \citenamefont {Panesi}}]{jaffe2018comparison}%
  \BibitemOpen
  \bibfield  {author} {\bibinfo {author} {\bibfnamefont {R.~L.}\ \bibnamefont
  {Jaffe}}, \bibinfo {author} {\bibfnamefont {M.}~\bibnamefont {Grover}},
  \bibinfo {author} {\bibfnamefont {S.}~\bibnamefont {Venturi}}, \bibinfo
  {author} {\bibfnamefont {D.~W.}\ \bibnamefont {Schwenke}}, \bibinfo {author}
  {\bibfnamefont {P.}~\bibnamefont {Valentini}}, \bibinfo {author}
  {\bibfnamefont {T.~E.}\ \bibnamefont {Schwartzentruber}}, \ and\ \bibinfo
  {author} {\bibfnamefont {M.}~\bibnamefont {Panesi}},\ }\bibfield  {title}
  {\enquote {\bibinfo {title} {Comparison of potential energy surface and
  computed rate coefficients for {N}2 dissociation},}\ }\href@noop {}
  {\bibfield  {journal} {\bibinfo  {journal} {Journal of Thermophysics and Heat
  Transfer}\ }\textbf {\bibinfo {volume} {32}},\ \bibinfo {pages} {869--881}
  (\bibinfo {year} {2018})}\BibitemShut {NoStop}%
\bibitem [{\citenamefont {Paukku}\ \emph {et~al.}(2013)\citenamefont {Paukku},
  \citenamefont {Yang}, \citenamefont {Varga},\ and\ \citenamefont
  {Truhlar}}]{paukku2013global}%
  \BibitemOpen
  \bibfield  {author} {\bibinfo {author} {\bibfnamefont {Y.}~\bibnamefont
  {Paukku}}, \bibinfo {author} {\bibfnamefont {K.~R.}\ \bibnamefont {Yang}},
  \bibinfo {author} {\bibfnamefont {Z.}~\bibnamefont {Varga}}, \ and\ \bibinfo
  {author} {\bibfnamefont {D.~G.}\ \bibnamefont {Truhlar}},\ }\bibfield
  {title} {\enquote {\bibinfo {title} {Global ab initio ground-state potential
  energy surface of {N}4},}\ }\href@noop {} {\bibfield  {journal} {\bibinfo
  {journal} {The Journal of {C}hemical physics}\ }\textbf {\bibinfo {volume}
  {139}},\ \bibinfo {pages} {044309} (\bibinfo {year} {2013})}\BibitemShut
  {NoStop}%
\bibitem [{\citenamefont {Paukku}\ \emph {et~al.}(2014)\citenamefont {Paukku},
  \citenamefont {Yang}, \citenamefont {Varga},\ and\ \citenamefont
  {Truhlar}}]{paukku2014erratum}%
  \BibitemOpen
  \bibfield  {author} {\bibinfo {author} {\bibfnamefont {Y.}~\bibnamefont
  {Paukku}}, \bibinfo {author} {\bibfnamefont {K.~R.}\ \bibnamefont {Yang}},
  \bibinfo {author} {\bibfnamefont {Z.}~\bibnamefont {Varga}}, \ and\ \bibinfo
  {author} {\bibfnamefont {D.~G.}\ \bibnamefont {Truhlar}},\ }\bibfield
  {title} {\enquote {\bibinfo {title} {Erratum:“global ab initio ground-state
  potential energy surface of {N4}”[{J}. {C}hem. {P}hys. 139, 044309
  (2013)]},}\ }\href@noop {} {\bibfield  {journal} {\bibinfo  {journal} {The
  Journal of {C}hemical physics}\ } (\bibinfo {year} {2014})}\BibitemShut
  {NoStop}%
\bibitem [{\citenamefont {Paukku}\ \emph {et~al.}(2017)\citenamefont {Paukku},
  \citenamefont {Yang}, \citenamefont {Varga}, \citenamefont {Song},
  \citenamefont {Bender},\ and\ \citenamefont {Truhlar}}]{paukku2017potential}%
  \BibitemOpen
  \bibfield  {author} {\bibinfo {author} {\bibfnamefont {Y.}~\bibnamefont
  {Paukku}}, \bibinfo {author} {\bibfnamefont {K.~R.}\ \bibnamefont {Yang}},
  \bibinfo {author} {\bibfnamefont {Z.}~\bibnamefont {Varga}}, \bibinfo
  {author} {\bibfnamefont {G.}~\bibnamefont {Song}}, \bibinfo {author}
  {\bibfnamefont {J.~D.}\ \bibnamefont {Bender}}, \ and\ \bibinfo {author}
  {\bibfnamefont {D.~G.}\ \bibnamefont {Truhlar}},\ }\bibfield  {title}
  {\enquote {\bibinfo {title} {Potential energy surfaces of quintet and singlet
  {O}4},}\ }\href@noop {} {\bibfield  {journal} {\bibinfo  {journal} {The
  Journal of {C}hemical Physics}\ }\textbf {\bibinfo {volume} {147}},\ \bibinfo
  {pages} {034301} (\bibinfo {year} {2017})}\BibitemShut {NoStop}%
\bibitem [{\citenamefont {Varga}, \citenamefont {Paukku},\ and\ \citenamefont
  {Truhlar}(2017)}]{O2OTruhlar}%
  \BibitemOpen
  \bibfield  {author} {\bibinfo {author} {\bibfnamefont {Z.}~\bibnamefont
  {Varga}}, \bibinfo {author} {\bibfnamefont {Y.}~\bibnamefont {Paukku}}, \
  and\ \bibinfo {author} {\bibfnamefont {D.~G.}\ \bibnamefont {Truhlar}},\
  }\bibfield  {title} {\enquote {\bibinfo {title} {Potential energy surfaces
  for {O} + {O}$_2$ collisions},}\ }\href {\doibase 10.1063/1.4997169}
  {\bibfield  {journal} {\bibinfo  {journal} {The Journal of {C}hemical
  Physics}\ }\textbf {\bibinfo {volume} {147}},\ \bibinfo {pages} {154312}
  (\bibinfo {year} {2017})},\ \Eprint
  {http://arxiv.org/abs/https://doi.org/10.1063/1.4997169}
  {https://doi.org/10.1063/1.4997169} \BibitemShut {NoStop}%
\bibitem [{\citenamefont {Varga}\ \emph {et~al.}(2016)\citenamefont {Varga},
  \citenamefont {Meana-Pa{\~n}eda}, \citenamefont {Song}, \citenamefont
  {Paukku},\ and\ \citenamefont {Truhlar}}]{varga2016potential}%
  \BibitemOpen
  \bibfield  {author} {\bibinfo {author} {\bibfnamefont {Z.}~\bibnamefont
  {Varga}}, \bibinfo {author} {\bibfnamefont {R.}~\bibnamefont
  {Meana-Pa{\~n}eda}}, \bibinfo {author} {\bibfnamefont {G.}~\bibnamefont
  {Song}}, \bibinfo {author} {\bibfnamefont {Y.}~\bibnamefont {Paukku}}, \ and\
  \bibinfo {author} {\bibfnamefont {D.~G.}\ \bibnamefont {Truhlar}},\
  }\bibfield  {title} {\enquote {\bibinfo {title} {Potential energy surface of
  triplet {N2}{O2}},}\ }\href@noop {} {\bibfield  {journal} {\bibinfo
  {journal} {The Journal of {C}hemical physics}\ }\textbf {\bibinfo {volume}
  {144}},\ \bibinfo {pages} {024310} (\bibinfo {year} {2016})}\BibitemShut
  {NoStop}%
\bibitem [{\citenamefont {Lin}\ \emph {et~al.}(2016)\citenamefont {Lin},
  \citenamefont {Varga}, \citenamefont {Song}, \citenamefont {Paukku},\ and\
  \citenamefont {Truhlar}}]{lin2016global}%
  \BibitemOpen
  \bibfield  {author} {\bibinfo {author} {\bibfnamefont {W.}~\bibnamefont
  {Lin}}, \bibinfo {author} {\bibfnamefont {Z.}~\bibnamefont {Varga}}, \bibinfo
  {author} {\bibfnamefont {G.}~\bibnamefont {Song}}, \bibinfo {author}
  {\bibfnamefont {Y.}~\bibnamefont {Paukku}}, \ and\ \bibinfo {author}
  {\bibfnamefont {D.~G.}\ \bibnamefont {Truhlar}},\ }\bibfield  {title}
  {\enquote {\bibinfo {title} {Global triplet potential energy surfaces for the
  {N2} (x$^1 \sum $)+ {O}($^3${P}) $\rightarrow $ {NO} (x$^2\pi$)+ {N} ($^4$
  {S}) reaction},}\ }\href@noop {} {\bibfield  {journal} {\bibinfo  {journal}
  {The Journal of {C}hemical physics}\ }\textbf {\bibinfo {volume} {144}},\
  \bibinfo {pages} {024309} (\bibinfo {year} {2016})}\BibitemShut {NoStop}%
\bibitem [{\citenamefont {Panesi}\ \emph {et~al.}(2014)\citenamefont {Panesi},
  \citenamefont {Munaf\`o}, \citenamefont {Magin},\ and\ \citenamefont
  {Jaffe}}]{panesi2014pre}%
  \BibitemOpen
  \bibfield  {author} {\bibinfo {author} {\bibfnamefont {M.}~\bibnamefont
  {Panesi}}, \bibinfo {author} {\bibfnamefont {A.}~\bibnamefont {Munaf\`o}},
  \bibinfo {author} {\bibfnamefont {T.~E.}\ \bibnamefont {Magin}}, \ and\
  \bibinfo {author} {\bibfnamefont {R.~L.}\ \bibnamefont {Jaffe}},\ }\bibfield
  {title} {\enquote {\bibinfo {title} {Nonequilibrium shock-heated nitrogen
  flows using a rovibrational state-to-state method},}\ }\href {\doibase
  10.1103/PhysRevE.90.013009} {\bibfield  {journal} {\bibinfo  {journal} {Phys.
  Rev. E}\ }\textbf {\bibinfo {volume} {90}},\ \bibinfo {pages} {013009}
  (\bibinfo {year} {2014})}\BibitemShut {NoStop}%
\bibitem [{\citenamefont {Macdonald}\ \emph
  {et~al.}(2018{\natexlab{a}})\citenamefont {Macdonald}, \citenamefont {Jaffe},
  \citenamefont {Schwenke},\ and\ \citenamefont {Panesi}}]{macdonald2018_QCT}%
  \BibitemOpen
  \bibfield  {author} {\bibinfo {author} {\bibfnamefont {R.}~\bibnamefont
  {Macdonald}}, \bibinfo {author} {\bibfnamefont {R.}~\bibnamefont {Jaffe}},
  \bibinfo {author} {\bibfnamefont {D.}~\bibnamefont {Schwenke}}, \ and\
  \bibinfo {author} {\bibfnamefont {M.}~\bibnamefont {Panesi}},\ }\bibfield
  {title} {\enquote {\bibinfo {title} {Construction of a coarse-grain
  quasi-classical trajectory method. i. theory and application to n2--n2
  system},}\ }\href@noop {} {\bibfield  {journal} {\bibinfo  {journal} {The
  Journal of {C}hemical physics}\ }\textbf {\bibinfo {volume} {148}},\ \bibinfo
  {pages} {054309} (\bibinfo {year} {2018}{\natexlab{a}})}\BibitemShut
  {NoStop}%
\bibitem [{\citenamefont {Andrienko}\ and\ \citenamefont
  {Boyd}(2018)}]{andrienko2018vibrational}%
  \BibitemOpen
  \bibfield  {author} {\bibinfo {author} {\bibfnamefont {D.~A.}\ \bibnamefont
  {Andrienko}}\ and\ \bibinfo {author} {\bibfnamefont {I.~D.}\ \bibnamefont
  {Boyd}},\ }\bibfield  {title} {\enquote {\bibinfo {title} {Vibrational energy
  transfer and dissociation in {O}2--{N}2 collisions at hyperthermal
  temperatures},}\ }\href@noop {} {\bibfield  {journal} {\bibinfo  {journal}
  {The Journal of {C}hemical physics}\ }\textbf {\bibinfo {volume} {148}},\
  \bibinfo {pages} {084309} (\bibinfo {year} {2018})}\BibitemShut {NoStop}%
\bibitem [{\citenamefont {Magin}\ \emph {et~al.}(2012)\citenamefont {Magin},
  \citenamefont {Panesi}, \citenamefont {Bourdon}, \citenamefont {Jaffe},\ and\
  \citenamefont {Schwenke}}]{magin2012coarse}%
  \BibitemOpen
  \bibfield  {author} {\bibinfo {author} {\bibfnamefont {T.~E.}\ \bibnamefont
  {Magin}}, \bibinfo {author} {\bibfnamefont {M.}~\bibnamefont {Panesi}},
  \bibinfo {author} {\bibfnamefont {A.}~\bibnamefont {Bourdon}}, \bibinfo
  {author} {\bibfnamefont {R.~L.}\ \bibnamefont {Jaffe}}, \ and\ \bibinfo
  {author} {\bibfnamefont {D.~W.}\ \bibnamefont {Schwenke}},\ }\bibfield
  {title} {\enquote {\bibinfo {title} {Coarse-grain model for internal energy
  excitation and dissociation of molecular nitrogen},}\ }\href@noop {}
  {\bibfield  {journal} {\bibinfo  {journal} {{C}hemical Physics}\ }\textbf
  {\bibinfo {volume} {398}},\ \bibinfo {pages} {90--95} (\bibinfo {year}
  {2012})}\BibitemShut {NoStop}%
\bibitem [{\citenamefont {Chaudhry}\ \emph {et~al.}(2018)\citenamefont
  {Chaudhry}, \citenamefont {Bender}, \citenamefont {Schwartzentruber},\ and\
  \citenamefont {Candler}}]{chaudhry2018qct}%
  \BibitemOpen
  \bibfield  {author} {\bibinfo {author} {\bibfnamefont {R.~S.}\ \bibnamefont
  {Chaudhry}}, \bibinfo {author} {\bibfnamefont {J.~D.}\ \bibnamefont
  {Bender}}, \bibinfo {author} {\bibfnamefont {T.~E.}\ \bibnamefont
  {Schwartzentruber}}, \ and\ \bibinfo {author} {\bibfnamefont {G.~V.}\
  \bibnamefont {Candler}},\ }\bibfield  {title} {\enquote {\bibinfo {title}
  {Quasiclassical trajectory analysis of nitrogen for high-temperature chemical
  kinetics},}\ }\href@noop {} {\bibfield  {journal} {\bibinfo  {journal}
  {Journal of Thermophysics and Heat Transfer}\ }\textbf {\bibinfo {volume}
  {32}},\ \bibinfo {pages} {833--845} (\bibinfo {year} {2018})}\BibitemShut
  {NoStop}%
\bibitem [{\citenamefont {Voelkel}, \citenamefont {Varghese},\ and\
  \citenamefont {Raman}(2017)}]{voelkel2017multitemperature}%
  \BibitemOpen
  \bibfield  {author} {\bibinfo {author} {\bibfnamefont {S.}~\bibnamefont
  {Voelkel}}, \bibinfo {author} {\bibfnamefont {P.~L.}\ \bibnamefont
  {Varghese}}, \ and\ \bibinfo {author} {\bibfnamefont {V.}~\bibnamefont
  {Raman}},\ }\bibfield  {title} {\enquote {\bibinfo {title} {Multitemperature
  dissociation rate of {N}$_2$+ {N}$2$ $\rightarrow$ {N} 2+ {N}+ {N} calculated
  using selective sampling quasi-classical trajectory analysis},}\ }\href@noop
  {} {\bibfield  {journal} {\bibinfo  {journal} {Journal of Thermophysics and
  Heat Transfer}\ }\textbf {\bibinfo {volume} {31}},\ \bibinfo {pages}
  {965--975} (\bibinfo {year} {2017})}\BibitemShut {NoStop}%
\bibitem [{\citenamefont {Singh}\ and\ \citenamefont
  {Schwartzentruber}(2020{\natexlab{b}})}]{singh2019nonboltzmann}%
  \BibitemOpen
  \bibfield  {author} {\bibinfo {author} {\bibfnamefont {N.}~\bibnamefont
  {Singh}}\ and\ \bibinfo {author} {\bibfnamefont {T.}~\bibnamefont
  {Schwartzentruber}},\ }\bibfield  {title} {\enquote {\bibinfo {title}
  {Non-boltzmann vibrational energy distributions and coupling to dissociation
  rate},}\ }\href {\doibase 10.1063/1.5142732} {\bibfield  {journal} {\bibinfo
  {journal} {The Journal of Chemical Physics}\ } (\bibinfo {year}
  {2020}{\natexlab{b}}),\ 10.1063/1.5142732}\BibitemShut {NoStop}%
\bibitem [{\citenamefont {Singh}\ and\ \citenamefont
  {Schwartzentruber}(2020{\natexlab{c}})}]{singh2019consistentII}%
  \BibitemOpen
  \bibfield  {author} {\bibinfo {author} {\bibfnamefont {N.}~\bibnamefont
  {Singh}}\ and\ \bibinfo {author} {\bibfnamefont {T.}~\bibnamefont
  {Schwartzentruber}},\ }\bibfield  {title} {\enquote {\bibinfo {title}
  {Consistent kinetic-continuum dissociation model {II}: Continuum formulation
  and verification},}\ }\href {\doibase 10.1063/1.5142754} {\bibfield
  {journal} {\bibinfo  {journal} {The Journal of Chemical Physics}\ } (\bibinfo
  {year} {2020}{\natexlab{c}}),\ 10.1063/1.5142754}\BibitemShut {NoStop}%
\bibitem [{\citenamefont {Singh}\ and\ \citenamefont
  {Schwartzentruber}(2020{\natexlab{d}})}]{singh2020aiaanon}%
  \BibitemOpen
  \bibfield  {author} {\bibinfo {author} {\bibfnamefont {N.}~\bibnamefont
  {Singh}}\ and\ \bibinfo {author} {\bibfnamefont {T.~E.}\ \bibnamefont
  {Schwartzentruber}},\ }\bibfield  {title} {\enquote {\bibinfo {title}
  {Non-boltzmann vibrational energy distribution model for shock-heated
  flows},}\ }in\ \href@noop {} {\emph {\bibinfo {booktitle} {AIAA Scitech 2020
  Forum}}}\ (\bibinfo {year} {2020})\ p.\ \bibinfo {pages} {1715}\BibitemShut
  {NoStop}%
\bibitem [{\citenamefont {Singh}\ and\ \citenamefont
  {Schwartzentruber}(2020{\natexlab{e}})}]{singh2020aiaaconsistent}%
  \BibitemOpen
  \bibfield  {author} {\bibinfo {author} {\bibfnamefont {N.}~\bibnamefont
  {Singh}}\ and\ \bibinfo {author} {\bibfnamefont {T.~E.}\ \bibnamefont
  {Schwartzentruber}},\ }\bibfield  {title} {\enquote {\bibinfo {title}
  {Consistent kinetic and continuum dissociation models for high-temperature
  air},}\ }in\ \href@noop {} {\emph {\bibinfo {booktitle} {AIAA Scitech 2020
  Forum}}}\ (\bibinfo {year} {2020})\ p.\ \bibinfo {pages} {1716}\BibitemShut
  {NoStop}%
\bibitem [{\citenamefont {Singh}\ and\ \citenamefont
  {Schwartzentruber}(2018)}]{Singhpnas}%
  \BibitemOpen
  \bibfield  {author} {\bibinfo {author} {\bibfnamefont {N.}~\bibnamefont
  {Singh}}\ and\ \bibinfo {author} {\bibfnamefont {T.}~\bibnamefont
  {Schwartzentruber}},\ }\bibfield  {title} {\enquote {\bibinfo {title}
  {Nonequilibrium internal energy distributions during dissociation},}\ }\href
  {\doibase 10.1073/pnas.1713840115} {\bibfield  {journal} {\bibinfo  {journal}
  {Proceedings of the National Academy of Sciences}\ }\textbf {\bibinfo
  {volume} {115}},\ \bibinfo {pages} {47--52} (\bibinfo {year} {2018})},\
  \Eprint {http://arxiv.org/abs/http://www.pnas.org/content/115/1/47.full.pdf}
  {http://www.pnas.org/content/115/1/47.full.pdf} \BibitemShut {NoStop}%
\bibitem [{\citenamefont {Andrienko}\ and\ \citenamefont
  {Boyd}(2015)}]{andrienko2015high}%
  \BibitemOpen
  \bibfield  {author} {\bibinfo {author} {\bibfnamefont {D.~A.}\ \bibnamefont
  {Andrienko}}\ and\ \bibinfo {author} {\bibfnamefont {I.~D.}\ \bibnamefont
  {Boyd}},\ }\bibfield  {title} {\enquote {\bibinfo {title} {High fidelity
  modeling of thermal relaxation and dissociation of oxygen},}\ }\href@noop {}
  {\bibfield  {journal} {\bibinfo  {journal} {Physics of Fluids}\ }\textbf
  {\bibinfo {volume} {27}},\ \bibinfo {pages} {116101} (\bibinfo {year}
  {2015})}\BibitemShut {NoStop}%
\bibitem [{\citenamefont {Kustova}\ \emph {et~al.}(2016)\citenamefont
  {Kustova}, \citenamefont {Nagnibeda}, \citenamefont {Oblapenko},
  \citenamefont {Savelev},\ and\ \citenamefont
  {Sharafutdinov}}]{kustova2016advanced}%
  \BibitemOpen
  \bibfield  {author} {\bibinfo {author} {\bibfnamefont {E.}~\bibnamefont
  {Kustova}}, \bibinfo {author} {\bibfnamefont {E.}~\bibnamefont {Nagnibeda}},
  \bibinfo {author} {\bibfnamefont {G.}~\bibnamefont {Oblapenko}}, \bibinfo
  {author} {\bibfnamefont {A.}~\bibnamefont {Savelev}}, \ and\ \bibinfo
  {author} {\bibfnamefont {I.}~\bibnamefont {Sharafutdinov}},\ }\bibfield
  {title} {\enquote {\bibinfo {title} {Advanced models for vibrational-chemical
  coupling in multi-temperature flows},}\ }\href@noop {} {\bibfield  {journal}
  {\bibinfo  {journal} {{C}hemical Physics}\ }\textbf {\bibinfo {volume}
  {464}},\ \bibinfo {pages} {1--13} (\bibinfo {year} {2016})}\BibitemShut
  {NoStop}%
\bibitem [{\citenamefont {Chaudhry}\ \emph {et~al.}(2020)\citenamefont
  {Chaudhry}, \citenamefont {Boyd}, \citenamefont {Torres}, \citenamefont
  {Schwartzentruber},\ and\ \citenamefont
  {Candler}}]{chaudhry2020implementation}%
  \BibitemOpen
  \bibfield  {author} {\bibinfo {author} {\bibfnamefont {R.~S.}\ \bibnamefont
  {Chaudhry}}, \bibinfo {author} {\bibfnamefont {I.~D.}\ \bibnamefont {Boyd}},
  \bibinfo {author} {\bibfnamefont {E.}~\bibnamefont {Torres}}, \bibinfo
  {author} {\bibfnamefont {T.~E.}\ \bibnamefont {Schwartzentruber}}, \ and\
  \bibinfo {author} {\bibfnamefont {G.~V.}\ \bibnamefont {Candler}},\
  }\bibfield  {title} {\enquote {\bibinfo {title} {Implementation of a chemical
  kinetics model for hypersonic flows in air for high-performance cfd},}\ }in\
  \href@noop {} {\emph {\bibinfo {booktitle} {AIAA Scitech 2020 Forum}}}\
  (\bibinfo {year} {2020})\ p.\ \bibinfo {pages} {2191}\BibitemShut {NoStop}%
\bibitem [{\citenamefont {Chaudhry}, \citenamefont {Boyd},\ and\ \citenamefont
  {Candler}(2020)}]{chaudhry2020vehicle}%
  \BibitemOpen
  \bibfield  {author} {\bibinfo {author} {\bibfnamefont {R.~S.}\ \bibnamefont
  {Chaudhry}}, \bibinfo {author} {\bibfnamefont {I.~D.}\ \bibnamefont {Boyd}},
  \ and\ \bibinfo {author} {\bibfnamefont {G.~V.}\ \bibnamefont {Candler}},\
  }\bibfield  {title} {\enquote {\bibinfo {title} {Vehicle-scale simulations of
  hypersonic flows using the mmt chemical kinetics model},}\ }in\ \href@noop {}
  {\emph {\bibinfo {booktitle} {AIAA AVIATION 2020 FORUM}}}\ (\bibinfo {year}
  {2020})\ p.\ \bibinfo {pages} {3272}\BibitemShut {NoStop}%
\bibitem [{\citenamefont {Levine}(1978)}]{levine1978information}%
  \BibitemOpen
  \bibfield  {author} {\bibinfo {author} {\bibfnamefont {R.}~\bibnamefont
  {Levine}},\ }\bibfield  {title} {\enquote {\bibinfo {title} {Information
  theory approach to molecular reaction dynamics},}\ }\href@noop {} {\bibfield
  {journal} {\bibinfo  {journal} {Annual Review of Physical Chemistry}\
  }\textbf {\bibinfo {volume} {29}},\ \bibinfo {pages} {59--92} (\bibinfo
  {year} {1978})}\BibitemShut {NoStop}%
\bibitem [{\citenamefont {Levine}(2009)}]{levine2009molecular}%
  \BibitemOpen
  \bibfield  {author} {\bibinfo {author} {\bibfnamefont {R.~D.}\ \bibnamefont
  {Levine}},\ }\href@noop {} {\emph {\bibinfo {title} {Molecular reaction
  dynamics}}}\ (\bibinfo  {publisher} {Cambridge University Press},\ \bibinfo
  {year} {2009})\BibitemShut {NoStop}%
\bibitem [{\citenamefont {Levine}\ and\ \citenamefont
  {Bernstein}(1971)}]{levine1971collision}%
  \BibitemOpen
  \bibfield  {author} {\bibinfo {author} {\bibfnamefont {R.~D.}\ \bibnamefont
  {Levine}}\ and\ \bibinfo {author} {\bibfnamefont {R.~B.}\ \bibnamefont
  {Bernstein}},\ }\bibfield  {title} {\enquote {\bibinfo {title}
  {Collision-induced dissociation: A simplistic optical model analysis},}\
  }\href@noop {} {\bibfield  {journal} {\bibinfo  {journal} {{C}hemical Physics
  Letters}\ }\textbf {\bibinfo {volume} {11}},\ \bibinfo {pages} {552--556}
  (\bibinfo {year} {1971})}\BibitemShut {NoStop}%
\bibitem [{\citenamefont {Macdonald}\ \emph
  {et~al.}(2018{\natexlab{b}})\citenamefont {Macdonald}, \citenamefont
  {Grover}, \citenamefont {Schwartzentruber},\ and\ \citenamefont
  {Panesi}}]{macdonald2018construction_DMS}%
  \BibitemOpen
  \bibfield  {author} {\bibinfo {author} {\bibfnamefont {R.}~\bibnamefont
  {Macdonald}}, \bibinfo {author} {\bibfnamefont {M.}~\bibnamefont {Grover}},
  \bibinfo {author} {\bibfnamefont {T.}~\bibnamefont {Schwartzentruber}}, \
  and\ \bibinfo {author} {\bibfnamefont {M.}~\bibnamefont {Panesi}},\
  }\bibfield  {title} {\enquote {\bibinfo {title} {Construction of a
  coarse-grain quasi-classical trajectory method. ii. comparison against the
  direct molecular simulation method},}\ }\href@noop {} {\bibfield  {journal}
  {\bibinfo  {journal} {The Journal of {C}hemical physics}\ }\textbf {\bibinfo
  {volume} {148}},\ \bibinfo {pages} {054310} (\bibinfo {year}
  {2018}{\natexlab{b}})}\BibitemShut {NoStop}%
\bibitem [{\citenamefont {LEE}(1984)}]{Lee1984}%
  \BibitemOpen
  \bibfield  {author} {\bibinfo {author} {\bibfnamefont {J.}~\bibnamefont
  {LEE}},\ }\bibfield  {title} {\enquote {\bibinfo {title} {Basic governing
  equations for the flight regimes of aeroassisted orbital transfer
  vehicles},}\ \ }(\bibinfo {year} {1984})\ p.\ \bibinfo {pages}
  {1729}\BibitemShut {NoStop}%
\bibitem [{\citenamefont {Jaffe}(1987)}]{jaffe1987}%
  \BibitemOpen
  \bibfield  {author} {\bibinfo {author} {\bibfnamefont {R.~L.}\ \bibnamefont
  {Jaffe}},\ }\bibfield  {title} {\enquote {\bibinfo {title} {The calculation
  of high-temperature equilibrium and nonequilibrium specific heat data for
  {N2}, {O2} and {NO}},}\ }in\ \href@noop {} {\emph {\bibinfo {booktitle} {22nd
  AIAA Thermophysics Conference Proceedings, Honolulu, Hawaii}}},\ Vol.\
  \bibinfo {volume} {-1633}\ (\bibinfo {year} {1987})\BibitemShut {NoStop}%
\bibitem [{\citenamefont {Tolman}(1979)}]{tolman1979principles}%
  \BibitemOpen
  \bibfield  {author} {\bibinfo {author} {\bibfnamefont {R.~C.}\ \bibnamefont
  {Tolman}},\ }\href@noop {} {\emph {\bibinfo {title} {The principles of
  statistical mechanics}}}\ (\bibinfo  {publisher} {Courier Corporation},\
  \bibinfo {year} {1979})\BibitemShut {NoStop}%
\bibitem [{\citenamefont {Lewis}(1925)}]{lewis1925new}%
  \BibitemOpen
  \bibfield  {author} {\bibinfo {author} {\bibfnamefont {G.~N.}\ \bibnamefont
  {Lewis}},\ }\bibfield  {title} {\enquote {\bibinfo {title} {A new principle
  of equilibrium},}\ }\href@noop {} {\bibfield  {journal} {\bibinfo  {journal}
  {Proceedings of the National Academy of Sciences of the United States of
  America}\ }\textbf {\bibinfo {volume} {11}},\ \bibinfo {pages} {179}
  (\bibinfo {year} {1925})}\BibitemShut {NoStop}%
\bibitem [{\citenamefont {Olejniczak}\ and\ \citenamefont
  {Candler}(1995)}]{olejniczak1995vibrational}%
  \BibitemOpen
  \bibfield  {author} {\bibinfo {author} {\bibfnamefont {J.}~\bibnamefont
  {Olejniczak}}\ and\ \bibinfo {author} {\bibfnamefont {G.~V.}\ \bibnamefont
  {Candler}},\ }\bibfield  {title} {\enquote {\bibinfo {title} {Vibrational
  energy conservation with vibration--dissociation coupling: General theory and
  numerical studies},}\ }\href@noop {} {\bibfield  {journal} {\bibinfo
  {journal} {Physics of Fluids}\ }\textbf {\bibinfo {volume} {7}},\ \bibinfo
  {pages} {1764--1774} (\bibinfo {year} {1995})}\BibitemShut {NoStop}%
\bibitem [{\citenamefont {Gimelshein}\ and\ \citenamefont
  {Wysong}(2017)}]{gimelshein2017dsmc}%
  \BibitemOpen
  \bibfield  {author} {\bibinfo {author} {\bibfnamefont {S.}~\bibnamefont
  {Gimelshein}}\ and\ \bibinfo {author} {\bibfnamefont {I.}~\bibnamefont
  {Wysong}},\ }\bibfield  {title} {\enquote {\bibinfo {title} {Dsmc modeling of
  flows with recombination reactions},}\ }\href@noop {} {\bibfield  {journal}
  {\bibinfo  {journal} {Physics of Fluids}\ }\textbf {\bibinfo {volume} {29}},\
  \bibinfo {pages} {067106} (\bibinfo {year} {2017})}\BibitemShut {NoStop}%
\bibitem [{\citenamefont {Boyd}\ and\ \citenamefont
  {Schwartzentruber}(2017)}]{boyd2017nonequilibrium}%
  \BibitemOpen
  \bibfield  {author} {\bibinfo {author} {\bibfnamefont {I.~D.}\ \bibnamefont
  {Boyd}}\ and\ \bibinfo {author} {\bibfnamefont {T.~E.}\ \bibnamefont
  {Schwartzentruber}},\ }\href@noop {} {\emph {\bibinfo {title} {Nonequilibrium
  Gas Dynamics and Molecular Simulation}}}\ (\bibinfo  {publisher} {Cambridge
  University Press},\ \bibinfo {year} {2017})\BibitemShut {NoStop}%
\bibitem [{Note1()}]{Note1}%
  \BibitemOpen
  \bibinfo {note} {$\epsilon _{v} \propto v $ and $\epsilon _{rot} \propto
  j(j+1)$}\BibitemShut {NoStop}%
\bibitem [{\citenamefont {Bender}\ \emph {et~al.}(2015)\citenamefont {Bender},
  \citenamefont {Valentini}, \citenamefont {Nompelis}, \citenamefont {Paukku},
  \citenamefont {Varga}, \citenamefont {Truhlar}, \citenamefont
  {Schwartzentruber},\ and\ \citenamefont {Candler}}]{bender2015improved}%
  \BibitemOpen
  \bibfield  {author} {\bibinfo {author} {\bibfnamefont {J.~D.}\ \bibnamefont
  {Bender}}, \bibinfo {author} {\bibfnamefont {P.}~\bibnamefont {Valentini}},
  \bibinfo {author} {\bibfnamefont {I.}~\bibnamefont {Nompelis}}, \bibinfo
  {author} {\bibfnamefont {Y.}~\bibnamefont {Paukku}}, \bibinfo {author}
  {\bibfnamefont {Z.}~\bibnamefont {Varga}}, \bibinfo {author} {\bibfnamefont
  {D.~G.}\ \bibnamefont {Truhlar}}, \bibinfo {author} {\bibfnamefont
  {T.}~\bibnamefont {Schwartzentruber}}, \ and\ \bibinfo {author}
  {\bibfnamefont {G.~V.}\ \bibnamefont {Candler}},\ }\bibfield  {title}
  {\enquote {\bibinfo {title} {An improved potential energy surface and
  multi-temperature quasiclassical trajectory calculations of {N2}+ {N2}
  dissociation reactions},}\ }\href@noop {} {\bibfield  {journal} {\bibinfo
  {journal} {The Journal of {C}hemical physics}\ }\textbf {\bibinfo {volume}
  {143}},\ \bibinfo {pages} {054304} (\bibinfo {year} {2015})}\BibitemShut
  {NoStop}%
\bibitem [{\citenamefont {Rubin}\ and\ \citenamefont
  {Shuler}(1956{\natexlab{a}})}]{rubin1956relaxation}%
  \BibitemOpen
  \bibfield  {author} {\bibinfo {author} {\bibfnamefont {R.~J.}\ \bibnamefont
  {Rubin}}\ and\ \bibinfo {author} {\bibfnamefont {K.~E.}\ \bibnamefont
  {Shuler}},\ }\bibfield  {title} {\enquote {\bibinfo {title} {Relaxation of
  vibrational nonequilibrium distributions. {I}. collisional relaxation of a
  system of harmonic oscillators},}\ }\href@noop {} {\bibfield  {journal}
  {\bibinfo  {journal} {The Journal of {C}hemical Physics}\ }\textbf {\bibinfo
  {volume} {25}},\ \bibinfo {pages} {59--67} (\bibinfo {year}
  {1956}{\natexlab{a}})}\BibitemShut {NoStop}%
\bibitem [{\citenamefont {Rubin}\ and\ \citenamefont
  {Shuler}(1956{\natexlab{b}})}]{rubin1956relaxationprob}%
  \BibitemOpen
  \bibfield  {author} {\bibinfo {author} {\bibfnamefont {R.~J.}\ \bibnamefont
  {Rubin}}\ and\ \bibinfo {author} {\bibfnamefont {K.~E.}\ \bibnamefont
  {Shuler}},\ }\bibfield  {title} {\enquote {\bibinfo {title} {Relaxation of
  vibrational nonequilibrium distributions. ii. the effect of the collisional
  transition probabilities on the relaxation behavior},}\ }\href@noop {}
  {\bibfield  {journal} {\bibinfo  {journal} {The Journal of {C}hemical
  Physics}\ }\textbf {\bibinfo {volume} {25}},\ \bibinfo {pages} {68--74}
  (\bibinfo {year} {1956}{\natexlab{b}})}\BibitemShut {NoStop}%
\bibitem [{\citenamefont {Bazley}\ \emph {et~al.}(1958)\citenamefont {Bazley},
  \citenamefont {Montroll}, \citenamefont {Rubin},\ and\ \citenamefont
  {Shuler}}]{bazley1958studies}%
  \BibitemOpen
  \bibfield  {author} {\bibinfo {author} {\bibfnamefont {N.~W.}\ \bibnamefont
  {Bazley}}, \bibinfo {author} {\bibfnamefont {E.~W.}\ \bibnamefont
  {Montroll}}, \bibinfo {author} {\bibfnamefont {R.~J.}\ \bibnamefont {Rubin}},
  \ and\ \bibinfo {author} {\bibfnamefont {K.~E.}\ \bibnamefont {Shuler}},\
  }\bibfield  {title} {\enquote {\bibinfo {title} {Studies in nonequilibrium
  rate processes. iii. the vibrational relaxation of a system of anharmonic
  oscillators},}\ }\href@noop {} {\bibfield  {journal} {\bibinfo  {journal}
  {The Journal of {C}hemical Physics}\ }\textbf {\bibinfo {volume} {28}},\
  \bibinfo {pages} {700--704} (\bibinfo {year} {1958})}\BibitemShut {NoStop}%
\bibitem [{\citenamefont {Montroll}\ and\ \citenamefont
  {Shuler}(1957)}]{montroll1957application}%
  \BibitemOpen
  \bibfield  {author} {\bibinfo {author} {\bibfnamefont {E.~W.}\ \bibnamefont
  {Montroll}}\ and\ \bibinfo {author} {\bibfnamefont {K.~E.}\ \bibnamefont
  {Shuler}},\ }\bibfield  {title} {\enquote {\bibinfo {title} {The application
  of the theory of stochastic processes to chemical kinetics},}\ }\href@noop {}
  {\bibfield  {journal} {\bibinfo  {journal} {Advances in {C}hemical Physics}\
  ,\ \bibinfo {pages} {361--399}} (\bibinfo {year} {1957})}\BibitemShut
  {NoStop}%
\bibitem [{Note2()}]{Note2}%
  \BibitemOpen
  \bibinfo {note} {At very high temperature and for higher quantum state this
  approximation may be inaccurate.}\BibitemShut {Stop}%
\bibitem [{Note3()}]{Note3}%
  \BibitemOpen
  \bibinfo {note} {The differences i.e. $\protect \langle \epsilon _v \protect
  \rangle ^* \protect \neq \protect \langle \epsilon _{rot} \protect \rangle ^*
  \protect \neq k_B T$ are the consequence of the diatomic energy variation
  with quantum state and the chosen vibration-prioritized \cite {jaffe1987}
  framework of separating internal energy into vibration and rotational
  mode.}\BibitemShut {Stop}%
\bibitem [{Note4()}]{Note4}%
  \BibitemOpen
  \bibinfo {note} {The focus of DMS so far has been to study the dissociation
  dominated conditions, observed in the regions immediately behind the shock
  waves. The addition of recombination in DMS for expanding flow studies is an
  ongoing effort.}\BibitemShut {Stop}%
\bibitem [{Note5()}]{Note5}%
  \BibitemOpen
  \bibinfo {note} {The distinction between thermal and chemical equilibrium is
  not meaningful when the chemical processes are coupled to thermal
  evolution.}\BibitemShut {Stop}%
\bibitem [{\citenamefont {Millikan}\ and\ \citenamefont
  {White}(1963)}]{millikan1963systematics}%
  \BibitemOpen
  \bibfield  {author} {\bibinfo {author} {\bibfnamefont {R.~C.}\ \bibnamefont
  {Millikan}}\ and\ \bibinfo {author} {\bibfnamefont {D.~R.}\ \bibnamefont
  {White}},\ }\bibfield  {title} {\enquote {\bibinfo {title} {Systematics of
  vibrational relaxation},}\ }\href@noop {} {\bibfield  {journal} {\bibinfo
  {journal} {The Journal of {C}hemical physics}\ }\textbf {\bibinfo {volume}
  {39}},\ \bibinfo {pages} {3209--3213} (\bibinfo {year} {1963})}\BibitemShut
  {NoStop}%
\end{thebibliography}%

\end{document}